\pdfoutput=1
\documentclass[a4paper,11pt]{article}

\usepackage[left=2.5cm, right=2.5cm, top=3.0cm, bottom=3.0cm]{geometry}

\usepackage{amsmath,amssymb,amsfonts,mathrsfs}

\usepackage{todonotes}
\usepackage{tikz}
\usepackage{tikz-cd}
\usetikzlibrary{cd}
\usepackage[all]{xy}
\usetikzlibrary{shapes,arrows}
\usetikzlibrary{shapes}
\usetikzlibrary{plotmarks}
\usepackage{tikz-3dplot}
\usetikzlibrary{calc}

\usepackage{jheppub}
\usepackage{verbatim}

\def\bZ{\mathbb{Z}}
\def\bR{\mathbb{R}}

\def\cC{\mathcal{C}}

\def\cN{\mathcal{N}}

\def\cC{\mathcal{C}}
\def\cQ{\mathcal{Q}}
\def\cT{\mathcal{T}}
\def\cP{\mathcal{P}}
\def\cZ{\mathcal{Z}}
\def\cU{\mathcal{U}}

\def\MCG{\mathrm{MCG}}

\def\tla{\Tilde{a}}
\def\tlb{\Tilde{b}}

\makeatletter
\newcommand\xleftrightarrow[2][]{%
  \ext@arrow 9999{\longleftrightarrowfill@}{#1}{#2}}
\newcommand\longleftrightarrowfill@{%
  \arrowfill@\leftarrow\relbar\rightarrow}
\makeatother

\def\centerarc[#1](#2)(#3:#4:#5)
    { \draw[#1] ($(#2)+({#5*cos(#3)},{#5*sin(#3)})$) arc (#3:#4:#5); }

\numberwithin{equation}{section} 

\title{Non-Invertible Surface Defects in 2+1d QFTs from Half Spacetime Gauging}
\author[1,2]{Wei Cui}
\author[2,1]{Babak Haghighat}
\author[2]{and Lorenzo Ruggeri}
\affiliation[1]{Beijing Institute of Mathematical Sciences and Applications (BIMSA), Huairou District, Beijing 101408, China}
\affiliation[2]{Yau Mathematical Sciences Center, Tsinghua University, Beijing, 100084, China}
\emailAdd{cwei@bimsa.cn}
\emailAdd{babakhaghighat@tsinghua.edu.cn}
\emailAdd{ruggeri@mail.tsinghua.edu.cn}
\preprint{}

\abstract{
We study duality defects in 2+1d theories with $\bZ^{(0)}_N\times\bZ^{(1)}_N$ global symmetry and trivial mixed 't Hooft anomaly. By gauging these symmetries simultaneously in half of the spacetime, we define duality defects for theories that are self-dual under gauging. We calculate the fusion rules involving duality defects and show that they obey a fusion 2-category. We also construct the corresponding symmetry topological field theory, obtained from a four-dimensional BF theory gauging a $\mathbb{Z}_4^\text{EM}$ electric-magnetic symmetry. Furthermore, we provide explicit examples of such duality defects in $U(1)\times U(1)$ gauge theories and in more general product theories. Finally, we find duality defects in non-Lagrangian theories obtained by compactification of 6d $\cN=(2,0)$ SCFTs of type $A_{N-1}$ on various three-manifolds.
}

\addtocontents{toc}{\protect\setcounter{tocdepth}{2}}

\begin{document}

\maketitle
\flushbottom

\section{Introduction}

The modern perspective on global symmetries in quantum field theories (QFTs) is that to each global symmetry there exists a topological symmetry defect associated to it \cite{Gaiotto:2014kfa}. In this language, ordinary ($0$-form) symmetries are generated by codimension-1 topological symmetry defects acting on local operators while, in general, $p$-form symmetries are generated by a set of codimension-$(p+1)$ topological defects acting on extended operators. See \cite{McGreevy:2022oyu,Cordova:2022ruw,Brennan:2023mmt,Bhardwaj:2023kri,Schafer-Nameki:2023jdn,Luo:2023ive,Shao:2023gho,Carqueville:2023jhb} for comprehensive reviews. These topological defects can fuse among each other and give rise to new defects. If the fusion rules follow a group structure, the symmetry associated with the defects is invertible. In general, however, the fusion of two defects can lead to a linear combination of more than one defect and certain symmetry defects may have no inverse. Topological defects with such fusion rules describe symmetries that are \emph{non-invertible} \cite{Verlinde:1988sn,Moore:1988qv,Moore:1989yh, Bhardwaj:2017xup,Chang:2018iay,Thorngren:2019iar,Komargodski:2020mxz,Thorngren:2021yso,Koide:2021zxj,Choi:2021kmx,Kaidi:2021xfk,Choi:2022zal,Arias-Tamargo:2022nlf,Hayashi:2022fkw,Roumpedakis:2022aik,Kaidi:2022uux,Choi:2022jqy,Cordova:2022ieu,Antinucci:2022eat,Bashmakov:2022jtl,Damia:2022rxw,Damia:2022bcd,Choi:2022rfe,Lu:2022ver,Bhardwaj:2022lsg,Lin:2022xod,Apruzzi:2022rei,GarciaEtxebarria:2022vzq, Benini:2022hzx, Wang:2021vki, Chen:2021xuc, DelZotto:2022ras,Bhardwaj:2022dyt,Brennan:2022tyl,Delmastro:2022pfo, Heckman:2022muc,Freed:2022qnc,Freed:2022iao,Niro:2022ctq,Mekareeya:2022spm,Antinucci:2022vyk,Chen:2022cyw,Karasik:2022kkq,Cordova:2022fhg,Decoppet:2022dnz,GarciaEtxebarria:2022jky,Choi:2022fgx,Yokokura:2022alv,Bhardwaj:2022kot,Bhardwaj:2022maz, Hsin:2022heo,Heckman:2022xgu,Antinucci:2022cdi,Apte:2022xtu,Garcia-Valdecasas:2023mis, Delcamp:2023kew, Bhardwaj:2023zix, Yu:2023nyn,Lawrie:2023tdz,Santilli:2024dyz,Perez-Lona:2023djo,Arbalestrier:2024oqg,Copetti:2024rqj,Li:2023knf,Braeger:2024jcj,Baume:2023kkf,Heckman:2024oot,Heckman:2024obe,Apruzzi:2024htg,Bhardwaj:2024kvy,Bhardwaj:2024wlr,Bhardwaj:2024qrf,Bhardwaj:2023bbf,Bhardwaj:2023fca,Bhardwaj:2023idu,Apruzzi:2023uma,Diatlyk:2023fwf,Damia:2023ses,Antinucci:2024zjp,Argurio:2024oym,Liu:2024znj,Ambrosino:2024ggh,Okada:2024qmk,Kaidi:2024wio,Bonetti:2024cjk,DelZotto:2024tae,Hasan:2024aow,Cordova:2024vsq,Antinucci:2023ezl,Copetti:2023mcq,Carqueville:2023jhb,Cordova:2023her,Cordova:2023jip,Cordova:2023qei,Cordova:2024ypu,Sun:2023xxv,Nardoni:2024sos,Brennan:2024fgj,Cordova:2023bja,Choi:2023vgk,Seifnashri:2024dsd,Seiberg:2024gek,Choi:2023pdp,Choi:2023xjw,Choi:2024rjm}.  

Non-invertible symmetries have been known for many years in the realm of 2d rational CFTs \cite{Verlinde:1988sn,Moore:1988qv,Moore:1989yh}. There, global symmetries are described by topological defect lines (TDLs) whose fusion rules are not described by group theory but rather by a fusion category. A typical example is the 2d Ising CFT where there is an invertible TDL generating a $\bZ_2$ symmetry and a non-invertible TDL realizing Kramers-Wannier duality \cite{Verlinde:1988sn,Aasen:2016dop,Freed:2018cec}. The latter can be understood as a topological interface separating the Ising model and its $\bZ_2$ gauged dual. Recently, this idea has been generalized to four dimensions \cite{Kaidi:2021xfk, Choi:2021kmx, Choi:2022zal} by gauging on half of the spacetime a $\mathbb{Z}_N^{(1)}$ symmetry. Examples studied so far include Maxwell theory and $\mathcal{N}=4$ $SU(2)$ super Yang-Mills. Sometimes a theory can be further stacked with an SPT phase which leads to different ways, related by discrete torsion, of gauging a symmetry. This has allowed to define triality and, more in general, $N$-ality defects.

An important tool in the study of higher form symmetries of a given $d$-dimensional theory $\mathcal{Q}$ is the \emph{symmetry topological field theory} (SymTFT), a $(d+1)$-dimensional TFT which captures the global forms, symmetries and 't Hooft anomalies of $\mathcal{Q}$ \cite{Freed:2012bs,Freed:2018cec,Gaiotto:2020iye,Apruzzi:2021nmk, Apruzzi:2022dlm,Burbano:2021loy,Freed:2022qnc,vanBeest:2022fss, Kaidi:2022cpf, Bashmakov:2022uek, Kaidi:2023maf, Lan:2018vjb,Kong:2019brm,Kong:2020cie,Kong:2020wmn,Kong:2020iek, Zhao:2022yaw,Kong:2024ykr}. A SymTFT is defined on a slab with topological boundary conditions at one end and non-topological boundary conditions capturing the dynamics of $\mathcal{Q}$ at the other end. Crucially, one can find all global variants, obtained acting with topological manipulations on $\mathcal{Q}$, upon choosing different topological boundary conditions and shrinking the slab. This implies that the SymTFT is invariant under topological manipulations (for example it is invariant under modular transformations for 2d theories).

Let us now discuss with more details the construction of a duality defect via gauging on half of the spacetime \cite{Kaidi:2021xfk, Choi:2021kmx, Choi:2022zal}. For a theory $\cQ$ with non-anomalous symmetry $\bZ^{(p)}_N$, gauging it in half of the spacetime creates a topological interface between $\cQ$ and $\cQ/\bZ^{(p)}_N$ with Dirichlet boundary conditions for $\bZ^{(p)}_N$ gauge fields on the interface. One finds a \emph{duality defect} if the theory is self-dual under such gauging, i.e. there is an isomorphism between the two theories $\cQ\simeq\cQ/\bZ^{(p)}_N$. In general, gauging $\bZ^{(p)}_N$ will induce a quantum symmetry $\widehat{\bZ}^{(d-p-2)}_N$ in the theory $\cQ/\bZ^{(p)}_N$. To satisfy the self-duality condition, one needs to at least match the global symmetries between the two theories, which constrains $p=\frac{d-2}{2}$. This constrains to even dimension the possibility of self-duality under gauging of a single $\bZ^{(p)}_N$ symmetry.

The goal of the current paper is to extend the construction of non-invertible duality and triality defects via half spacetime gauging to three-dimensional theories\footnote{The construction of invertible surface defects in 3d TFTs was initiated in the pioneering work \cite{Kapustin:2010hk}. Besides the duality defects studied here via half spacetime gauging, non-invertible symmetries in 3d theories have appeared previously. For example non-invertible 1-form symmetries generated by line operators, or anyons, have been extensively studied for 3d TQFTs \cite{Witten:1988hf}. More recently, they have been considered via higher gauging \cite{Roumpedakis:2022aik} and in theories with a 2-group symmetry \cite{Damia:2022seq}. Finally, as we will discus below, non-invertible 0-form symmetries in general 2+1d QFTs have recently been studied via gauging invertible symmetries with mixed anomalies \cite{Kaidi:2021xfk, Mekareeya:2022spm}.}. Following the discussion in the previous paragraph, a first requirement is that a theory needs to have at least two symmetries, a 1-form symmetry which is mapped to a 0-form symmetry under gauging and a 0-form symmetry which is mapped to a 1-form symmetry. Starting from a 3d theory $\cQ$ with $\bZ_N^{(0)}\times \bZ_N^{(1)}$ global symmetry and trivial mixed 't Hooft anomaly, and subsequently gauging $\bZ_N^{(0)}\times \bZ_N^{(1)}$, the gauged theory has a new quantum symmetry $\widehat{\bZ}_N^{(1)}\times \widehat{\bZ}_N^{(0)}$. This defines a topological interface between two theories which are, a priori, inequivalent. Whenever $\mathcal{Q}$ is self-dual, the gauging procedure produces a non-invertible duality defect. Stacking with an SPT phase it is possible to obtain also triality defects.

Duality defects in 3d have been discussed in \cite{Kaidi:2021xfk} for theories with $\bZ_{2,1}^{(0)}\times\bZ_{2,2}^{(0)}\times \bZ_2^{(1)}$ symmetry and mixed 't Hooft anomaly among them, by gauging an anomaly-free $\bZ_{2,1}^{(0)}\times \bZ_2^{(1)}$ subgroup. After gauging, the symmetry $\mathbb{Z}_{2,2}^{(0)}$ needs to be stacked with a non-invertible TFT to preserve gauge invariance and consequently it becomes a non-invertible symmetry. Such defects are denoted \emph{non-intrinsic} non-invertible\footnote{Often in the literature they are denoted group theoretical \cite{gelaki2009centers}.} as they can be related to invertible defects in a theory obtained via a topological manipulation \cite{Sun:2023xxv}. Instead, in our setup, the duality and triality defects we study can in principle be intrinsically non-invertible defects as we do not rely on the existence of a mixed 't Hooft anomaly. We furthermore determine the complete fusion rules involving both duality and triality defects together with invertible line and surface defects generating the $\bZ_N^{(1)}\times\bZ_N^{(0)}$ symmetry. These fusion rules are naturally described by a fusion 2-category \cite{Inamura:2023qzl}, the Tambara-Yamagami fusion 2-category \cite{Decoppet:2023bay} which we denote as $\text{TY}(\bZ^{(0)}_N\times \bZ^{(1)}_N )$. We believe our construction can be applied also to five-dimensional theories with $\bZ_N^{(1)}\times \bZ_N^{(2)}$ global symmetry and which are self-dual under gauging.

We also study the corresponding four-dimensional SymTFT and, first without assuming invariance of the theory $\mathcal{Q}$ under gauging, we find that it is described by a 4d BF theory with two copies of $\bZ_N^{(1)} \times \bZ_N^{(2)}$ as global symmetries. While Dirichlet boundary conditions give rise to a theory $\cQ$ upon shrinking the slab, Neumann boundary conditions give rise to $\mathcal{Q}/(\bZ_N^{(0)} \times \bZ_N^{(1)})$. The BF theory has an electric-magnetic exchange symmetry\footnote{For $N=2$ the electric-magnetic symmetry is $\mathbb{Z}_2^\text{EM}$.} $\mathbb{Z}_4^\text{EM}$ implemented by a \emph{condensation defect} which precisely exchanges between the two boundary conditions when acting on the topological boundary. Following \cite{Kaidi:2022cpf}, this defect is constructed by one-gauging $\bZ_N^{(1)}\times\bZ_N^{(2)}$ on a three-dimensional hypersurface in the 4d bulk. We also define \emph{twist defects}, obtained considering condensation defects on hypersurfaces with boundaries and imposing Dirichlet boundary conditions. This creates a higher duality interface, that is a duality interface associated with one-gauging. Placing the twist defect parallel to the boundaries and shrinking the slab, we find an interface in the three-dimensional theory $\mathcal{Q}$. 

To find the SymTFT describing duality defects in the three-dimensional theory, we need to gauge the $\mathbb{Z}_4^\text{EM}$ symmetry in the BF theory. Hence, the bulk of the twist defects becomes transparent and its boundary becomes a codimension-2 symmetry operator. Upon shrinking the slab, it gives rise to a codimension-1 duality defect in the three-dimensional theory $\mathcal{Q}$. Since the symmetry on the boundary is given by a fusion 2-category, we expect this SymTFT to be a Douglas-Reutter TQFT \cite{douglas2018fusion}.

We also provide explicit examples of 3d theories admitting  duality defects. First, we consider a $U(1)\times U(1)$ gauge theory which has two copies of the electric one-form symmetry $U(1)^{(1)}_1\times U(1)^{(1)}_2$ and of the magnetic zero-form symmetry $U(1)^{(0)}_1\times U(1)^{(0)}_2$. Moreover, the theory admits an $SL(2,\bZ)$ duality acting on the coupling constants of the two $U(1)$ gauge factors. Hence, the duality defect is obtained gauging a $\mathbb{Z}^{(1)}_{N,1}\times \mathbb{Z}_{N,2}^{(0)}$ subgroup of the electric and magnetic symmetries and acting with the $SL(2,\bZ)$ duality. We also find the worldsheet action for this duality defect, which we use to explicitly reproduce the fusion rules. More generally, starting from a theory $\mathcal{Q}$ with a symmetry $\mathbb{Z}_N^{(0)}$ without 't Hooft anomaly, the \emph{product theory} $\mathcal{T}=\mathcal{Q}\times(\mathcal{Q}/\mathbb{Z}_N^{(0)})$ always admits a duality defect. Furthermore, for duality defects preserved along the RG flow, we give, as an application, constraints on the existence of trivially gapped theories. 

Another class of examples is given by compactification of six-dimensional superconformal field theories (SCFTs) of type $A_{N-1}$ on a three-dimensional manifold $X_3$, along the lines of the recent studies of compactifications on 2-manifolds \cite{Tachikawa:2013hya,Bashmakov:2022jtl,Bashmakov:2022uek} and on 4-manifolds \cite{Chen:2022vvd,Bashmakov:2023kwo}. Twisted compactifications on $X_3$ lead to a large class of three-dimensional theories, which we denote by $T_N[X_3]$. Whenever $X_3$ has nontrivial $H_1(X_3,\bZ_N)$ and $H_2(X_3,\bZ_N)$, $T_N[X_3]$ is a relative field theory \cite{Witten:1998wy,Witten:2009at,Freed:2012bs,Tachikawa:2013hya} which requires a choice of polarization $\Lambda$ to define an absolute theory. In this paper we mostly focus on $N=p$ prime and consider three-manifolds without torsional one/two-cycles, as $S^2\times S^1$, connected sums of $S^2\times S^1$ and $T^3$. Different choices of polarizations $\Lambda$ give rise to different absolute theories labeled by $T_{N,\Lambda}[X_3]$ which generally have different symmetries. Gauging a subgroup of these symmetries transforms absolute theories into each other. Similar to class S theories, the mapping class group (MCG) of $X_3$ implies non-trivial dualities in the three-dimensional theories. As global forms $T_{p,\Lambda}[X_3]$ have non-trivial couplings parametrized by geometric moduli of $X_3$, we study how the absolute theories and the couplings transform under gauging of $\bZ_p^{(0)}\times\bZ_p^{(1)}$ and under dualities associated to $\MCG(X_3)$. In some cases, we find that combining gauging and duality keeps a given absolute theory invariant, thus realizing a non-invertible duality defect in $T_{p,\Lambda}[X_3]$. This allows us to construct duality defects for several non-Lagrangian theories. 


The organization of this paper is as follows. In \autoref{sec.2}, we define both the duality defect obtained by gauging $\bZ_N^{(0)} \times \bZ_N^{(1)}$ in half of the spacetime and the triality defect obtained by gauging and stacking with an SPT phase. We also compute the fusion rules of these defects. The (3+1)d SymTFT describing the duality defects is introduced in \autoref{sec.3}. In \autoref{sec.4}, we provide examples of theories admitting duality defects, such as the $U(1)\times U(1)$ gauge theory and more general product theories. Moreover, we discuss constraints imposed by the existence of duality defects on trivially gapped phases. In \autoref{sec.5}, we study the absolute theories, symmetries and, in particular, duality defects in 3d QFTs obtained by compactification of 6d $\cN=(2,0)$ SCFTs of type $A_{N-1}$ on various 3-manifolds. Finally, in \autoref{sec.6}, we summarize our results and highlight possible future directions.

\paragraph*{Note added:} During the preparation of this work, a preprint \cite{Choi:2024rjm} appeared whose intent is also to study duality defects in three-dimensional theories via half spacetime gauging. Despite their focus being mainly on 2+1d lattice gauge theories, they have comments about the continuum limit and they mention the product theories we discussed in the introduction.

\section{Non-Invertible Topological Surface Defects}\label{sec.2}

In this section, we consider a (2+1)d QFTs theory $\mathcal{Q}$ and study duality and triality defects defined by gauging simultaneously a 0-form symmetry $\bZ^{(0)}_N$ and a 1-form symmetry $\bZ^{(1)}_N$ in half of the spacetime. For theories $\mathcal{Q}$ that are invariant under such gauging, we determine the fusion rules involving duality defect and triality defect together with the other invertible topological operators in the theory.

\subsection{Duality Defects}

Consider a 3d quantum field theory $\cQ$ on an orientable three manifold $M_3$ with 0-form discrete symmetry $\bZ^{(0)}_N$ and 1-form symmetry $\bZ^{(1)}_N$. We will assume that there are no 't Hooft anomaly for both $\bZ^{(0)}_N$, $\bZ^{(1)}_N$ and that there is no mixed 't Hooft anomaly between them. Similar to the previous works in even dimensions \cite{Choi:2021kmx,Choi:2022zal}, by gauging $\bZ^{(0)}_N \times \bZ^{(1)}_N$ on half spacetime of $M_3$, one can construct duality defects when $\cQ$ is self-dual\footnote{Notice that $\cQ$ cannot be self-dual under gauging, separately, either a 0- or a 1-form symmetry as the gauged theory has different global symmetries than the ungauged one.}.

Let $Z_{\cQ}[A,B]$ be the partition function on $M_3$ with $A$ and $B$ the background gauge fields for $\bZ^{(0)}_N $ and $ \bZ^{(1)}_N$. After gauging $\mathbb{Z}_N^{(0)}\times\mathbb{Z}^{(1)}_N$, an operation denoted by $\sigma\mathcal{Q}\equiv\mathcal{Q}/(\mathbb{Z}_N^{(0)}\times\mathbb{Z}^{(1)}_N)$, the partition function becomes 
\begin{equation} \label{Eq:sigmaG}
    Z_{\sigma \cQ}[A,B] =  \frac{1}{|H^1(M_3;\mathbb{Z}_N)|}
    \sum_{\substack{a \in H^1(M_3;\mathbb{Z}_N),\\
    b \in H^2(M_3;\mathbb{Z}_N)}} Z_{\cQ}[a,b] \exp{\left(\frac{2\pi i}{N} \int_{M_3} a \cup B + b \cup A  \right)}
     .
\end{equation}
where now $A$ and $B$ denote the background fields for the quantum symmetry obtained after gauging. The choice of this normalization is explained in \autoref{app.A}. The quantum symmetries $\widehat{\bZ}^{(1)}_N$ and $\widehat{\bZ}^{(0)}_N$ are generated by 
\begin{equation}\label{eq.Qeta1}
\eta_1(M_1) = \exp\left(\frac{2\pi i}{N}\oint_{M_1} a\right)~,
\end{equation}
\begin{equation}\label{eq.Qeta0}
\eta_0(M_2) = \exp\left(\frac{2\pi i}{N}\oint_{M_2} b\right)~
\end{equation}
which are the Wilson line and surface operators. 
 
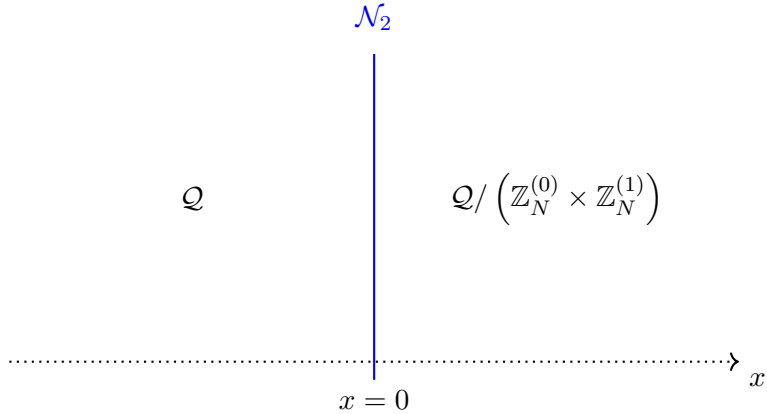
\begin{figure}[!tbp]
\centering
\begin{tikzpicture}[baseline=19,scale=1.2]
\draw[blue, thick] (4,-0.)--(4,3.6);
\draw[ thick,dotted,->] (0,0.2) -- (8,0.2);
\node[right] at (8,0) {$x$};
\node[blue] at (4,4) {$\cN_2$};
\node[below] at (4,0) {$x=0$};
\node at (2,2) {$\cQ$};
\node at (6,2) {$\cQ/\left(\bZ^{(0)}_N \times \bZ^{(1)}_N\right)$};
\end{tikzpicture}
\caption{Duality defect defined by gauging $\bZ^{(0)}_N \times \bZ^{(1)}_N$.}
\label{Fig:dualityD}
\end{figure}

Suppose that spacetime is of the form $M_3 = M_2 \times I$ and denote $x$ the coordinate along the interval. As shown in \autoref{Fig:dualityD}, we consider a setup where the theory $\mathcal{Q}$ lives in the left half of $M_3$ with $x<0$ while its gauged dual $\sigma\mathcal{Q}$ lives in right half of $M_3$ with $x>0$. Imposing Dirichlet boundary conditions for the gauge fields $a$ and $b$ at $x=0$ defines a topological interface $\cN_2$ between $\cQ$ and $\sigma \cQ$. When the theory is self-dual under gauging of $\bZ^{(0)}_N \times \bZ^{(1)}_N$, i.e. $Z_{\cQ}[A,B] = Z_{\sigma\cQ}[A,B]$ (up to possible dualities), the interface $\cN_2$ is a duality defect. 


The gauging operation obeys $\sigma^2 = \cU,$
%
%
where $\cU$ is the charge conjugation operation given by 
\begin{equation} \label{eq.chargeConj}
Z_{\cU \cQ}[A,B] = Z_{\cQ}[-A,-B].
\end{equation}
%
The orientation-reversal of the duality defect is 
%
\begin{align} \label{eqn:orienRN}
    \overline{\cN}_2  =  \cU \times \cN_2 =\cN_2 \times \cU
\end{align}
which is the same as  $\cN_2$ for $N=2$.

\subsection{Triality Defects}\label{sec.2.1}



Assuming the theory $\cQ$ to have $Z^{(0)}_N \times \bZ^{(1)}_N$ symmetries, one can stack $\cQ$ with the following SPT phase as follows \cite{Wan:2018bns}
\begin{equation}
   \qquad Z_{ \tau \cQ}[A,B] = Z_{\cQ}[A,B]  \exp{\left( \frac{2\pi i}{N}\int_{M_3} A \cup B \right)}.
\end{equation}
We denote this topological manipulation by $\tau$. Combining it with the gauging of $Z^{(0)}_N \times \bZ^{(1)}_N$, the partition function becomes 
\begin{equation} \label{Eq:sigmaG2}
    Z_{\sigma \tau \cQ}[A,B] =  \frac{1}{|H^1(M_3;\mathbb{Z}_N)|}
    \sum_{\substack{a \in H^1(M_3;\mathbb{Z}_N),\\
    b \in H^2(M_3;\mathbb{Z}_N)}} Z_{\cQ}[a,b] \exp{\left(\frac{2\pi i}{N} \int_{M_3} a \cup B + b \cup A+ a \cup b  \right)}
     .
\end{equation}
One can show that $Z_{(\sigma \tau)^2 \cT}[A,B]$ is 
\begin{align*}
    &= 
    \frac{1}{|H^1(M_3;\mathbb{Z}_N)|^2}
    \sum_{\substack{a_1,a_2 \in H^1(M_3;\mathbb{Z}_N),\\
    b_1,b_2 \in H^2(M_3;\mathbb{Z}_N)}} Z_{\cT}[a_1,b_1] \exp{\left(\frac{2\pi i}{N} \int_{M_3} a_1 b_1+a_1  b_2 + b_1  a_2  + a_2 b_2+a_2  B + b_2 A   \right)} \\
    &=  \frac{1}{|H^1(M_3;\mathbb{Z}_N)|}
    \sum_{\substack{a_1 \in H^1(M_3;\mathbb{Z}_N),\\
    b_1\in H^2(M_3;\mathbb{Z}_N)}} Z_{\cT}[a_1,b_1] \exp{\left(\frac{2\pi i}{N} \int_{M_3} -b_1 A -a_1 B - AB  \right)} \\
    &=Z_{(\tau^{-1}\cU\sigma)\cT}[-A,-B],
\end{align*}
where in the second line we have integrated out $a_2$ and imposed $a_2=-a_1-A$ on $M_3$. Hence, we find that
\begin{equation}
  \sigma^2=\cU, \qquad \tau^N=1,\qquad   (\sigma \tau)^3 = 1.
\end{equation}
Thus, combining the gauging operation $\sigma$ and stacking with an SPT phase $\tau$ generates the group $SL(2,\bZ_N)$. Similar to the discussion in \cite{Choi:2021kmx,Choi:2022zal}, different combinations of these topological manipulations generate a web of global variants of $\cQ$ \cite{Kaidi:2022uux, Bashmakov:2022uek, Chen:2023qnv, Bashmakov:2023kwo}. The transformation between these global variants are determined by the group theoretic properties of $SL(2,\bZ_N)$.


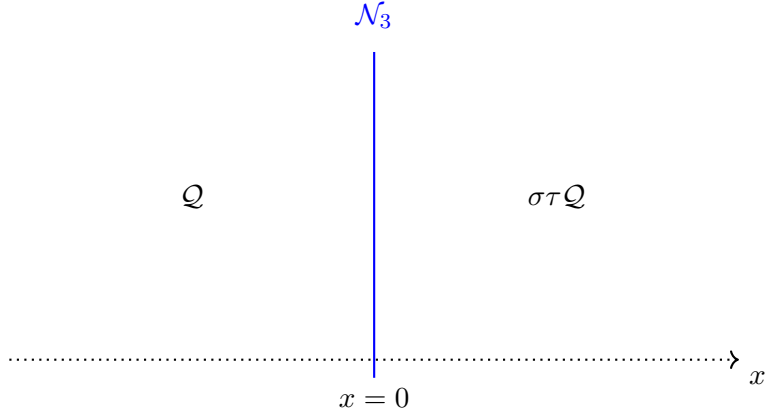
\begin{figure}[!tbp]
\centering
\begin{tikzpicture}[baseline=19,scale=1.2]
\draw[blue, thick] (4,-0.)--(4,3.6);
\draw[ thick,dotted,->] (0,0.2) -- (8,0.2);
\node[right] at (8,0) {$x$};
\node[blue] at (4,4) {$\cN_3$};
\node[below] at (4,0) {$x=0$};
\node at (2,2) {$\cQ$};
\node at (6,2) {$\sigma \tau \cQ$};
\end{tikzpicture}
\caption{Triality defect defined by gauging $\bZ^{(0)}_N \times \bZ^{(1)}_N$, an operation denoted by $\sigma$, and stacking with an SPT phase, denoted by $\tau$. }
\label{Fig:trialityD}
\end{figure}

Suppose the theory $\cQ$ is self-dual under the combination of topological manipulations $\sigma$ and $\tau$, i.e $\cQ \simeq \sigma \tau \cQ$. Then, the configuration in \autoref{Fig:trialityD} defines a triality defect $\cN_3$ in $\cQ$. 
Moreover, the orientation-reversal of the triality interface $\overline{\cN}_3$ is defined as the interface between $\cQ$ and $(\tau^{-1} \cU \sigma) \cQ$. On the two sides of this interface, the partition functions changes as follows
\begin{equation} \label{Eq:sigmaG3}
    Z_{ (\tau^{-1} \cU \sigma) \cQ}[A,B] =  \frac{1}{|H^1(M_3;\mathbb{Z}_N)|}
    \sum_{\substack{a \in H^1(M_3;\mathbb{Z}_N),\\
    b \in H^2(M_3;\mathbb{Z}_N)}} Z_{\cQ}[a,b] \exp{\left(-\frac{2\pi i}{N} \int_{M_3}  a \cup B + b \cup A + A \cup B  \right)}.
\end{equation}

\subsection{Condensation Defects}

Instead of gauging the $\bZ^{(0)}_N \times \bZ^{(1)}_N$ symmetries of $\cQ$ in half of the spacetime, one can also gauge them separately along a higher codimension submanifold. This operation is called higher gauging \cite{Roumpedakis:2022aik} and the resulting topological
defect is denoted condensation defect. In our case, we will consider condensation defects given by the one-gauging of $\bZ^{(1)}_N$ and $\bZ^{(0)}_N$ along a codimension-1 manifold $M_2$ in the 2+1d spacetime $M_3$. 

Suppose that both $\bZ^{(1)}_N$ and $\bZ^{(0)}_N$ are one-gaugeable. A condensation defect defined by one-gauging $\bZ^{(1)}_N$ on a surface $M_2\subset M_3$ is given by 
\begin{equation} \label{eq.condW1}
    C_{\bZ^{(1)}_N}(M_2) = \frac{1}{|H^0(M_2, \bZ_N)|}  \sum_{M_1 \in H_1(M_2,\bZ_N)} \eta_1(M_1) . 
\end{equation}
Up to a Euler counterterm $\chi^{1/2}$ in the normalization, this is the same as the one defined in \cite{Roumpedakis:2022aik}. Note that the orientation reversal of the condensation defect $\overline{C}$ is equivalent to $C$. Similarly, a condensation defect associated with the one-gauging of $\bZ^{(0)}_N$ is defined as summing over insertions of $\eta_0$ on a $2$-dimensional manifold $M_2\subset M_3$. Since $\eta_0$ is a surface operator, it is equivalent to insert all topological defects generated by $\eta_0$ along $M_2$. The corresponding condensation defect is given by  
\begin{equation} \label{eq.condW2}
    \cC_{\bZ^{(0)}_N}(M_2) =  \sum_{i=0}^{N-1} \eta_0^i(M_2). 
\end{equation}

\begin{figure}[!tbp]
\centering
\[{
\begin{tikzpicture}[baseline=50,scale=1.25]
\draw[blue, thick] (2,-0)--(2,3);
\draw[blue, thick] (4,-0)--(4,3);
\draw[ thick,dotted] (0,0.2) -- (6,0.2);
\node at (1,1.5) {$\cQ$};
\node at (3,1.5) {$\cQ/G$};
\node at (5,1.5) {$\cQ/G/\widehat{G}$};
\end{tikzpicture}}
\quad\Longrightarrow\quad
{\begin{tikzpicture}[baseline=50,scale=1.25]
\draw[blue, thick] (2,-0)--(2,3.0);
\draw[ thick,dotted] (0,0.2) -- (4,0.2);
\node[blue] at (2,3.3) {$\cC_G$};
\node at (1,1.5) {$\cQ$};
\node at (3,1.5) {$\cQ$};
\end{tikzpicture}}\]
\caption{Condensation defect $\cC_G$ associate to one-gauging $G$ along a codimension-1 hypersurface in a 3d QFT $\cQ$. This can be constructed by shrinking the region in the middle and colliding the two interfaces implementing the gauging of $G$ and $\widehat{G}$. Notice that $\cQ/G/\widehat{G}=\cQ$ up to a charge conjugation $\cU$. }
\label{Fig:condenSlab}
\end{figure}
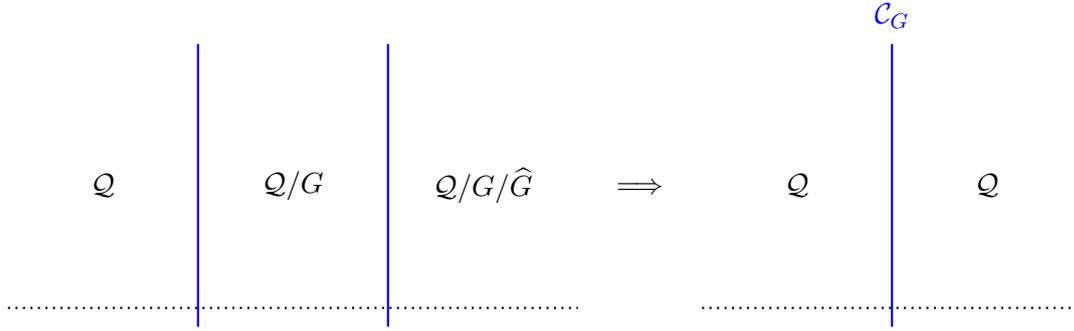

It is useful to have an alternate definition of these condensation defects in terms of their actions on the bulk theory $\cQ$. Thus, we give a bulk realization of the condensation defects associated with one-gauging of a higher-form symmetries $G$ and show how the partition function and Lagrangians of $\cQ$ are affected when passing through these condensation defects. Consider a theory $\mathcal{Q}$ with a global symmetry $G$. As shown in \autoref{Fig:condenSlab}, we can first gauge $G$ and then gauge also the associated quantum symmetry $\widehat{G}$. This splits spacetime in three regions divided by two interfaces. After shrinking the region in the middle, where $G$ has been gauged, the two interfaces collapse and give rise to a defect implementing one-gauging of $G$ along a codimension-1 region up to charge conjugation. This exactly produces the condensation defect $\cC_G$ in $\cQ$. 

With this interpretation we can compute the action of the condensation defect on the partition function of $\mathcal{Q}$. Let us take $G=\bZ^{(1)}_N$. The partition on the left and right in \autoref{Fig:condenSlab} are related by 
\begin{equation}
Z_{\cU \cQ/\bZ_N^{(1)}/\widehat{\bZ}_N^{(0)}}[B] = \sum_{\substack{a \in H^1(M_3, \bZ_N)\\ b \in H^2(M_3, \bZ_N) }} 
Z_{\cQ}[M_3,b]\, \exp{\left({2 \pi i \over N}\int_{M_3} b a - a B \right)},
\end{equation}
where $b \in H^2(M_3,\bZ_N)$ and $a \in H^1(M_3,\bZ_N)$ are the gauge fields of $\bZ^{(1)}_N$ and of the quantum symmetry $\widehat{\bZ}_N^{(0)}$. In terms of the Lagrangian, one has 
\begin{equation} \label{eq.condenL1}
    \cC_{\bZ^{(1)}_N}:\quad L[B] \to L[b] + \frac{2 \pi i}{N}\left(\int_{M_3} ba-aB \right)
\end{equation}
For $G=\bZ^{(0)}_N$, by a similar calculation, one has the Lagrangians across the defect $\cC_{\bZ^{(0)}_N}$ differ by 
\begin{equation} \label{eq.condenL2}
    \cC_{\bZ^{(0)}_N}: \quad L[A] \to L[a] + \frac{2 \pi i}{N}\left(\int_{M_3} ab-bA \right)
\end{equation}
These relations between actions will be important when we will study the fusion rules involving duality and triality defects below. 

Finally, notice that unlike the condensation defect obtained from one-gauging $\bZ^{(1)}_N$ on $M_2$, $\cC_{\bZ^{(0)}_N}$ can be expressed entirely in terms of $\eta_0$. Thus, in the rest of the paper we will not use $\cC_{\bZ^{(0)}_N}$, but rather simply $\eta_0$. Moreover, we will label $\cC=\cC_{\bZ^{(1)}_N}$ in the rest of the paper.

\subsection{Fusion Rules}\label{sec.2.4}

In this subsection, we will derive fusion rules involving either a duality defect $\cN_2$ or a triality defect $\cN_3$ in a 2+1d theory $\cQ$ introduced above. By construction, $\cQ$ has also both codimension-2 line defects $\eta_1$ \eqref{eq.Qeta1} and codimension-1 surface defects $\eta_0$ \eqref{eq.Qeta0} generating, respectively, non-anomalous $\bZ^{(1)}_N$ and $ \bZ^{(0)}_N$ symmetries. There are three types of fusion involving different objects, i.e. the fusion of two line defects, the fusion of a surface and a line defect and the fusion of two surface defects. Line defects $\eta_1$ are invertible and they generate the $\bZ^{(1)}_N$ symmetry. Thus, their fusion rules obey a group law. We will mainly focus on the other two types of fusions. 

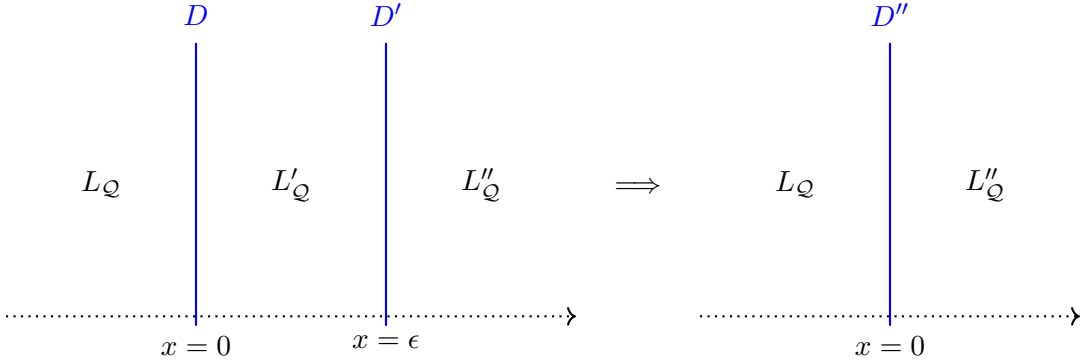
\begin{figure}[!tbp]
\centering
\[{
\begin{tikzpicture}[baseline=50,scale=1.25]
\draw[blue, thick] (2,-0)--(2,3);
\draw[blue, thick] (4,-0)--(4,3);
\node[blue] at (2,3.3) {$D$};
\node[blue] at (4,3.3) {$D'$};
\node[below] at (2,0) {$x=0$};
\node[below] at (4,0) {$x=\epsilon$};
\draw[thick,dotted,->] (0,0.1) -- (6,0.1);
\node at (1,1.5) {$L_{\cQ}$};
\node at (3,1.5) {$L'_{\cQ}$};
\node at (5,1.5) {$L''_{\cQ}$};
\end{tikzpicture}}
\quad\Longrightarrow\quad
{\begin{tikzpicture}[baseline=50,scale=1.25]
\draw[blue, thick] (2,-0)--(2,3.0);
\draw[ thick,dotted,->] (0,0.1) -- (4,0.1);
\node[blue] at (2,3.3) {$D''$};
\node[below] at (2,0) {$x=0$};
\node at (1,1.5) {$L_{\cQ}$};
\node at (3,1.5) {$L''_{\cQ}$};
\end{tikzpicture}}\]
\caption{Fusion of two codimension-1 topological defects $D$ and $D'$. }
\label{Fig:fusion1}
\end{figure}

First, let us consider the fusion of two codimension-1 defects $D$ and $D'$, independently on them being duality/triality defects, a condensation defect or a defect generating $\mathbb{Z}_N^{(0)}$. They are supported on a codimension-1 submanifold which we denote $M_2$. We write spacetime locally as $M_3=M_2\times I$ where $I$ is an interval labeled by a coordinate $x \in \bR$. As shown in \autoref{Fig:fusion1}, we place $D$ at $x=0$ and $D'$ at $x=\epsilon$. The Lagrangian of $\cQ$ is denoted by $L_{\cQ}$ in the region $x<0$, $L'_{\cQ}$ in the region $x \in (0,\epsilon)$ and $L''_{\cQ}$ for $x>\epsilon$ \footnote{Note that since $D$ and $D'$ are topological defects, their insertion does not change the theory $\cQ$. Different expressions for the Lagrangian are just a convenient way to keep track the action of $D$ and $D'$.}. The fusion of $D$ and $D'$ is implemented by taking $\epsilon \to 0$. As we will see later, the fusion rule can be read from the Lagrangian $L''_{\cQ}$ in the $x>0$ region. The general form of the fusion rule is given by
\begin{equation}
    D(M_2) \times D'(M_2) = T(M_2) D''(M_2),
\end{equation}
where the fusion coefficient $T(M_2)$ is in general the partition function of a TQFT on $M_2$. 

The other type of fusion is between a codimension-1 defect $D$ supported on $M_2$ and the line defect $\eta_1(M_1)$ supported on $M_1$. Let us take again the neighborhood around $M_2$ to be $M_2 \times I$. We place the 1-form symmetry defect $\eta_1(M_1)$ parallel to $M_2$ separated with it by a small distance $\epsilon$ along $x$. The fusion is defined by taking $\epsilon \to 0$ where $M_1$ is embedded into $M_2$, and it defines a topological line operator living inside $D(M_2)$. The fusion rule is 
%
\begin{equation}
    D(M_2) \times \eta_1(M_1) = D(M_2).
\end{equation}
%
It implies that the symmetry defect $\eta_1(M_1)$ becomes a trivial line on $D(M_2)$ when brought inside $M_2$. To simplify the notation, we will suppress the $M_2$ and $M_1$ dependence of the topological defects in the fusion rules below. 



\subsubsection{Fusion Rules of Duality Defects}

Following \cite{Choi:2021kmx,Choi:2022zal}, we calculate fusion rule involving duality defects in $\cQ$. The results are summarized by the following relations
\begin{equation}\label{eq.fusion.mathcalN}
\begin{split}
    &\cN_2 \times \overline{\cN}_2 =  \overline{\cN}_2 \times \cN_2 =  \cC \left( \sum_{i=0}^{N-1} \eta_0^i \right),\quad \cN_2 \times \cN_2 = \cU \cC \left( \sum_{i=0}^{N-1} \eta_0^i \right) \\
    & \cN_2 \times \cC = \cC \times \cN_2 = (\cZ_N) \cN_2,\quad \cC \times \cC = (\cZ_N) \cC\\
    &  \cC \times \eta_0 = \eta_0 \times \cC =  \cC, \quad \cC \times \eta_1 = \eta_1 \times \cC =   \cC\\
    & \eta_0 \times \cN =  \cN \times \eta_0 = \cN,\quad \eta_1 \times \cN =  \cN \times \eta_1 = \cN \\
     &  \eta_1^N= \eta_0^N=1, \quad \eta_0 \times \eta_1 = \eta_1 \times \eta_0 
\end{split}
\end{equation}
Here, $\cZ_N$ is the partition function of a (1+1)d $\bZ_N$ gauge theory on $M_2$. Its Lagrangian is given by 
\begin{equation} \label{eq.lagZn}
    L = \frac{N}{2\pi} \int_{M_2}  \phi da 
\end{equation}
where $\phi \in H^0(M_2,\bZ_N)$ is a scalar and $a \in H^1(M_2,\bZ_N)$ is a gauge field. We now provide the details to obtain the relations in \eqref{eq.fusion.mathcalN}.

\paragraph{Fusion of Two Surface Defects.} 

\begin{itemize}
    \item $\cN_2$ and $\overline{\cN}_2$

The Lagrangian in the $x>0$ region is 
\begin{equation} \label{eq.fus20}
\begin{split}
    \cN_2 \times \overline{\cN}_2: \qquad  & L[a,b] + \frac{2\pi i}{N} \left(b \Tilde{a}+a\Tilde{b} - \Tilde{b} A-\Tilde{a}B\right) \\
     = & \; L[a,b] + \frac{2\pi i}{N} \left(b \Tilde{a}- \Tilde{a} B \right) + \frac{2\pi i}{N} (a \tlb - \tlb A).
\end{split}\end{equation} 
Similarly, we find that 
\begin{equation} \begin{split}
     \overline{\cN}_2 \times \cN_2: \qquad  & L[a,b] + \frac{2\pi i}{N} \left(b \Tilde{a}+a\Tilde{b} - \Tilde{b} A-\Tilde{a}B\right) \\
     = & \; L[a,b] + \frac{2\pi i}{N} \left(b \Tilde{a}- \Tilde{a} B \right) + \frac{2\pi i}{N} (a \tlb - \tlb A),
\end{split}\end{equation} 
which is the same as the Lagrangian in \eqref{eq.fus20}. By comparing with the Lagrangian description of condensation defects in \eqref{eq.condenL1} and \eqref{eq.condenL2}, we conclude that the fusion between a duality defect and its orientation-reversal defect on $M_2$ is given by:
\begin{equation} \label{eq.fusionD1}
     \cN_2 \times \overline{\cN}_2  = \cC  \sum_{i=0}^{N-1} \eta_0^{i}.
\end{equation}
The same result can be obtained for the fusion $\overline{\cN}_2 \times \cN_2$. With the method developed in \cite{Kaidi:2022cpf}, we re-derive these fusion rules in \autoref{app.A}.

\item $\cC$ and $\cC$

The Lagrangian in the region $x>0$ obtained from the fusion of two condensation defects is 
\begin{equation}\begin{split}\label{eq.appendixB}
    \cC \times \cC:\qquad  & L[b] + \frac{2\pi i}{N} \left(ba-a\Tilde{b} + \Tilde{b} \Tilde{a}-\Tilde{a}B\right) \\
     = & \;  \frac{2\pi i}{N} (a'b') + L[b] + \frac{2\pi i}{N} \left(b\Tilde{a}-\Tilde{a}B \right) 
\end{split}\end{equation} 
with $b'=b-\tlb$ and $a'=a-\tla$. 
The first term is a decoupled topological action. After shrinking the interval $\epsilon \to 0$, we impose Dirichlet boundary conditions for $b'$ and $a'$ at $x=0$. As studied in detail in \autoref{app.ta}, the boundary theory is a (1+1)d $\bZ_N$ gauge theory on $M_2$. 

From \eqref{eq.condenL1}, the rest of the terms in the Lagrangian are equivalent to the insertion of a condensation defect $\cC$. Thus, the fusion rule is 
\begin{equation}
    \cC \times \cC = (\cZ_N) \cC.
\end{equation}
This fusion rule was first obtained in \cite{Roumpedakis:2022aik} from the definition \eqref{eq.condW1} of condensation defect. Here, we re-derived it from a bulk perspective.

\item $\eta_0$ and $\cN_2$ 

The $\bZ^{(0)}_N$ symmetry defect $\eta_0$ can be written as 
\begin{equation} \label{eq:fusZ0}
    \eta_0(M_2) = \exp{\left( \frac{2 \pi i }{N}\int_{M_2} b \right)} = \exp{\left( \frac{2 \pi i }{N}\int_{M_3 }b \cup A \right)} 
\end{equation}
where $b\in H^2(M_2,\bZ_N)$ is a dynamical gauge field and $M_2 = \text{PD}(A)$, where $\text{PD}$ represents the Poincar\'e dual. The fusion of $\cN_2$ and $\eta_0$ is determined simply by placing $\eta_0$ at the boundary at $x=0$. Due to the Dirichlet boundary conditions at the insertion locus $x=0$ of the duality defect, $a|_{M_2}=b|_{M_2}=0$, on $M_2$ the symmetry defect gets absorbed. We then have  
\begin{align}
    & \cN_2 \times \eta_0  = \cN_2, \quad \overline{\cN}_2 \times \eta_0  = \overline{\cN}_2.
\end{align}

\item $\eta_0$ and $\cC$

Unlike the previous case, the condensation defect $\cC$ does not absorb $\eta_0$. Since $\cC$ is defined by the sum of $\eta_1$ over a codimension-1 space $M_2$ \eqref{eq.condW1}, and $\eta_0$ and $\eta_1$ are independent, the fusion is 
\begin{align}
    &  \cC \times \eta_0 = \eta_0 \times \cC = \eta_0 \cC ,
\end{align}
which is different from the fusion with the duality defect $\cN_2$.

\end{itemize}

\paragraph{Fusion of $\eta_1$ and Codimension-1 Defects.}

Similar to the $\bZ^{(0)}_N$ symmetry defect $\eta_0$, the $\bZ^{(1)}_N$ symmetry defect can be written as 
\begin{equation} \label{eq:fusZ1}
    \eta_1(M_1) = \exp{\left( \frac{2 \pi i }{N}\int_{M_1} a \right)} = \exp{\left( \frac{2 \pi i }{N}\int_{M_3 }a \cup B \right)} 
\end{equation}
where $M_1 = \text{PD}(B)$ is a one dimensional submanifold in $M_3$. We now study the fusion rules in details.

\begin{itemize}
    
\item  $\eta_0$ and $\eta_1$

Since $\eta_1$ and $\eta_0$ are independent, their fusion is simply given by
\begin{equation}
    \eta_0 \times \eta_1 = \eta_1 \times \eta_0 =\eta_1 \eta_0
\end{equation}
which gives the same line defect $\eta_1$ inside the codimension-1 defect $\eta_0$. 

\item $\cN_2$ and $\eta_1$ 

Due to the Dirichlet boundary conditions of $a|_{M_2}=b|_{M_2}=0$ in the definition of $\cN_2$.  
The symmetry defect $\eta_1$ just gets absorbed when brought inside $\cN_2$ or $\overline{\cN}_2$. The fusion rules are
\begin{align}
    &  \cN_2 \times \eta_1 = \cN_2 ,\quad \overline{\cN}_2 \times \eta_1 = \overline{\cN}_2.
\end{align}

\item $\cC$ and $\eta_1$ 

The bulk definition of the condensation defect $\cC(M_2)$ in \eqref{eq.condenL1}
also imposes Dirichlet boundary conditions $a|_{M_2}=0$ and $b|_{M_2}=0$. By the same argument, the symmetry defect $\eta_1$ gets absorbed when brought inside $\cC$. The fusion rules are then
\begin{align}
    & \eta_1 \times \cC =  \cC \times \eta_1 = \cC.
\end{align}
The same fusion rules have also been derived in \cite{Roumpedakis:2022aik} directly from the definition in \eqref{eq.condW1}.

\end{itemize}

\subsubsection{Fusion Rules of Triality Defects}

The fusion rules involving the triality defects are summarized as follows  
\begin{equation} \label{eq:fusTriality}
\begin{split}
    &\cN_3 \times \overline{\cN}_3 =  \overline{\cN}_3 \times \cN_3 =  \cC \left( \sum_{i=0}^{N-1} \eta_0^i \right), \quad \cN_3 \times \cN_3 = (\cZ_N) \overline{\cN}_3 \\
    & \cN_3 \times \cC = \cC \times \cN_3 = (\cZ_N) \cN_3, \quad \eta_0 \times \cN_3 =  \cN_3 \times \eta_0 = \cN_3,\\
    &\eta_1 \times \cN_3 =  \cN_3 \times \eta_1 = \cN_3 
\end{split}
\end{equation}
As before, we will derive these fusion rules below. 

\begin{itemize}
    
\item $\cN_3$ and $\overline{\cN}_3$:

The Lagrangian $L[A,B]$ after passing through $\cN_3$ and $\overline{\cN}_3$ becomes 
\begin{equation} \label{eq.fuT1}
\begin{split}
    \cN_3 \times \overline{\cN}_3: \qquad  & L[a,b] + \frac{2\pi i}{N} \left(ab+a\tlb+b\tla - \tlb A -\tla B - AB\right) \\
     = & \; L[a,b] + \frac{2\pi i}{N} \left( a b'-b' A  \right)  + \frac{2\pi i}{N}  \left(
     b a' -a' B
     \right)
\end{split}
\end{equation} 
with $b'=\tlb + \frac{b+B}{2}$ and $a'=\tla + \frac{a+A}{2}$. If we switch the order of $\cN_3$ and $\overline{\cN}_3$, the Lagrangian in the $x>0$ region is 
\begin{equation} \begin{split}
     \overline{\cN}_3 \times \cN_3: \qquad  & L[a,b] + \frac{2\pi i}{N} \left( - a\tlb - b\tla -\tla \tlb + \tla \tlb + \tlb A +\tla B \right) \\
     = & \; L[a,b] + \frac{2\pi i}{N} \left(\tlb A - a \tlb \right) + \frac{2\pi i}{N} (\tla B- b \tla) 
\end{split}\end{equation} 
which is the same as the Lagrangian in \eqref{eq.fuT1}. Comparing with \eqref{eq.condenL1} and \eqref{eq.condenL2}, the fusion rule is 
\begin{equation}
    \cN_3 \times \overline{\cN}_3 = C \sum_{i=0}^{N-1} \eta^i_0.
\end{equation}

\item $\cN_3$ and $\cN_3$

The Lagrangian after passing through a pair of defects $\cN_3$ is given by
\begin{equation} 
\begin{split}
    \cN_3 \times \cN_3:\qquad  & L[a,b] + \frac{2\pi i}{N} \left(ab+a\tlb+b\tla+\tla\tlb+\tla B + \tlb A\right) \\
     = & \; \frac{2\pi i}{N} (a'b') + L[a,b] + \frac{2\pi i}{N} \left(-AB-aB-bA \right)  
\end{split}
\end{equation} 
with $a'=a+\tla+A$ and $b'=b+\tlb+B$. Again, the first term gives a (1+1)d $\bZ_N$ gauge theory and the remaining two terms, for \eqref{Eq:sigmaG3}, are equivalent to the insertion of $\overline{\cN}_3$ at $x=0$. Thus, the fusion rules are
\begin{equation}
    \cN_3 \times \cN_3 = (\cZ_N) \overline{\cN}_3.
\end{equation}

\item $\cN_3$ and $\cC$

The Lagrangian in the $x>0$ region is 
\begin{equation} \label{eq.fusT2}
\begin{split}
    \cN_3 \times \cC:\qquad  & L[a,b] + \frac{2\pi i}{N} \left(ab+a\tlb+b A+\tla\tlb - \tla B\right) \\
     = & \;  \frac{2\pi i}{N}(a'b')  + L[a,b] + \frac{2\pi i}{N} \left(ab+aB+bA\right) 
\end{split}
\end{equation} 
with $b'=\tlb-B$ and $a'=a+\tla$. Switching the order of $\cN$ and $\cC$, the Lagrangian becomes 
\begin{equation} 
\begin{split}
    \cC \times \cN_3:\qquad  & L[\tla,b] + \frac{2\pi i}{N} \left(ba-a\tlb+\tla \tlb + \tlb A+ \tla B\right) \\
     = & \;  \frac{2\pi i}{N} (a'b') + L[\tla,b] + \frac{2\pi i}{N} \left(\tla b+\tla B+bA\right) 
\end{split}
\end{equation} 
with $b'=b-\tlb$ and $a'=a-\tla-A$. Up to a field redefinition, it is the same as the Lagrangian in \eqref{eq.fusT2}. The first term gives a (1+1)d $\bZ_N$ gauge theory. The remaining terms give the defect $\cN_3$ from \eqref{Eq:sigmaG2}. The fusion rules are 
\begin{equation}
     \cN_3 \times \cC =   \cC \times \cN_3 = (\cZ_N) \cN_3.
\end{equation}

\item  $\cN_3$ and $\eta_0$ (or $\eta_1$)

The derivation of the fusion rules involving $\cN_3$ and $\eta_0$ or$\cN_3$ and $\eta_1$ are almost identical. Thus, we discuss them together. First, let us consider the fusions $\cN_3 \times \eta_0$ and $\cN_3 \times \eta_1$. From the definition of $\eta_0$ in \eqref{eq:fusZ0} and $\eta_1$ in \eqref{eq:fusZ1}, it is straightforward to obtain
\begin{equation}
    \cN_3 \times \eta_0 = \cN_3, \qquad \cN_3 \times \eta_1 = \cN_3
\end{equation}
from the Dirichlet boundary conditions $a|_{M_2}=b|_{M_2}=0$ imposed on $\cN_3$. However, the fusion of $\overline{\cN}_3 $ with $ \eta_0$ or $ \eta_1$ does not follow from this analysis automatically. After imposing the Dirichlet boundary condition, there is an extra SPT phase left in region $x>0$
\begin{equation}
    \exp{\left( \frac{2\pi i}{N}\int_{M_3:x>0} A\cup B\right)},
\end{equation}
where $A \in H^1(M_3,M_2,\bZ_N)$ and $B \in H^2(M_3,M_2,\bZ_N)$. With the isomorphism $H^3(M_2\times I,M_2,\bZ_N) \cong H^2(M_2,\bZ_N)$, one can show that this extra factor vanishes when restricted on $M_2$ \cite{Kaidi:2021xfk}. Thus, we derived the fusion rules in \eqref{eq:fusTriality}. 

\end{itemize}

\section{Symmetry TFT for Duality Defects}\label{sec.3}

In the previous sections we considered a 3d theory $\mathcal{Q}$ with non-invertible symmetry defects obtained gauging $\mathbb{Z}_N^{(0)}\times\mathbb{Z}^{(1)}_N$ symmetries. We now want to understand how the defects originate from the SymTFT perspective \cite{Freed:2012bs,Freed:2018cec,Gaiotto:2020iye,Apruzzi:2021nmk, Apruzzi:2022dlm,Burbano:2021loy,Freed:2022qnc,vanBeest:2022fss, Kaidi:2022cpf, Bashmakov:2022uek, Kaidi:2023maf}. We do so expanding the 3d theory into a four-dimensional BF theory \cite{Horowitz:1989ng,Blau:1989bq,Cattaneo:1995tw} on a slab with topological and dynamical boundary conditions at the two boundaries, first without assuming invariance under gauging of $\mathcal{Q}$. The global symmetry content of this theory is\footnote{As we will show momentarily, for $N=2$ the last factor will turn out to be $\mathbb{Z}_2^\text{EM}$.} $\mathbb{Z}_N^{(2)}\times\mathbb{Z}_N^{(2)}\otimes\mathbb{Z}_N^{(1)}\times\mathbb{Z}_N^{(1)}\times\mathbb{Z}_4^\text{EM}$. The first four factors are generated by Wilson surfaces and lines while the latter corresponds to the electric-magnetic exchange of the four gauge fields. In general, it is known that, given a $d$-dimensional QFT $\cQ$ with (higher) categorical symmetry $\cC$, the SymTFT is expected to be the Turaev-Viro theory on $\cC$, or the Reshetikhin-Turaev theory on the Drinfeld center $\cZ(\cC)$. 

In the second part of the section, we require the 3d theory $\mathcal{Q}$ to be invariant under simultaneous gauging of $\mathbb{Z}_N^{(0)}\times\mathbb{Z}^{(1)}_N$. Based on earlier works for 2d theories \cite{Izumi:2001mi,gelaki2009centers,Barkeshli:2014cna,Teo:2015xla}, the authors of \cite{Kaidi:2022cpf} have shown that for even-dimensional theories $\mathcal{Q}$, the symmetry group is (a generalization of) the Tambara-Yamagami fusion category $\text{TY}(\mathbb{Z}_N)$. As an example, for $Q$ two-dimensional the corresponding SymTFT is a $\bZ_2^\text{EM}$ gauging of the  3d BF theory (or $\mathbb{Z}_N$ gauge theory). Here, following their approach, we find that the 4d SymTFT of the 3d theory $\mathcal{Q}$ invariant under gauging is the $\bZ_4^\text{EM}$ gauging of the 4d BF theory.

\subsection{4d BF Theory}
We start constructing the SymTFT for the 3d theory $\mathcal{Q}$ without assuming invariance under gauging of $\mathbb{Z}_4^\text{EM}$.

\paragraph{Action and Boundary Conditions.}
We consider a 4d BF theory \cite{Horowitz:1989ng,Blau:1989bq,Cattaneo:1995tw} with action   
\begin{equation}\label{eq.SymTFTaction}
    S_{4d} = \frac{4\pi}{N} \int_{M_4} \left(\delta a_1\cup b_1 + \delta a_2\cup b_2\right),
\end{equation}
where $a_1,a_2$ are are dynamical $\bZ_N$-valued 1-cochains and $b_1,b_2$ are are dynamical $\bZ_N$-valued 2-cochains. Moreover, 4d spacetime is a product $M_4=M_3\times I_{[0,\epsilon]}$ where $x$ is the coordinate along the interval and the two boundaries are denoted $M_3|_\epsilon$ and $M_3|_0$.
\begin{figure}[!tbp]
\centering
\[\begin{tikzpicture}[baseline=50,scale=1.25]
\draw[blue, thick] (2,-0)--(2,2.0);
\node[blue] at (2,2.3) {$\mathcal{Q}$};
\node[below] at (2,0) {$x=0$};
\draw[thick,dotted,<-] (3,1) -- (4,1);
\draw[blue, thick] (5,0)--(5,2);
\draw[blue, thick] (9,0)--(9,2);
\node[blue] at (5,2.3) {$\langle D(A,B)|$};
\node[blue] at (9,2.3) {$|\mathcal{Q}\rangle$};
\node[below] at (5,0) {$x=0$};
\node[below] at (9,0) {$x=\epsilon$};
\node at (7,1) {$\frac{4\pi}{N} \int_{M_4} \delta a_1  b_1 + \delta a_2  b_2$};
\filldraw[draw=blue,fill=blue!20,opacity=0.5]       
    (5,0) -- (9,0) -- (9,2) -- (5,2) -- cycle;
\end{tikzpicture}\]
\caption{The 3d theory $Q$ is obtained choosing Dirichlet boundary conditions on the topological boundary and shrinking the 4d slab where the SymTFT is defined. }
\label{Fig:SymTFT1}
\end{figure}
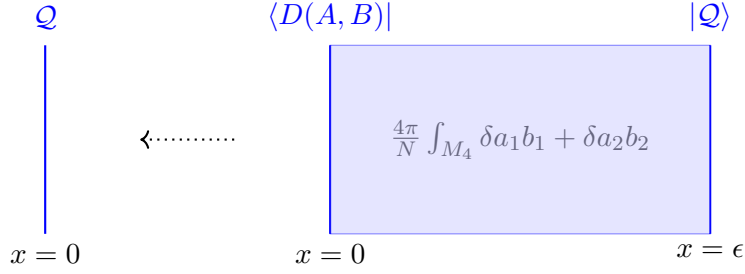
\begin{figure}[!tbp]
\centering
\[\begin{tikzpicture}[baseline=50,scale=1.25]
\draw[blue, thick] (2,-0)--(2,2.0);
\node[blue] at (2,2.3) {$\mathcal{Q}/(\bZ_N^{(0)}\times\bZ_N^{(1)})$};
\node[below] at (2,0) {$x=0$};
\draw[thick,dotted,<-] (3,1) -- (4,1);
\draw[blue, thick] (5,0)--(5,2);
\draw[blue, thick] (9,0)--(9,2);
\node[blue] at (5,2.3) {$\langle N(A,B)|$};
\node[blue] at (9,2.3) {$|\mathcal{Q}\rangle$};
\node[below] at (5,0) {$x=0$};
\node[below] at (9,0) {$x=\epsilon$};
\node at (7,1) {$\frac{4\pi}{N} \int_{M_4} \delta a_1  b_1 + \delta a_2  b_2$};
\filldraw[draw=blue,fill=blue!20,opacity=0.5]       
    (5,0) -- (9,0) -- (9,2) -- (5,2) -- cycle;
\end{tikzpicture}\]
\caption{Choosing Neumann boundary conditions on the topological boundary gives, upon shrinking the slab, a 3d theory where where $\bZ_N^{(0)}\times\bZ_N^{(1)}$ is gauged.}
\label{Fig:SymTFT2}
\end{figure}
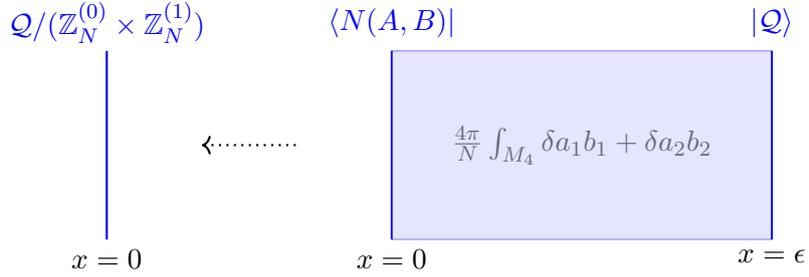

On one side of the slab ($x=\epsilon$) we impose dynamical boundary conditions
\begin{equation}\label{eq.dynamicalbc}
    |\cQ\rangle = 
    \sum_{\substack{a_1 \in H^1(M_3|_{\epsilon};\mathbb{Z}_N),\\
    b_1 \in H^2(M_3|_{\epsilon};\mathbb{Z}_N)}}
    Z_{\cQ}[M_3|_{\epsilon},a_1,b_1] |a_1,b_1\rangle,
\end{equation}
while on the opposite side ($x=0$) we impose topological boundary conditions, which can be either Dirichlet boundary conditions 
\begin{equation}\label{eq.Dirichletbc}
    \langle D(A,B)| =\sum_{\substack{a_1 \in H^1(M_3|_{\epsilon};\mathbb{Z}_N),\\
    b_1 \in H^2(M_3|_{\epsilon};\mathbb{Z}_N)}}
    \langle a,b| \delta(a_1-A) \delta(b_1-B)
\end{equation}
or Neumann boundary condition
\begin{equation}\label{eq.Neumannbc}
    \langle N(A,B)| = \sum_{\substack{a_1\in H^1(M_3|_{\epsilon};\mathbb{Z}_N),\\
    b_1\in H^2(M_3|_{\epsilon};\mathbb{Z}_N)}}
    \langle a_1,b_1| \exp{\left( \frac{2\pi i}{N} \int_{M_3|0} a_1 B + b_1A\right)}.
\end{equation}
The two choices of boundary conditions are shown, respectively, in \autoref{Fig:SymTFT1} and \autoref{Fig:SymTFT2}. It is straightforward to check that switching from Dirichlet to Neumann boundary condition corresponds to gauge $\bZ^{(0)}_N \times \bZ^{(1)}_N$ on the boundary\footnote{We assumed that individual and mixed 't Hoof anomalies trivialize so that there is no obstruction to gauging. In cases when there is a 't Hooft anomaly, the corresponding SymTFT is generalized Dijkgraaf-Witten theory and Neumann boundary conditions are not allowed.}. A crucial property is that, for any two theories $\mathcal{Q}$ and $\mathcal{Q}'$ related by topological manipulation, the corresponding SymTFT is the same.

\paragraph{Topological Defects in the SymTFT.}

The BF theory has 2-form and 1-form symmetries generated by, respectively, line operators
\begin{equation}\label{eq.SymTFT.line}
    L_{(l_1,l_2)}(\gamma)=\exp{\left( \frac{2\pi i}{N} \oint_{\gamma} l_1 a_1 \right)}\exp{\left( \frac{2\pi i}{N} \oint_{\gamma} l_2 a_2 \right)}, \quad (l_1,l_2) \in \bZ_N \times \bZ_N,
\end{equation}
and surface operators
\begin{equation}\label{eq.SymTFT.surface}
    S_{(s_1,s_2)}(\sigma)=\exp{\left( \frac{2\pi i}{N} \oint_{\sigma} s_1 b_1 \right)} \exp{\left( \frac{2\pi i}{N} \oint_{\sigma} s_2 b_2 \right)},
    \quad (s_1,s_2) \in \bZ_N \times \bZ_N.
\end{equation}
Thus, there are in total $2N^2$ topological operators which generate the symmetry group $\mathbb{Z}_N^{(2)}\times\mathbb{Z}_N^{(2)}\otimes\mathbb{Z}_N^{(1)}\times\mathbb{Z}_N^{(1)}$. Lines $L_{(1,0)}$ and $L_{(0,1)}$ generate, respectively, the \emph{electric} and \emph{magnetic} two-form symmetries. The same holds for electric and magnetic surface operators $S_{(1,0)}$ and $S_{(0,1)}$.

The fusion rules for line and surface operators are 
\begin{align}\label{eq.fusionLS}
\begin{split}
    &L_{(l_1,l_2)}(\gamma) \times L_{(l'_1,l'_2)}(\gamma) = L_{(l_1+l'_1,l_2+l'_2)}(\gamma), \\
    &S_{(s_1,s_2)}(\sigma) \times S_{(s'_1,s'_2)}(\sigma) = S_{(s_1+s'_1,s_2+s'_2)}(\sigma). 
\end{split}
\end{align}
This shows that these topological operators are invertible. Moreover, since in four dimensions the linking number of two one-cycles (two-cycles) is trivial, the dependence of correlation functions on pair of line operators (surface operators) is trivial. Moreover, when pushing the corresponding cycles, either one-cycles $\gamma,\gamma'$ or two cycles $\sigma,\sigma'$, to a 3d hypersurface inside $M_4$, also the equal-time commutation relation \cite{Gaiotto:2014kfa} between pairs of curves or surfaces are trivial due to the vanishing of their intersection number. Therefore, the quantum torus algebra is simply:
\begin{align} \label{eq.qtaLS}
\begin{split}
    & L_{(l_1,l_2)}(\gamma) L_{(l_1,l_2)}(\gamma') = L_{(l_1,l_2)}(\gamma+\gamma'), \\
    & S_{(s_1,s_2)}(\sigma) S_{(s_1,s_2)}(\sigma') = S_{(s_1,s_2)}(\sigma+\sigma').
\end{split}
\end{align}
Compare this with odd-dimensional SymTFT considered in \cite{Kaidi:2022cpf}. There the linking number of two curves (surfaces) in 3d (5d) can be non-zero, giving non-trivial dependence when a pair of line (surface) operators is inserted in a correlation function. Moreover, considering the 3d case, the linking number of the curves when restricted on a 2d plane, and thus their equal-time commutation relation, is non-trivial.
%
%

What is peculiar for four-dimensional SymTFTs is that a line and a surface can have nontrivial linking number between each other\footnote{A similar phenomenon would occur for a 6d SymTFT where the linking number between a two-cycle and a three-cycle can be non-trivial. These could be topological operators generating, respectively, three and two form symmetries.}. Hence, the correlation functions involving $ L_{(l_1,l_2)}(\gamma)$ and $ S_{(s_1,s_2)}(\sigma)$ are 
\begin{align}\label{eq.correlation.sigmagamma}
    & \langle L_{(l_1,l_2)}(\gamma) S_{(s_1,s_2)}(\sigma) \cdots \rangle = \exp{\left(-\frac{2\pi i}{N}(l_1s_1+l_2s_2) \text{link}(\gamma, \sigma)\right)}\langle\cdots\rangle.
\end{align}
Here, $\text{link}(\gamma, \sigma)$ is the linking number of $\sigma$ and $\gamma$ in $M_4$. The derivation \eqref{eq.correlation.sigmagamma} is performed explicitly in \autoref{app.B}. After canonical quantization, the equal-time commutation relations are obtained restricting $\gamma,\sigma$ on a 3d hypersurface
\begin{equation}\label{eq.comuteLS}
    L_{(l_1,l_2)}(\gamma) S_{(s_1,s_2)}(\sigma) = 
    \exp{\left(- \frac{2\pi i}{N} (l_1s_1+l_2s_2) \langle \gamma,\sigma\rangle\right)}
    S_{(s_1,s_2)}(\sigma)  L_{(l_1,l_2)}(\gamma)
\end{equation}
where $\langle \gamma, \sigma \rangle$ is the intersection number between $\sigma$ and $\gamma$. These relations will be important in the computation of the fusion rules in the rest of the section.

\paragraph{Condensation Defect.}
The SymTFT action \eqref{eq.SymTFTaction} is invariant\footnote{In odd dimensions \cite{Kaidi:2022cpf} the action is invariant only on a closed manifold while in our case invariance holds also in the presence of boundaries. This because there is no need to integrate by parts in our setup to prove the equivalence.} under the exchange of the electric and magnetic gauge fields 
\begin{equation}\label{eq.emsymmetry}
    (a_1,a_2) \to (-a_2,a_1), \qquad  (b_1,b_2) \to (-b_2,b_1).
\end{equation}
This operation generates a $\bZ^\text{EM}_4$ symmetry for $N>2$. In particular, the square of the action transforms all the gauge fields by a minus sign, thus generating charge conjugation symmetry. When $N=2$, charge conjugation acts trivially and \eqref{eq.emsymmetry} only generates a $\bZ^\text{EM}_2$ symmetry. 

The line and surface operators under this operation transform as
\begin{equation} \label{eq.emls}
    L_{(l_1,l_2)}(\gamma) \to L_{(-l_2,l_1)}(\gamma),\quad S_{(s_1,s_2)}(\sigma) \to S_{(-s_2,s_1)}(\sigma).
\end{equation}
The symmetry $\mathbb{Z}_4^\text{EM}$ (or $\bZ^\text{EM}_2$ symmetry for $N=2$) is implemented by a codimension-1 condensation defect. To understand what are the operators condensing on it we employ the folding trick \cite{Kaidi:2022cpf} for a symmetry operator entering the defect from the left and leaving from the right. First, we create a hole in the defect so that the right-hand side on the two relations in \eqref{eq.emls}, let it be line or a surface, can pass through it and go back to the left of the defect. We further manipulate the system such that the intersection point of the two symmetry operators coincides. Reversing the orientation of the topological operator, we find that all operators of the form 
\begin{equation}
    L_{(l_1+l_2,l_2-l_1)}=L_{(1,-1)}^{\otimes e}\otimes L^{\otimes m}_{(1,1)},\qquad S_{(s_1+s_2,s_2-s_1)}=S_{(1,-1)}^{\otimes e}\otimes S^{\otimes m}_{(1,1)},
\end{equation}
can be absorbed. Therefore the symmetry defect is a condensate of two line operators $L_{(1,\pm 1)}$ and two surface operators $S_{(1,\pm 1)}$ and it is obtained by one-gauging the higher-form symmetries on a codimension-1 hypersurface \cite{Roumpedakis:2022aik}. From the point of view of the hypersurface, we are (0-)gauging both a one and a zero form symmetry.

The definition of the condensation defect depends on the value of $N$. For odd $N$, $L_{(1,\pm 1)}$ and $S_{(1,\pm 1)}$ generate $\bZ^{(2)}_N \times \bZ^{(2)}_N\times\bZ^{(1)}_N\times\bZ^{(1)}_N$ symmetry in the 4d SymTFT. When $N>2$ is even, not the entire symmetry group $\bZ^{(2)}_N\times\bZ^{(2)}_N\times\bZ^{(1)}_N\times\bZ^{(1)}_N$ is generated as the following relations among symmetry operators hold
\begin{equation}
    L_{(N/2,N/2)}=L_{(N/2,-N/2)},\qquad S_{(N/2,N/2)}=S_{(N/2,-N/2)}.
\end{equation}
Therefore, the symmetry group is reduced to
\begin{equation}
\frac{\bZ^{(2)}_N\times\bZ^{(2)}_N}{\mathbb{Z}_2}\times\frac{\bZ^{(1)}_N \times \bZ^{(1)}_N}{\mathbb{Z}_2}.
\end{equation}
To simplify notations, let us introduce
\begin{equation} \label{eq.condesN}\Xi^i\equiv(\Xi^1,\Xi^2),\quad\Xi^1=\mathbb{Z}_N\times\mathbb{Z}_N,\quad\Xi^2=(\mathbb{Z}_N\times\mathbb{Z}_N)/\mathbb{Z}_2
\end{equation}
Then, the condensation defect is written as follows:
\begin{equation}\label{eq.condefect.N}
\begin{split}
    D^i(M_3)=\frac{1}{|H^1(M_3,\Xi^i)|}\sum_{\substack{(\gamma,\gamma')\in H_1(M_3,\Xi^i),\\
    (\sigma,\sigma')\in H_2(M_3,\Xi^i)}}&\exp{\left(-\frac{2\pi i}{N} (\langle \sigma,\gamma-\gamma'\rangle+\langle \sigma',\gamma+\gamma'\rangle)\right)} 
    \\
    &\times S_{(1,-1)}(\sigma)S_{(1,1)}(\sigma')L_{(1,-1)}(\gamma)L_{(1,1)}(\gamma'), 
\end{split}
\end{equation}
where $i=1$ for $N$ odd and $i=2$ for $N$ even. Moreover, $M_3$ is a codimension-1 hypersurface and the discrete torsion is introduced to produce the correct fusion rule. 

Finally, for $N=2$ the lines $L_{(1,-1)}$ and $L_{(1,1)}$ are identified and they generate only a $\bZ^{(2)}_N$ symmetry. Similarly, $S_{(1,-1)}$ and $S_{(1,1)}$ generate only a $\bZ^{(1)}_N$ symmetry. The condensation defect is given by:
\begin{equation}\label{eq.condefect.2}
    D^{N=2}(M_3)=\frac{1}{|H^1(M_3,\bZ_2)|}\sum_{\substack{\gamma \in H_1(M_3,\mathbb{Z}_2),\\
    \sigma\in H_2(M_3,\mathbb{Z}_2)}} e^{-\pi i\langle\sigma,\gamma\rangle} \;S_{(1,-1)}(\sigma)L_{(1,-1)}(\gamma).
\end{equation}
Again, a discrete torsion is included to have the correct fusion rule. 

We calculate the fusion rules involving the condensation defects $D(M_3)$ defined above. The results are summarised as
\begin{align}
    &L_{(l_1,l_2)}(M_1) \times D(M_3)=D(M_3)\times L_{(-l_2,l_1)}(M_1),\label{eq.FusionD1}\\
    &S_{(s_1,s_2)}(M_2) \times D(M_3)=D(M_3)\times S_{(-s_2,s_1)},(M_2)\label{eq.fusionD2}\\
    & D(M_3)\times D(M_3)=C(M_3),\label{eq.fusionD3}\\
    & C(M_3)\times D(M_3)=\overline{D}(M_3), \;\label{eq.fusionD4} \\
    & \overline{D}(M_3)\times D(M_3)=D(M_3)^4=1\;\label{eq.fusionD5}. 
\end{align}
Note that, whenever confusion cannot arise, we will drop the superscripts\footnote{The same applies to the charge conjugation operator $C^i(M_3)$ and the twist defect $V^i_{(0,0)}(M_3,M_2)$ that we will define momentarily.} $D^1,D^2$ and $D^{N=2}$. In the fusion rules above, for $N\neq 2$, $C(M_3)$ we defined the charge conjugation defect as
\begin{equation}\begin{split}\label{eq.symTFTC}
    C^i(M_3)=\frac{1}{|H^1(M_3,\Xi^i)|}\sum_{\substack{(\gamma,\gamma') \in H_1(M_3,\Xi^i),\\
    (\sigma,\sigma') \in H_2(M_3,\Xi^i)}}&\exp{\left(-\frac{2\pi i}{N}(\langle\sigma,\gamma\rangle+\langle\sigma',\gamma'\rangle)\right)}\\
    &\times S_{(1,-1)}(\sigma)S_{(1,1)}(\sigma')L_{(1,-1)}(\gamma)L_{(1,1)}(\gamma'),
\end{split}\end{equation}
which is instead trivial for $N=2$, and $\overline{D}(M_3)$ is the orientation reversal of $D(M_3)$ 
\begin{equation}\begin{split}\label{eq.barD}
    \overline{D}^i(M_3)=\frac{1}{|H^1(M_3,\Xi)|}\sum_{\substack{(\gamma,\gamma') \in H_1(M_3,\Xi^i),\\
    (\sigma,\sigma') \in H_2(M_3,\Xi^i)}}  & \exp{\left(-\frac{2\pi i}{N} (\langle \sigma,\gamma+\gamma' \rangle + \langle \sigma', \gamma'-\gamma \rangle ) \right)}\\
    &\times S_{(1,-1)}(\sigma)S_{(1,1)}(\sigma')L_{(1,-1)}(\gamma)L_{(1,1)}(\gamma') 
\end{split}\end{equation}
For $N=2$, $\overline{D}^{N=2}(M_3)=D^{N=2}$ and thus $D^{N=2}(M_3)\times D^{N=2}(M_3)=1$ in \eqref{eq.fusionD5} as expected given that the electric-magnetic symmetry \eqref{eq.emsymmetry} is $\mathbb{Z}_2^\text{EM}$. The detailed derivation of these fusion rules is performed in \autoref{app.fusionCon}.

\subsection{Twist Defects and Fusion Rules}
While line and surface symmetry operators cannot be given boundaries due to gauge invariance\footnote{Unless the boundary consists in a non-trivial junction where other symmetry operators ends. We leave the study of such junctions, and the corresponding F-symbols, for future investigation.}, we can instead impose topological boundary conditions for the condensation defect just introduced. Thus, following \cite{Barkeshli:2014cna,Teo:2015xla,Kaidi:2022cpf}, we now introduce a condensation operator with topological boundaries which will give rise to duality interfaces in the theory $\mathcal{Q}$ upon shrinking the slab. As the definition of $D(M_3)$ \eqref{eq.condefect.N}-\eqref{eq.condefect.2} depends on whether $N=2$ or $N\neq 2$, we will treat the two cases separately.

\paragraph{Twist Defect.}
We start considering $N\neq 2$. The only difference with the previous construction of condensation operators is that we one-gauge both $L_{(1,\pm 1)}$ and $S_{(1,\pm 1)}$ on an hypersurface $M_3$ such that $\partial M_3=M_2$ and we impose Dirichlet boundary conditions on $M_2$. Thus, we simply have to replace the absolute cohomology with the relative cohomology\footnote{Note that, unlike in $d=2,4$ where, via Lefschetz duality the homology in the summand remains absolute as $H^{d/2}(M_{d/2},M_{d/2-1},\mathbb{Z}_N)\simeq H_{d/2}(M_{d/2},\mathbb{Z}_N)\simeq H^{d/2}(M_{d/2},\mathbb{Z}_N)$, this is not the case in od$d$-dimensions.}. We define the minimal twist defect as follows:
\begin{equation}\begin{split}\label{eq.twistdef.N}
    V^i_{(0,0)}(M_3,M_2)=\frac{1}{|H^1(M_3,M_2,\Xi^i)|}&\sum_{\substack{\gamma,\gamma'\in H_1(M_3,M_2,\Xi^i),\\
    \sigma,\sigma'\in H_2(M_3,M_2,\Xi^i)}}\exp{\left(-\frac{2\pi i}{N}(\langle\sigma,\gamma-\gamma'\rangle +\langle\sigma',\gamma+\gamma'\rangle)\right)} 
    \\
    &  \times  S_{(1,-1)}(\sigma) S_{(1,1)}(\sigma')L_{(1,-1)}(\gamma) L_{(1,1)}(\gamma'), 
\end{split}\end{equation}
Given that we are imposing Dirichlet boundary conditions on $M_2$, if we fuse with lines $L_{(1,\pm 1)}$ or surfaces $S_{(1,\pm 1)}$, the twist defect is not affected
\begin{equation}\begin{split}\label{eq.twistabsorb.Nodd}
    &L_{(1,1)}(\gamma)V_{(0,0)}(M_2,M_3)=V_{(0,0)}(M_2,M_3),\quad L_{(1,-1)}(\gamma')V_{(0,0)}(M_2,M_3)=V_{(0,0)}(M_2,M_3),\\
    &S_{(1,1)}(\sigma)V_{(0,0)}(M_2,M_3)=V_{(0,0)}(M_2,M_3),\quad S_{(1,-1)}(\sigma')V_{(0,0)}(M_2,M_3)=V_{(0,0)}(M_2,M_3).
\end{split}\end{equation}
Moreover, as the commutations relations between two pairs of lines $\gamma,\gamma'$ and surfaces $\sigma,\sigma'$ are trivial it is immediate to satisfy, separately, the first and the second line. The same is not true for the compatibility of operation involving both line and surface operators, due to \eqref{eq.comuteLS}. To make sure that $\langle\gamma,\sigma\rangle$ is trivial we regularize the intersection between them inside $M_2$ to be the intersection between $\gamma$ in $M_2$ and $\sigma$ in $M'_2$, the parallel transport of $M_2$ inside $M_3$ \cite{Kaidi:2022cpf}. Because of this, $\gamma$ and $\sigma$ do not have a submanifold in common and their intersection vanishes.

To understand whether we can construct new twist defects fusing $V_{(0,0)}$ with symmetry operators, we decompose, using the fusion rules \eqref{eq.fusionLS}, a generic line $L_{(l_1,l_2)}$ and surface $S_{(s_1,s_2)}$ operator in terms of its basic constituents. Here, we will find the main difference between $N$ odd and even. Let us now look at all the possibilities, starting from line operators and $l_1\pm l_2\in 2\mathbb{Z}$
\begin{equation}\label{eq.decompL.N1}
    L_{(1,-1)}^{\otimes\frac{l_1-l_2}{2}}(\gamma)\times L_{(1,1)}^{\otimes\frac{l_1+l_2}{2}}(\gamma), 
\end{equation}
for both $N$ even and odd. Then, for $l_1\pm l_2\in 2\mathbb{Z}+1$, we find
\begin{equation}\label{eq.decompL.N2}
    L_{(l_1,l_2)}(\gamma)=
    \begin{cases}
      L_{(1,-1)}^{\otimes\frac{l_1-l_2+N}{2}}(\gamma)\times L_{(1,1)}^{\otimes\frac{l_1+l_2+N}{2}}(\gamma), & N\text{ odd}, \\
      L_{(1,0)}(\gamma)\times L_{(1,-1)}^{\otimes\frac{l_1-l_2+N}{2}}(\gamma)\times L_{(1,1)}^{\otimes\frac{l_1+l_2+N}{2}}(\gamma), & N\text{ even}.
    \end{cases}
\end{equation}
Those for surfaces operators are identical given that they obey the same fusion rules. For $s_1\pm s_2\in 2\mathbb{Z}$
\begin{equation}\label{eq.decompS.N1}
    S_{(1,-1)}^{\otimes\frac{s_1-s_2}{2}}(\gamma)\times S_{(1,1)}^{\otimes\frac{s_1+s_2}{2}}(\gamma), 
\end{equation}
and for $s_1\pm s_2\in 2\mathbb{Z}+1$
\begin{equation}\label{eq.decompS.N2}
    S_{(s_1,s_2)}(\sigma)=
    \begin{cases}
      S_{(1,-1)}^{\otimes\frac{s_1-s_2+N}{2}}(\sigma)\times S_{(1,1)}^{\otimes\frac{s_1+s_2+N}{2}}(\sigma), & N\text{ odd}, \\
      S_{(1,0)}(\sigma)\times S_{(1,-1)}^{\otimes\frac{s_1-s_2+N}{2}}(\sigma)\times S_{(1,1)}^{\otimes\frac{s_1+s_2+N}{2}}(\sigma), & N\text{ even}.
    \end{cases}
\end{equation}
Therefore, we find that for $N$ odd all topological operators in the 4d SymTFT can be absorbed by the minimal twist defect \eqref{eq.twistdef.N}. Instead, for $N$ even\footnote{To stress that these exist only for $N$ even we reinstate the superscript $i=2$ in $V^i$.} we can fuse minimal twist defect with $L_{(1,0)}$ and $S_{(1,0)}$
\begin{equation}\begin{split}\label{eq.twistdef2.Neven}
    &V^2_{(1,0)}(M_2,M_3)=L_{(1,0)}(\gamma)\times V^2_{(0,0)}(M_2,M_3),\\
    &V^2_{(0,1)}(M_2,M_3)=S_{(1,0)}(\sigma)\times V^2_{(0,0)}(M_2,M_3),\\
    &V^2_{(1,1)}(M_2,M_3)=S_{(1,0)}(\sigma)\times L_{(1,0)}(\gamma)\times V^2_{(0,0)}(M_2,M_3).
\end{split}\end{equation}

The twist defects just constructed are interpreted as higher duality interfaces, that is the analog of half spacetime gauging defects \cite{Choi:2021kmx,Kaidi:2021xfk} for higher gauging \cite{Roumpedakis:2022aik}. In particular, in this case we are one-gauging $\mathbb{Z}_N^{(1)}\times\mathbb{Z}_N^{(2)}$ imposing Dirichlet boundary conditions on the codimension-2 boundary. As we will show monetarily via en explicit computation of the fusion rules, the twist defects are non-invertible as the corresponding defects associated to half spacetime (0-)gauging.

We define the orientation reversal of the twist defect starting from $\overline{D}^i(M_3)$ \eqref{eq.barD} and considering the relative (co)homology  
\begin{equation}\begin{split}\label{eq.ortwistdef.N}
    \overline{V}^i_{(0,0)}(M_3,M_2)=\frac{1}{|H^1(M_3,M_2,\Xi^i)|}&\sum_{\substack{(\gamma,\gamma')\in H_1(M_3,M_2,\Xi^i),\\
    (\sigma,\sigma')\in H_2(M_3,M_2,\Xi^i)}}\exp{\left(-\frac{2\pi i}{N}(\langle\sigma,\gamma-\gamma'\rangle-\langle\sigma',\gamma+\gamma'\rangle)\right)}
    \\
    &\times S_{(1,-1)}(\sigma)S_{(1,1)}(\sigma')L_{(1,-1)}(\gamma)L_{(1,1)}(\gamma'),
\end{split}\end{equation}
which unlike for odd-dimensional SymTFTs, it is not the Hermitian conjugate of $V_{(0,0)}$.

We now repeat the same discussion for twist defects for $N=2$. We will be brief highlighting only the differences with the $N\neq 2$ case. Let us recall that, in this case, electric-magnetic symmetry \eqref{eq.emsymmetry} is $\mathbb{Z}_2^\text{EM}$. As before we define the minimal twist defect from the condensation defect \eqref{eq.condefect.2} on a manifold $M_3$ with boundary $M_2$:
\begin{equation}\begin{split}\label{eq.twistdef.N2}
    V^{N=2}_{(0,0)}(M_3,M_2)=\frac{1}{|H^1(M_3,M_2,\bZ_2)|}\sum_{\substack{\gamma \in H_1(M_3,M_2,\mathbb{Z}_2),\\
    \sigma\in H_2(M_3,M_2,\mathbb{Z}_2)}} e^{-\pi i\langle\sigma,\gamma\rangle} \;S_{(1,-1)}(\sigma)L_{(1,-1)}(\gamma).
\end{split}\end{equation}
Due to the choice of Dirichlet boundary conditions the following operators are absorbed by the twist defect
\begin{equation}\label{eq.twistabsorb.N2}
    L_{(1,-1)}(\gamma)V_{(0,0)}(M_2,M_3)=V_{(0,0)}(M_2,M_3),\quad S_{(1,-1)}(\sigma)V_{(0,0)}(M_2,M_3)=V_{(0,0)}(M_2,M_3).
\end{equation}
This leaves the possibility of fusing \eqref{eq.twistdef.N2} with $L_{(1,0)}$ and $S_{(1,0)}$, creating three new twist defects, exactly as for $N$ even \eqref{eq.twistdef2.Neven}
\begin{equation}\begin{split}\label{eq.twistdef2.N2}
    &V^{N=2}_{(1,0)}(M_2,M_3)=L_{(1,0)}(\gamma)\times V^{N=2}_{(0,0)}(M_2,M_3),\\
    &V^{N=2}_{(0,1)}(M_2,M_3)=S_{(1,0)}(\sigma)\times V^{N=2}_{(0,0)}(M_2,M_3),\\
    &V^{N=2}_{(1,1)}(M_2,M_3)=S_{(1,0)}(\sigma)\times L_{(1,0)}(\gamma)\times V^{N=2}_{(0,0)}(M_2,M_3).
\end{split}\end{equation}
It is immediate to check that fusing with $L_{(0,1)}$ and $S_{(0,1)}$ produces the same twist defects.

As before, we define the orientation reversal of the twist defect as follows
\begin{equation}\label{eq.ortwistdef.N2}
    \overline{V}^{N=2}_{(0,0)}(M_3,M_2)=\frac{1}{|H^1(M_3,M_2,\bZ_2)|}\sum_{\substack{\gamma \in H_1(M_3,M_2,\mathbb{Z}_2),\\
    \sigma\in H_2(M_3,M_2,\mathbb{Z}_2)}} e^{\pi i\langle\sigma,\gamma\rangle} \;S_{(1,-1)}(-\sigma)L_{(1,-1)}(-\gamma).
\end{equation}

\paragraph{Fusion Rules.}
Starting from $N\neq 2$, we compute the fusion rules involving twist defects. Those with line and surface operators can be immediately derived from the decomposition of generic topological operators \eqref{eq.decompL.N1}-\eqref{eq.decompS.N2} and the fusion rules of $L_{(1,\pm 1)}$ and $S_{(1,\pm 1)}$ with the minimal twist defect \eqref{eq.twistabsorb.Nodd}. For $N$ odd the fusion rules confirm that generic line and surface operators can be absorbed by the minimal twist defect
\begin{equation}\begin{split}\label{eq.fusionLSV.Nodd}
    &L_{(l_1,l_2)}(\gamma)\times V_{(0,0)}(M_2,M_3)=V_{(0,0)}(M_2,M_3),\\
    &S_{(l_1,l_2)}(\sigma)\times V_{(0,0)}(M_2,M_3)=V_{(0,0)}(M_2,M_3),
\end{split}\end{equation}
For $N$ even we find
\begin{equation}\begin{split}\label{eq.fusionLSV.Neven}
    &L_{(l_1,l_2)}(\gamma)\times V_{(\rmd{l}{2},\rmd{s}{2})}(M_2,M_3)=V_{(\rmd{l+l_1-l_2}{2},\rmd{s}{2})}(M_2,M_3),\\
    &S_{(l_1,l_2)}(\sigma)\times V_{(\rmd{l}{2},\rmd{s}{2})}(M_2,M_3)=V_{(\rmd{l}{2},\rmd{s+s_1-s_2}{2})}(M_2,M_3),
\end{split}\end{equation}
where we defined $\rmd{n}{2}$ as the remainder of the division of $n$ by 2.

Next, we calculate the fusion between the minimal twist defect and its orientation reversal \eqref{eq.ortwistdef.N}. The geometry of $M_3$ near the boundary can be written as $M_2 \times \bR_+$. Let $x$ be the coordinate on $\bR_+$. Insert $ V_{(0,0)}(M_2,M_3)$ along the region $M_3^{\geq 0}$ and $\overline{V}_{(0,0)}(M_2,M_3)$ along the region $M_3^{\geq \epsilon}$. The fusion of them is 
\begin{equation}\begin{split}
    &V_{(0,0)}(M_2,M_3)\times\overline{V}_{(0,0)}(M_2,M_3)
     = \frac{|H^1(M_3^{\geq \epsilon},M_2,\Xi)|^{-1}}{|H^1(M_3^{\geq 0},M_2,\Xi)|}\sum_{\substack{(\gamma_1,\gamma'_1) \in H_1(M_3^{\geq 0},M_2,\Xi),\\
     (\sigma_1,\sigma_1') \in H_2(M_3^{\geq 0},M_2,\Xi)
    }}\sum_{\substack{
    (\gamma_2,\gamma'_2) \in H_1(M_3^{\geq \epsilon},M_2,\Xi),\\
    (\sigma_2,\sigma_2') \in H_2(M_3^{\geq \epsilon},M_2,\Xi)}}\\  
    & S_{(1,-1)}(\sigma_1) S_{(1,1)}(\sigma'_1)  L_{(1,-1)}(\gamma_1) L_{(1,1)}(\gamma'_1)S_{(1,-1)}(\sigma_2) S_{(1,1)}(\sigma'_2) L_{(1,-1)}(\gamma_2)L_{(1,1)}(\gamma'_2)\\
    & \exp{\left(-\frac{2\pi i}{N} ( \langle \sigma_1,\gamma_1-\gamma'_1 \rangle + \langle \sigma'_1, \gamma_1+\gamma'_1 \rangle + 
    \langle \sigma_2,\gamma_2+\gamma'_2 \rangle + \langle \sigma'_2, \gamma'_2-\gamma_2 
    \rangle) \right)}
\end{split}\end{equation}
This can be computed directly using a method similar to that in \autoref{app.A}. Thus, we convert the sums over relative cohomology into sums over cochains by introducing additional fields and appropriate BF terms. Integrating out $(\gamma_2,\gamma'_2)$ and $(\sigma_2,\sigma_2')$, we find
\begin{equation}\begin{split}
     & \frac{1}{|H^1(M_3^{[0,\epsilon] },M_2|_{0} \cup M_2|_{\epsilon},\Xi)|}
    \sum_{\substack{
    (\gamma_1,\gamma'_1) \in H_1(M_3^{[0,\epsilon]},M_2|_{0} \cup M_2|_{\epsilon},\Xi)
    ,\\
    (\sigma_1,\sigma_1') \in H_2(M_3^{[0,\epsilon]},M_2|_{0} \cup M_2|_{\epsilon},\Xi)}} 
    S_{(1,-1)}(\sigma_1) S_{(1,1)}(\sigma'_1)  L_{(1,-1)}(\gamma_1) L_{(1,1)}(\gamma'_1) \\
    & \hspace{6.5cm} \times  \exp{\left(-\frac{2\pi i}{N} ( \langle \sigma_1,\gamma_1-\gamma'_1 \rangle + \langle \sigma'_1, \gamma_1+\gamma'_1 \rangle ) \right)},
\end{split}\end{equation}
where $M_3^{[0,\epsilon]}=M_2\times I^{[0,\epsilon]}$ is the product of an oriented surface and an interval. On the interval we can regularize lines and surfaces to have trivial intersection pairing (this will hold also after shrinking the slab $\epsilon \to 0$). Finally, the fusion becomes 
\begin{equation}\begin{split}\label{eq.VV}
 & V_{(0,0)}(M_2,M_3)\times\overline{V}_{(0,0)}(M_2,M_3)=  \\
     &  \hspace{3.5cm}  \frac{1}{|H^0(M_2,\Xi)|}
    \sum_{\substack{
    (\gamma,\gamma') \in H_1(M_2,\Xi)
    ,\\
    (\sigma,\sigma') \in H_2(M_2,\Xi)}} 
    S_{(1,0)}(\sigma) S_{(0,1)}(\sigma')  L_{(1,0)}(\gamma) L_{(0,1)}(\gamma'),
\end{split}\end{equation}
where we used $|H^1(M_3^{[0,\epsilon]},M_2|_{0}\cup M_2|_{\epsilon},\bZ_N)|=1$. With similar computations one can compute the fusion rule of the minimal twist defect with itself.

For $N=2$, the fusion rules between line and surface operators and twist defects in \eqref{eq.twistdef.N2} and \eqref{eq.twistdef2.N2} are
\begin{equation}\begin{split}
    &L_{(l_1,l_2)}(\gamma)\times V_{(l,s)}(M_2,M_3)=V_{(l+l_1-l_2,s)}(M_2,M_3),\\
    &S_{(l_1,l_2)}(\sigma)\times V_{(l,s)}(M_2,M_3)=V_{(l,s+s_1-s_2)}(M_2,M_3),
\end{split}\end{equation}
and follow from those for $N$ even \eqref{eq.fusionLSV.Neven}.

The fusion between two twist defects for $N=2$ can be evaluated with a similar approach as the one for $N\neq 2$. One can consider a a manifold $M_3=M_2 \times \bR_+$ and place $V_{(0,0)}(M_3,M_2)$ in $M_3^{\geq 0}$ while $V_{(0,0)}(M_3,M_2)$ in $M_3^{\geq \epsilon}$. Their fusion is given by 
\begin{equation}\begin{split}
    &V_{(0,0)}(M_3,M_2)\times\overline{V}_{(0,0)}(M_3,M_2)=\frac{1}{|H^1(M_3^{\geq 0},M_2|_0,\bZ_2)|} \frac{1}{|H^1(M_3^{\geq \epsilon},M_2|_{\epsilon},\bZ_2)|}  \; \times \\\\
    &
    \sum_{\substack{\gamma \in H_1(M_3^{\geq 0},M_2,\mathbb{Z}_2),\\
    \sigma\in H_2(M_3^{\geq 0},M_2,\mathbb{Z}_2)}} \sum_{\substack{\gamma' \in H_1(M_3^{\geq \epsilon},M_2,\mathbb{Z}_2),\\
    \sigma'\in H_2(M_3^{\geq \epsilon},M_2,\mathbb{Z}_2)}}
    e^{\pi i(\langle\sigma,\gamma\rangle+\langle\sigma',\gamma'\rangle)} \;S_{(1,-1)}(\sigma)L_{(1,-1)}(\gamma)S_{(1,-1)}(\sigma')L_{(1,-1)}(\gamma').
\end{split}\end{equation}
Similar to the calculations in \autoref{app.A}, one can show that  
\begin{equation}\begin{split}
    &\frac{1}{|H^1(M_3^{[0,\epsilon]},M_2|_0 \cup M_2|_{\epsilon},\bZ_2)|}   \; \times 
    \sum_{\substack{\gamma \in H_1(M_3^{[0,\epsilon]},\mathbb{Z}_2),\\
    \sigma\in H_2(M_3^{[0,\epsilon]},\mathbb{Z}_2)}} 
    e^{\pi i\langle\sigma,\gamma\rangle} \;S_{(1,-1)}(\sigma)L_{(1,-1)}(\gamma).
\end{split}\end{equation}
Since the intersection paring is trivial on the interval $M_3^{[0,\epsilon]}$ (and it will hold also after shrinking the slab), we obtain 
\begin{equation}\begin{split}\label{eq.VV2}
    &V_{(0,0)}(M_3,M_2)\times\overline{V}_{(0,0)}(M_3,M_2)= \frac{1}{|H^0(M_2,\bZ_2)|}   
    \sum_{\substack{\gamma \in H_1(M_2,\mathbb{Z}_2),\\
    \sigma\in H_2(M_2,\mathbb{Z}_2)}} 
    \;S_{(1,-1)}(\sigma)L_{(1,-1)}(\gamma).
\end{split}\end{equation}

\subsection{Duality Interfaces from Twist Defects}\label{sec.3.3}
We have defined line operators $L_{(l_1,l_2)}$, surface operators $S_{(s_1,s_2)}$ and the minimal twist defect $V^i_{(0,0)}$ in the interior of the slab. Our goal now is to show that these operators give rise, upon shrinking the slab where the 4d SymTFT is defined, to the different (topological and non-topological) operators in the 3d theory $\mathcal{Q}$ discussed in \autoref{sec.2}. In this subsection, we assume that on the topological boundary we pick Dirichlet boundary conditions \eqref{eq.Dirichletbc}. The generalization to Neumann boundary conditions \eqref{eq.Neumannbc} is straightforward and exchanges the role of electric and magnetic objects.

\paragraph{Electric Lines and Surfaces.}
As we imposed Dirichlet conditions, the electric gauge fields vanish at the topological boundary at $x=0$
\begin{equation}
    a_1|_{M_2|_0}=b_1|_{M_2|_0}=0.
\end{equation}
Thus, an electric line $L_{(1,0)}$ or surface $S_{(1,0)}$ operator survives only if we place it perpendicularly to the boundary, with the other endpoint (endline for surface operators) at $x=\epsilon$ attached to the dynamical boundary conditions \eqref{eq.dynamicalbc}. Because of the latter boundary conditions, they become non-topological operators upon shrinking the slab. Line operators give rise to point-like operators $\mathcal{O}_0$ in the 3d theory, while surface operators to line operators $\mathcal{O}_1$. We will show momentarily that these objects are charged, respectively, under the $\mathbb{Z}^{(0)}_N\times\mathbb{Z}_N^{(1)}$ symmetry and serve as order parameters.

\paragraph{Magnetic Lines and Surfaces.}
On the contrary, magnetic operators $L_{(0,1)}$ and $S_{(0,1)}$ cannot end neither on the Dirichlet boundary nor be absorbed by it. Thus, we place these operators parallel to the topological boundary and when we shrink the slab, they remain topological defects. In particular $L_{(0,1)}$ gives rise to the topological line $\eta_1$ \eqref{eq.Qeta1} generating the $\mathbb{Z}_N^{(1)}$ symmetry while $S_{(0,1)}$ to the topological surface $\eta_0$ \eqref{eq.Qeta0} generating $\mathbb{Z}_N^{(0)}$. As we have observed earlier, in the 4d SymTFT a curve (surface) has trivial linking number with other curves (surfaces). Instead, $L_{(1,0)}$ and $S_{(0,1)}$ can have non-trivial linking number and thus correlations functions involving $\eta_0$ and $\mathcal{O}_0$ \eqref{eq.correlation.sigmagamma} determine the charges of the order parameter $\mathcal{O}_0$ under $\mathbb{Z}_N^{(0)}$ through their linking numbers. Similarly, one can determine the charge of $\mathcal{O}_1$ under the $\mathbb{Z}_N^{(1)}$ symmetry generated by $\eta_1$.

\paragraph{Twist Defects.}
The effect of inserting a condensation defect $D(M_3)$ \eqref{eq.condefect.N} parallel to the two boundaries is that of exchanging the topological boundary from Dirichlet boundary conditions \eqref{eq.Dirichletbc} to Neumann boundary conditions \eqref{eq.Neumannbc}
\begin{equation}\begin{split}
    \langle D(A,B)|\xrightarrow{D}&\sum_{\substack{a_2\in H^1(M_3|_{\epsilon};\mathbb{Z}_N),\\
    b_2\in H^2(M_3|_{\epsilon};\mathbb{Z}_N)}}
    \langle a_2,b_2| \delta(a_2-A) \delta(b_2-B)\\
    =&\sum_{\substack{a_1\in H^1(M_3|_{\epsilon};\mathbb{Z}_N),\\
    b_1\in H^2(M_3|_{\epsilon};\mathbb{Z}_N)}}
    \langle a_1,b_1| \exp{\left( \frac{2\pi i}{N} \int_{M_3|0} a_1 B + b_1A\right)}\\
    =&\langle N(A,B)|.
\end{split}\end{equation}
In the second line we used the fact that electric-magnetic dual bases are related by a discrete Fourier transformation.
\begin{figure}[!tbp]
\centering
\[\begin{tikzpicture}[baseline=50,scale=1.25]
\draw[blue, thick] (2,1)--(2,2.0);
\draw[red, thick] (2,1)--(2,0);
\node[blue] at (2,2.3) {$\mathcal{Q}$};
\node[blue] at (2,-0.3) {$\mathcal{Q}/(\mathbb{Z}^{(0)}_N\times\mathbb{Z}^{(1)}_N)$};
\draw[fill=red] (2,1) circle [radius=0.05cm];
\node[red] at (1.5,1) {$\mathcal{N}_2$};
\draw[thick,dotted,<-] (3,1) -- (4,1);
\draw[blue, thick] (5,0)--(5,2);
\draw[blue, thick] (9,0)--(9,2);
\node[blue] at (5,2.3) {$\langle D(A,B)|$};
\node[blue] at (9,2.3) {$|\mathcal{Q}\rangle$};
\node[below] at (5,0) {$x=0$};
\node[below] at (9,0) {$x=\epsilon$};
\draw[thick,red] (7,0) -- (7,1);
\draw[fill=red] (7,1) circle [radius=0.05cm];
\node[red] at (7,1.3) {$V_{(0,0)}$};
\filldraw[draw=blue,fill=blue!20,opacity=0.5]       
    (5,0) -- (9,0) -- (9,2) -- (5,2) -- cycle;
\end{tikzpicture}\]
\caption{The interface $\mathcal{N}_2$ exchanging Dirichlet and Neumann boundary conditions is obtained, upon shrinking the slab, from a twist defect $V_{(0,0)}$ inserted parallel to the boundary.}
\label{Fig:SymTFT3}
\end{figure}
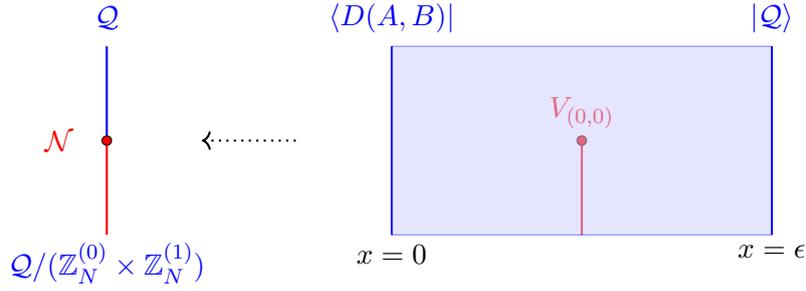

Similarly, as shown in \autoref{Fig:SymTFT3}, we can insert a twist defect $V_{(0,0)}(M_2,M_3)$ \eqref{eq.twistdef.N} and study its behaviour upon shrinking the slab. As the defect now has a boundary itself, it can be chosen so that it acts only on half of the topological boundary. Hence, the effect is that of exchanging Dirichlet with Neumann boundary conditions only in this half of the 3d boundary. The theory living on this half of spacetime is $\mathcal{Q}/(\mathbb{Z}_N^{(0)}\times\mathbb{Z}_N^{(1)})$ with the partition function given by \eqref{Eq:sigmaG}. Thus, the twist defect gives rise to the duality interface\footnote{For $N$ even there exist also non-minimal twist defects \eqref{eq.twistdef2.Neven}. However, due to the choice of Dirichlet boundary conditions, they all give rise to the same duality interface.} $\mathcal{N}_2$. Momentarily, we will assume that the 3d theory $\mathcal{Q}$ is invariant under the simultaneous gauging of $\mathbb{Z}_N^{(0)}\times\mathbb{Z}_N^{(1)}$, then the twist defect will become a duality defect upon shrinking the slab.

To summarize, we identify operators in the SymTFT with those of the theory $\mathcal{Q}$ of \autoref{sec.2}, as follows:
\begin{equation}\begin{split}\label{eq.idenSymtft}
     L_{(1,0)}(\gamma),\,S_{(1,0)}(\sigma)&\longleftrightarrow\mathbb{Z}_N^{(1)}\times\mathbb{Z}_N^{(0)}\text{ order parameters }\mathcal{O}_0,\,\mathcal{O}_1,\\
     L_{(0,1)}(\gamma),\,S_{(0,1)}(\sigma)&\longleftrightarrow\mathbb{Z}_N^{(1)}\times\mathbb{Z}_N^{(0)}\text{ symmetry defects }\eta_1(\gamma),\,\eta_0(\sigma),\\
     V_{(0,0)}(M_3,M_2) &\longleftrightarrow \text{Duality interface }\cN_2(M_3,M_2)
\end{split}\end{equation}

\paragraph{Fusion Rules.}
We can reproduce the fusion rules of the duality interface \eqref{eq.fusion.mathcalN} from those of the twist defects just computed. The fusion rules
\begin{equation}
    \eta_i\times\mathcal{N}_2=\mathcal{N}_2
\end{equation}
follow immediately from \eqref{eq.fusionLSV.Nodd}-\eqref{eq.fusionLSV.Neven}. More interesting are those involving $\mathcal{N}_2$ and its orientation reversal. For odd $N$, one can verify that the fusion rule \eqref{eq.VV} becomes 
\begin{equation} 
\begin{split}
    \cN_2(M_3,M_2)\times \overline{\cN}_2 (M_3,M_2)  
    &= \frac{1}{|H^0(M_2,\bZ_N)|}
    \sum_{\substack{
    \gamma' \in H_1(M_2,\bZ_N)
    ,\\
    \sigma' \in H_2(M_2,\bZ_N)}} 
     S_{(0,1)}(\sigma')  L_{(0,1)}(\gamma')\\
    & = \left(\sum_{i=0}^{N-1} \eta_0^i(M_2) \right) \left(\frac{1}{|H^0(M_2,\bZ_N)|} 
    \sum_{\substack{\gamma'\in H_1(M_2,\mathbb{Z}_N)}}\eta_1(\gamma')\right)
\end{split}
\end{equation}
since, upon shrinking the slab, $\gamma\in H_1(M_2,M_2,\bZ_N)$ and $\sigma\in H_2(M_2,M_2,\bZ_N)$ are trivial as $|H_1(M_2,M_2,\bZ_N)|=|H_2(M_2,M_2,\bZ_N)|=1$. 

Similarly, we consider even $N$. Recall that the coefficient in homology is $\Xi^2 = (\bZ_N \times \bZ_N)/\bZ_2$ for even $N$. Thus, the summation over ${\gamma,\gamma'} \in (\bZ_N \times \bZ_N)/\bZ_2$ is equivalent to sums $\gamma \in H_1(M_2,\bZ_{N/2})$ and $\gamma' \in H_1(M_2,\bZ_{N})$. By shrinking the slab, $\gamma \in H_1(M_2,M_2,\bZ_{N/2})$ and similarly for $\sigma$. Thus, the fusion rule \eqref{eq.VV} becomes 
\begin{equation} 
\begin{split}
    \cN_2(M_3,M_2)\times \overline{\cN}_2 (M_3,M_2) 
    &= \frac{1}{|H^0(M_2,\bZ_N)|}
    \sum_{\substack{
    \gamma' \in H_1(M_2,\bZ_N)
    ,\\
    \sigma' \in H_2(M_2,\bZ_N)}} 
     S_{(0,1)}(\sigma')  L_{(0,1)}(\gamma')\\
    & = \left(\sum_{i=0}^{N-1} \eta_0^i(M_2) \right) \left(\frac{1}{|H^0(M_2,\bZ_N)|} 
    \sum_{\substack{\gamma \in H_1(M_2,\mathbb{Z}_N)}} \eta_1(\gamma)  \right)
\end{split}
\end{equation}
The contributions from $\gamma$ and $\sigma$ vanish as $|H_1(M_2,M_2,\bZ_N)|=|H_2(M_2,M_2,\bZ_N)|=1$.

For $N=2$, the way line and surface operators in the 4d SymTFT behave upon shrinking the slab is identical to the $N>2$ case. Minimal twist defects give rise to duality interfaces on the boundary. Given that generic twist defects can be obtained fusing $L_{(1,0)}$ or $S_{(1,0)}$ with $V_{(0,0)}$ \eqref{eq.twistdef2.N2}, and these electric lines and surfaces can be absorbed by the boundary, all twist defects give rise to the same interface duality on the boundary.

With the identification in \eqref{eq.idenSymtft}, one can verify that the fusion rule \eqref{eq.VV2} becomes 
\begin{equation} 
\begin{split}
    \cN_2(M_3,M_2)\times \cN_2 (M_3,M_2) =\left(
    1+
   \eta_0\right) \left(\frac{1}{|H^0(M_2,\bZ_2)|} 
    \sum_{\substack{\gamma \in H_1(M_2,\mathbb{Z}_2)}}\eta_1(\gamma)\right)
\end{split}
\end{equation}

\subsection{SymTFT for Duality Defects}
We now require that the 3d theory is invariant under the simultaneous gauging of $\mathbb{Z}^{(0)}_N\times\mathbb{Z}^{(1)}_N$, that is $\mathcal{Q}\cong\mathcal{Q}/(\mathbb{Z}_N^{(0)}\times\mathbb{Z}_N^{(1)})$. Thus, the first consequence will be that the duality interfaces just described become duality defects \eqref{Eq:sigmaG}. To obtain the correct SymTFT we need to gauge the\footnote{For $N=2$ we need to gauge $\mathbb{Z}_2^\text{EM}$. Moreover, to obtain triality defects studied in \autoref{sec.2.1} we would need to do a twisted gauging by stacking with an SPT phase. We will not consider these cases in our work.} $\mathbb{Z}_4^\text{EM}$ symmetry \eqref{eq.emsymmetry} implemented by the condensation defect $D(M_3)$ \eqref{eq.condefect.N}-\eqref{eq.condefect.2} and consider the $\widehat{\mathbb{Z}}^{(2)}_4$ quantum symmetry originating from such gauging. The full symmetry of the SymTFT is a higher categorical analog of the Tambara-Yamagami category $\text{TY}(\mathbb{Z}_N^{(0)}\times\mathbb{Z}_N^{(1)})$.

\paragraph{Topological Operators.}
We now consider the topological operators studied in \autoref{sec.3.3} and explain how they behave upon gauging $\mathbb{Z}_4^\text{EM}$. As it is standard when gauging a symmetry, we need to keep only gauge invariant objects in the theory. Moreover, in the gauged SymTFT there is an $\widehat{\mathbb{Z}}_4^{(2)}$ ($\widehat{\mathbb{Z}}_2^{(2)}$ for $N=2$) quantum symmetry generated by a line operator $K$, such that $K^4=1$. Thus, if the operators we find in the new theory arise from $\mathbb{Z}_4^\text{EM}$-invariant in the pregauged theory, they can be fused with $K^q$, $q=1,2,3$, and thus be assigned a representation of $\mathbb{Z}_4^\text{EM}$.

Recalling that $\mathbb{Z}_4^\text{EM}$ acts on line and surface operators as in \eqref{eq.emls}
\begin{equation}
    \mathbb{Z}_4^{\text{EM}}:\qquad L_{(l_1,l_2)}\rightarrow L_{(-l_2,l_1)},\quad S_{(s_1,s_2)}\rightarrow S_{(-s_2,s_1)},
\end{equation}
the only simple line or surface gauge invariant operators are:
\begin{equation}\begin{split}\label{eq.ob1}
    N\mbox{ odd:}\quad&\widehat{L}_{(0,0)}\equiv L_{(0,0)},\quad\widehat{S}_{(0,0)}\equiv S_{(0,0)},\\
    N=2\text{:}\quad&\widehat{L}_{(0,0)}\equiv L_{(0,0)},\quad\widehat{L}_{(1,1)}\equiv L_{(1,1)},\quad\widehat{S}_{(0,0)}\equiv S_{(0,0)},\quad\widehat{S}_{(1,1)}\equiv S_{(1,1)},\\
    N\geq 2\mbox{ even:}\quad&\widehat{L}_{(0,0)}\equiv L_{(0,0)},\quad\widehat{L}_{(N/2,N/2)}\equiv L_{(N/2,N/2)},\\
    &\widehat{S}_{(0,0)}\equiv S_{(0,0)},\quad\widehat{S}_{(N/2,N/2)}\equiv S_{(N/2,N/2)}.
\end{split}\end{equation}
These can be fused with the line generating the quantum $\widehat{\mathbb{Z}}_4^{(2)}$ symmetry as follows
\begin{equation}\label{eq.ob2}
    \widehat{L}_{(l_1,l_2)}^q=K^q\times \widehat{L}_{(l_1,l_2)},\qquad\widehat{S}^q_{(s_1,s_2)}=K^q\times \widehat{S}_{(s_1,s_2)},
\end{equation}
where $q=0,...,3$ and $l_1,l_2,s_1,s_2=0$ for $N$ odd and $l_1,l_2,s_1,s_2=0,N/2$ for $N$ even.

In addition to these lines and surfaces we can also consider combinations of either lines $L_{l_1,l_2}\oplus L_{l_1',l_2'}$ or surfaces $S_{s_1,s_2}\oplus S_{s_1',s_2'}$ which are not simple in the ungauged theory but become so after gauging. From the action of $\mathbb{Z}_4^\text{EM}$, the conditions on the charges are
\begin{equation}
    (-l_2,l_1)=(l_1',l_2'),\qquad(-l_2',l_1')=(l_1,l_2)
\end{equation}
and similarly for $s_1,s_2,s_1',s_2'$. As we need to restrict to combinations of operators which are not gauge invariant by themselves, we do not find any new operator for $N$ odd. Instead, for $N$ even we find:
\begin{equation}\label{eq.ob3}
    N\mbox{ even:}\qquad\widehat{L}_{\{N/2,0\}}\equiv L_{(N/2,0)}\oplus L_{(0,N/2)},\quad\widehat{S}_{\{N/2,0\}}\equiv S_{(N/2,0)}\oplus S_{(0,N/2)}.
\end{equation}
Similarly one can consider the sum of either four line operators or surface operators
\begin{equation}\begin{split}\label{eq.ob4}
    &\widehat{L}_{[l_1,l_2]}\equiv L_{(l_1,l_2)}\oplus L_{(-l_2,l_1)}\oplus L_{(-l_1,-l_2)}\oplus L_{(l_2,-l_1)},\\
    &\widehat{S}_{[s_1,s_2]}\equiv S_{(s_1,s_2)}\oplus S_{(-s_2,s_1)}\oplus S_{(-s_1,-s_2)}\oplus S_{(s_2,-s_1)}.
\end{split}\end{equation}
For $N$ odd there $(N^2-1)/2$ such lines and surfaces are while for $N$ even there $N^2/2-2$ of them. In particular there are none for $N=2$. All these operators are not invariant under $\mathbb{Z}_N^\text{EM}$ in the pregauged theory, thus, they absorb the lines $K^q$.

\begin{figure}[!tbp]
\centering
\[\begin{tikzpicture}[baseline=50,scale=1.25]
\draw[blue, thick] (2,-0)--(2,2.0);
\node[blue] at (2,2.3) {$\mathcal{Q}$};
\node[blue] at (2,-0.3) {$\mathcal{Q}/(\mathbb{Z}^{(0)}_N\times\mathbb{Z}^{(1)}_N)\cong\mathcal{Q}$};
\draw[fill=red] (2,1) circle [radius=0.05cm];
\node[red] at (1.5,1) {$\mathcal{N}_2$};
\draw[thick,dotted,<-] (3,1) -- (4,1);
\draw[blue, thick] (5,0)--(5,2);
\draw[blue, thick] (9,0)--(9,2);
\node[blue] at (5,2.3) {$\langle D(A,B)|$};
\node[blue] at (9,2.3) {$|\mathcal{Q}\rangle$};
\node[below] at (5,0) {$x=0$};
\node[below] at (9,0) {$x=\epsilon$};
\draw[thick,red,dashed] (7,0) -- (7,1);
\draw[fill=red] (7,1) circle [radius=0.05cm];
\node[red] at (7,1.3) {$V_{(0,0)}$};
\filldraw[draw=blue,fill=blue!20,opacity=0.5]       
    (5,0) -- (9,0) -- (9,2) -- (5,2) -- cycle;
\end{tikzpicture}\]
\caption{The bulk of the twist defect $V_{(0,0)}$ becomes transparent in 4d. Upon shrinking the slab it gives rise to the topological defect $\mathcal{N}_2$ for the non-invertible symmetry.}
\label{Fig:SymTFT4}
\end{figure}
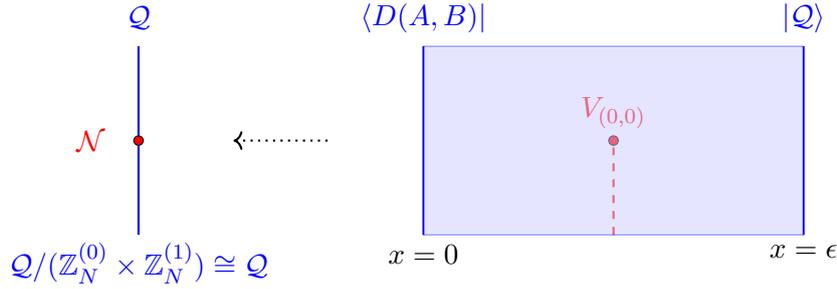

Moreover, let us consider the minimal twist defect $V_{(0,0)}(M_3,M_2)$. These were constructed from condensation defects $D(M_3)$ on three-manifolds with boundaries. Upon gauging, the bulk of the twist defect becomes transparent (as we are restricting to objects invariant under $D(M_3)$), see \autoref{Fig:SymTFT4}. Hence, $V_{(0,0)}(M_3,M_2)$ becomes a surface operator in the gauged SymTFT. Upon shrinking the slab the twist defect gives rise to the duality defect $\mathcal{N}_2$ introduced in \autoref{sec.2}. Note that for $N$ even we found also non-minimal twist defects $V_{(1,0)},V_{(0,1)},V_{(1,1)}$. All these operators are invariant under $\mathbb{Z}_4^\text{EM}$ in the pregauged theory. Thus, we find:
\begin{equation}\begin{split}\label{eq.ob5}
    N\mbox{ odd:}\quad&\widehat{V}_{(0,0)}\equiv K^q V_{(0,0)},\\
    N\mbox{ even:}\quad&\widehat{V}_{(0,0)}\equiv K^qV_{(0,0)},\quad\widehat{V}_{(1,0)}\equiv K^qV_{(1,0)},\quad\widehat{V}_{(0,1)}\equiv K^qV_{(0,1)}\quad\widehat{V}_{(1,1)}\equiv K^qV_{(1,1)},
\end{split}\end{equation}
for $q=0,...,3$.

Finally, we should also include non-genuine line operators living on the surface operators we just described
\begin{equation}\begin{split}\label{eq.nongenuineline}
    N\mbox{ odd:}\quad &\mathfrak{J}^q_1(\widehat{V}_{(0,0)}),\quad\mathfrak{J}^q_1(\widehat{S}_{(0,0)}),\quad\mathfrak{J}_1(\widehat{S}_{[s_1,s_2]})\\
    N\mbox{ even:}\quad &\mathfrak{J}^q_1(\widehat{V}_{(0,0)}),\quad\mathfrak{J}^q_1(\widehat{V}_{(1,0)}),\quad\mathfrak{J}^q_1(\widehat{V}_{(0,1)}),\quad\mathfrak{J}^q_1(\widehat{V}_{(1,1)}),\quad\mathfrak{J}^q_1(\widehat{S}_{(0,0)}),\quad\mathfrak{J}^q_1(\widehat{S}_{(N/2,N/2)}),\\
    &\mathfrak{J}^p_1(\widehat{S}_{\{N/2,0\}}),\quad\mathfrak{J}_1(\widehat{S}_{[s_1,s_2]}).
\end{split}\end{equation}
We need to explain this notation. First, $\mathfrak{J}_1^0(\cdot)$ denotes the identity operator living on each two-manifold. Then, as $\widehat{S}_{(0,0)}$ is the trivial surface operator, $\mathfrak{J}^0_1(\widehat{S}_{(0,0)})$ is a trivial genuine line. Furthermore, following the same logic as in previous cases, we can stack the line operator $K$ with these line operators and this results in the superscript $q$ in \eqref{eq.nongenuineline}. Note that $\mathfrak{J}^0_1(\widehat{S}_{\{N/2,0\}})$ cannot absorb $K$ but can absorb $K^2$ \cite{Kaidi:2022cpf}. This explains why there exist only $p=0,1$ of them.

In conclusion, the operators in \eqref{eq.ob1},\eqref{eq.ob2},\eqref{eq.ob3},\eqref{eq.ob4},\eqref{eq.ob5} and \eqref{eq.nongenuineline} give all the morphisms of the higher category of the 4d $\mathbb{Z}_4^\text{EM}$ gauged SymTFT.

\paragraph{Fusion Rules.}
The fusions rules of genuine operators in the gauged SymTFT descend from those of the ungauged SymTFT we derived earlier. The only novelty here is to determine the distribution of lines $K$ and thus the dependence on $\widehat{\mathbb{Z}}^{(2)}_4$. Moreover, fusion rules involving non-genuine operators can be computed following \cite{Kaidi:2022cpf}. We leave an explicit computations of these fusion rules to future work.

\section{Examples and Applications}\label{sec.4}

In this section, we provide some explicit examples of theories admitting non-invertible symmetry defects defined in \autoref{sec.2} and whose fusion rules have been computed in \autoref{sec.2.4}. First, let us mention that one such example is already present in \cite{Kaidi:2021xfk} for $SO(N)_K$ gauge theories with $N_f$ fermions. As we shall briefly discuss below, this example relies on a mixed 't Hooft anomaly among two 0-form symmetries and a 1-form symmetry and it is thus non-intrinsically non-invertible \cite{Sun:2023xxv}. Instead, the main novelty of our examples is that they are independent on the existence of 't Hooft anomalies.






\subsection{$U(1)\times U(1)$ Gauge Theory}

Let us first look at 2+1d Maxwell theory. It has an electric 1-form symmetry $U(1)^{(1)}$ and a magnetic 0-from symmetry $U(1)^{(0)}$. However, if we want to gauge subgroups $\mathbb{Z}^{(1)}_N$ and $\mathbb{Z}^{(0)}_{N'}$ there is a mixed 't Hooft anomaly whenever $\gcd(N,N')\neq 1$. Therefore, one cannot realize the duality defects studied above which require $N=N'$. This motivates us to consider a $U(1)\times U(1)$ gauge theory. Since the two copies of the $U(1)$ gauge theory are completely decoupled, it is allowed to gauge simultaneously a subgroup of the electric symmetry of the first $U(1)$ and a subgroup of the magnetic theory in the second $U(1)$. 

Hence, we consider $U(1) \times U(1)$ gauge theory on 2+1d space $M_3$. The action is  
\begin{equation}
    S = \frac{1}{2e_1^2}\int_{M_3} F_1 \wedge \star F_1 + \frac{1}{2e_2^2}\int_{M_3}  F_2 \wedge \star F_2
\end{equation}
where $e_i$ are the gauge couplings and $F_i=dA_i$ are the field strength of the two dynamical 1-form gauge fields $A_i$ for $i=1,2$. The gauge transformations are $A_i \to A_i + d \alpha_i$. By 3d electric-magnetic duality acting on each $U(1)$ factor, this theory can be equivalently described by two compact scalars with action 
\begin{equation}
    S = \frac{e_1^2}{8\pi^2}\int_{M_3} d\phi_1 \wedge \star d\phi_1 + \frac{e_2^2}{8\pi^2}\int_{M_3}  d\phi_2 \wedge \star d\phi_2
\end{equation}
where $\phi_i = \phi_i+2\pi$. As it is well known, the field strength of $A_i$ and the dual photon $\phi_i$ are related by 
\begin{equation}
    d\phi_i = \frac{2\pi i}{e_i^2} \star F_i.
\end{equation}
Thus, we recover a sigma model on $M_3$ with target space $T^2$ with the standard $SL(2,\mathbb{Z})$ group acting on the torus. Because of this, the dual theory written in terms of the gauge field $A_i$ also has a $SL(2,\bZ)$ duality. Let $\tau = ie_1/e_2$ be a dimensionless coupling constant. The $S$ and $T$ transformation act on $\tau$ as %
\begin{equation} \label{eq:maxwell2ST}
 S:\;\; \tau \to -\tau^{-1},  \quad T:\;\; \tau \to \tau+1
\end{equation}
which satisfies $S^2 = (ST)^3 = 1$. 

Moreover, the $U(1) \times U(1)$ gauge theory has two 0-form and 1-form symmetries 
\begin{equation}
    \Gamma_0= U(1)_1^{(0)} \times U(1)_2^{(0)}, \quad \Gamma_1 = U(1)_1^{(1)} \times U(1)_2^{(1)}.
\end{equation}
%
The topological defects generating $U(1)^{(0)}_i$ and $U(1)^{(1)}_i$ are
\begin{equation} \label{eq:symTmax}
    \eta^{(0)}_i(\Sigma) = \exp{ \left( i \int_{\Sigma} F_i \right)}, \quad  \eta^{(1)}_i(\gamma) = \exp{ \left(i\int_{\gamma} \star F_i \right)}.
\end{equation}
The charged operator are, respectively, the monopole operators and the Wilson lines 
\begin{equation}
    M_{m_i}(p) = e^{im_i\phi_i(p)}, \quad W_{q_i}(L) = \exp{\left(iq_i\oint_{L} A_i\right)},
\end{equation}
where $q_i$ is the electric charge and $m_i$ is the magnetic charge.

\paragraph{Gauging $\bZ_N^{(0)} \subset U(1)^{(0)}_1 $ and  $ \bZ_N^{(1)} \subset U(1)^{(1)}_2$.}

Similarly to four dimensions \cite{Choi:2021kmx,Choi:2022zal,Cordova:2023ent}, gauging this symmetry changes the gauge fields to $A_1 \to A_1N$ and $A_2 \to  A_2 /N$. Equivalently, we can redefine the gauge couplings to be $e_1\to e_1 N $ and $e_2 \to e_2/N$. The gauging operation defines a topological interface. As we bring the minimal monopole operators and Wilson lines across the interface, they become
\begin{equation}
\begin{split}
 e^{i\phi_1(p)} & \to e^{\frac{i}{N} \phi_1(p)} =  \exp{\left(\frac{i}{N}\int_{\gamma} d\phi_1\right)}\\
\exp{ \left( i\oint_L A_2 \right)} &\to \exp{\left( \frac{i}{N}\oint_L A_2\right)} = \exp{\left( \frac{i}{N}\int_{\Sigma} F_2 \right)} \end{split}
\end{equation}
where $\partial \gamma = p$ and $\partial \Sigma = L$. We find that the minimal monopole operator becomes a line operator and the minimal Wilson line becomes a surface operator generating, respectively, the quantum symmetry $\widehat{\bZ}_N^{(1)}$ and $\widehat{\bZ}_N^{(0)}$. On the other hand, open surface and line operators after going through the interface become
\begin{equation}
\begin{split}
    \exp{ \left( \frac{i}{N}\int_{\Sigma} F_1 \right) } =  \exp{\left( \frac{i}{N}\oint_L A_1 \right) } &\to \exp{\left( i\oint_L A_1 \right)} \\
    \exp{\left(\frac{i}{N}\int_{\gamma} d\phi_2 \right)}= e^{\frac{i}{N} \phi_2(p)} & \to e^{\phi_2(p)} .
\end{split}
\end{equation}
These, instead, become the minimal Wilson lines and monopole operators charged under the quantum symmetries. 


In general, two theories before and after gauging $\bZ_N^{(0)} \times \bZ_N^{(1)}$ are different. Combining gauging and the $S$-transformation defined in \eqref{eq:maxwell2ST}\footnote{This transformation does not need to be confused with S-duality in three-dimensional Maxwell theory \cite{Gaiotto:2008ak,Kapustin:2009av}.}, the gauge couplings transform as 
\begin{equation}
    \tau \xrightarrow{\quad  \sigma \quad } \tau N^2 \xrightarrow{\quad S \quad} -N^2\tau^{-1}
\end{equation}
If one takes $\tau=iN$, i.e. $e_1=e_2N$, then the theory before and after gauging is the same and the composition of gauging $\bZ_N^{(0)} \times \bZ_N^{(1)}$ and S-transformation defines a duality defect $\cN_2$ in a $U(1) \times U(1)$ gauge theory.

\paragraph{Fusion of Duality Defects.}

We have seen that a $U(1) \times U(1)$ gauge theory has a duality defect $\cN_2$ when the gauge coupling satisfies $e_1 = Ne_2$. Similar to the work of \cite{Choi:2021kmx,Roumpedakis:2022aik, Choi:2022zal}, we can find the worldsheet action describing this duality defect. Consider the insertion of a duality defect $\cN_2$ along a codimension-1 surface $M_2$. The worldsheet action of $\cN_2$ is 
\begin{equation}\label{eq:wsN}
    S = \frac{iN}{2\pi}\int_{M_2}   d\phi^L_1 \wedge A^R_2+  \phi^R_1 \wedge  dA^L_2  
\end{equation}
where $A^L_2$, $A^R_2$ and $\phi_1^L$, $\phi_1^R$ are, respectively, the gauge potential $A_2$ and the dual photon of $A_1$ to the left and to the right of $\cN_2$.   

As before, let us write spacetime locally as $M_3 = M_2 \times I$ with $x$ the coordinate along $I$. Assume that the duality defect is placed at $x=0$. The action describing this configuration is given by 
\begin{equation}
\begin{split}
    S = &\frac{1}{2e_2^2} \int_{x<0} \left(\frac{1}{N^2} F^L_{1} \wedge \star F^L_1 +  F^L_2 \wedge \star F^L_2 \right) + \frac{1}{2e_2^2} \int_{x>0} \left(  \frac{1}{N^2} F^R_1 \wedge \star F_1^R +  F_2^R \wedge \star F_2^R \right)\\
    & +\frac{iN}{2\pi}\int_{x=0}   \left( d\phi^L_1 \wedge A^R_2+  \phi^R_1 \wedge  dA^L_2 \right).
\end{split}
\end{equation}
The equations of motion along $M_2$ give
\begin{equation}
    dA_1^L|_{M_2} = NdA_2^R|_{M_2}, \quad dA_2^L|_{M_2} = \frac{1}{N} dA_1^R|_{M_2},
\end{equation}
which corresponds to gauge the $\bZ_N^{(0)}$ subgroup in the first $U(1)^{(0)}_1$, gauge $\bZ_N^{(1)}$ subgroup in the second $ U(1)^{(1)}_2$, then perform the S-duality defined in \eqref{eq:maxwell2ST} to switch $A_1^R$ with $A_2^R$.






Let's insert the duality defect $\cN_2$ at $x=0$ and its orientation reversal $\overline{\cN}_2$ at $x=\epsilon$. The action describing this configuration is  
\begin{equation}
    \begin{split}
        S = & \left(\int_{x<0} 
                L_{b}[A_1^L]+ 
                L_{b}[A_2^L]\right)
                +
                \left(\int_{0<x<\epsilon}
                L_{b}[A_1^I]+
                L_{b}[A_2^I]
                \right)+
                \left(
                \int_{x>\epsilon} 
                L_{b}[A_1^R]+
                L_{b}[A_2^R] \right)\\
                &+ \left(\int_{x=0} 
                 L_{\cN_2}[A_1^L,A_2^I] +  L_{\cN_2}[A_2^L,A_1^I]
                 \right)
                +
                \left(\int_{x=\epsilon} 
                 L_{\overline{\cN}_2}[A_1^I,A_2^R] +  L_{\overline{\cN}_2}[A_2^I,A_1^L]
                 \right)
    \end{split}
\end{equation}
where $L_{b}$ is the Lagrangian in the bulk while $L_{\cN_2}$ and $L_{\overline{\cN}_2}$ are the worldvolume Lagrangians on the defects. To fuse $\cN_2$ with $\overline{\cN}_2$, we take $\epsilon \to 0$ where the gauge fields $A_1^I$ and $A_2^I$ are constrained on the codimension-one surface $M_2$ placed at $x=0$. The worldsheet action becomes 
\begin{equation}
\begin{split}
& \quad \int_{M_2} L_{\cN_2}[A_1^L,A_2^I] +  L_{\cN_2}[A_2^L,A_1^I] +
L_{\overline{\cN}_2}[A_1^I,A_2^R] +  L_{\overline{\cN}_2}[A_2^I,A_1^L]\\
&= \frac{iN}{2\pi}\int_{M_2}   d\phi^L_1 \wedge A^I_2+  \phi^I_1 \wedge  dA^L_2  - d\phi^I_1 \wedge A^R_2-  \phi^R_1 \wedge  dA^I_2  \\
&=  \frac{iN}{2\pi}\int_{M_2}    \phi   (dA^L_2 - dA^R_2) + \frac{iN}{2\pi}\int_{M_2}   (d\phi^L_1-d\phi^R_1) \wedge a ,
\end{split}
\end{equation}
where $\phi=\phi^I_1$ and $a=A^I_2$ are defined to be living only on $M_2$. The first term is the worldsheet action for the condensation defect $\cC$ obtained by gauging $\bZ_N^{(1)} \subset U(1)^{(1)}_2$ along $M_2$ \cite{Roumpedakis:2022aik}. The second term shows the one-gauging of $\bZ_N^{(0)} \subset U(1)^{(0)}_1$ on $M_2$, i.e. $\sum_{i=0}^{N-1}(\eta^{(0)}_1)^i$. Indeed, integrating out $A_2^I$ constrains $\phi^L_1-\phi^R_1$ to be $\bZ_N$-valued corresponding to gauging a $\bZ^{(0)}_N$ 0-form symmetry along the defect. Thus, we recover the expected fusion rule from the explicit worldsheet action \eqref{eq:wsN} in a $U(1) \times U(1)$ gauge theory.

\paragraph{General Gauging Operation.}

As stated above, in Maxwell theory one can gauge $\bZ_r^{(0)}  \subset U(1)^{(0)} $ and $ \bZ_s^{(1)} \subset U(1)^{(1)}$ simultaneously when $\text{gcd}(r,s)=1$ \cite{Niro:2022ctq,Cordova:2023ent}. For a $U(1) \times U(1)$ gauge theory considered here, the most general gauging operation is then given by
\begin{equation} \label{eq:genGauging}
    \bZ_{r_1}^{(0)} \times \bZ_{s_1}^{(1)} \subset U(1)^{(0)}_1 \times U(1)^{(1)}_1, \quad    \bZ_{r_2}^{(0)} \times \bZ_{s_2}^{(1)} \subset U(1)^{(0)}_2 \times U(1)^{(1)}_2
\end{equation}
with $r_1,s_1$ and $r_2,s_2$ coprime. The gauge couplings combining this general gauging $\sigma$ and an S-transformation change as follows
\begin{equation}
    (e_1,e_2) \xrightarrow{\quad  \sigma \quad }  \left(\frac{r_1}{s_1}e_1,\frac{r_2}{s_2}e_2\right) \xrightarrow{\quad  S \quad } \left(\frac{r_1}{s_1}e_2,\frac{r_2}{s_2}e_1\right).
\end{equation}
Hence, invariant gauge couplings under $\sigma$ and $S$ need to satisfy the condition 
\begin{equation} \label{eq:invCoupling}
\frac{e_1}{e_2} = \frac{r_1}{s_1}= \frac{s_2}{r_2} , 
\end{equation}
which has solutions only if $r_1 r_2 =  s_1 s_2$. 

If the above condition is true, one can realize the duality defect $\cN_2$ by performing a general gauging operation in \eqref{eq:genGauging} followed by an S-transformation. The fusion rule for such duality defect is 
\begin{equation} \label{eq:fusU1U1}
\begin{split}
    \cN_{2}  \times \overline{\cN}_{2} & = \left( \sum_{k=0}^{r_1-1} (\eta_1^{(0)})^k \right)
    \left(
    \frac{1}{|H^0(M_2, \bZ_{s_1})|}  \sum_{\gamma \in H_1(M_2,\bZ_{s_1})}
    \eta_1^{(1)}(\gamma) \right)  \\
    &\quad \left( \sum_{k=0}^{r_2-1} (\eta_2^{(0)})^k \right) \left(
    \frac{1}{|H^0(M_2, \bZ_{s_2})|}  \sum_{\gamma \in H_1(M_2,\bZ_{s_2})}
    \eta_2^{(1)}(\gamma) \right), 
\end{split}
\end{equation}
where $\eta_i^{(1)}$ and $\eta_i^{(0)}$ for $i=1,2$ are the symmetry defects generating the invertible symmetries $\bZ_{r_i}^{(0)}$ and $\bZ_{s_i}^{(1)}$ defined in \eqref{eq:symTmax}. 

We find the worldsheet action of duality defect $\cN_{2}$ to be 
\begin{equation}
\begin{split}
    S =   
    &  \frac{i}{2\pi}\int_{M_2} \left( s_1 a_1 \wedge d\sigma_1 - r_1 a_1 \wedge d\phi^L_1 + d\sigma_1 \wedge A^R_2 \right) \\
    \;\;+& \frac{i}{2\pi}\int_{M_2} \left(   r_2 d a_2 \wedge \sigma_2 - s_2 da_2 \wedge \phi^R_1 + \sigma_2 \wedge dA^L_2 \right), 
\end{split}
\end{equation}
where $a_i$ are gauge fields and $\sigma_i$ are compact scalars defined only on $M_2$. Inserting such duality defect at $x=0$, we find that the action describing this configuration is given by 
\begin{equation}
\begin{split}
    S = &\int_{x<0} \left(\frac{1}{2e_1^2}  F^L_{1} \wedge \frac{1}{2e_2^2}  \star F^L_1 +  F^L_2 \wedge \star F^L_2 \right) +  \int_{x>0} \left( \frac{1}{2e_1^2}   F^R_1 \wedge \star F_1^R + \frac{1}{2e_2^2}  F_2^R \wedge \star F_2^R \right)\\
    & +\frac{i}{2\pi}\int_{x=0} \left( s_1 a_1 \wedge d\sigma_1 - r_1 a_1 \wedge d\phi^L_1 + d\sigma_1 \wedge A^R_2 \right) \\
    & + \frac{i}{2\pi}\int_{x=0} \left(   r_2 d a_2 \wedge \sigma_2 - s_2 da_2 \wedge \phi^R_1 + \sigma_2 \wedge dA^L_2 \right).
\end{split}
\end{equation}
At the invariant gauge coupling value \eqref{eq:invCoupling}, the equations of motion along $M_2$ at $x=0$ give
\begin{equation}
    dA_1^L|_{M_2} = \frac{r_1}{s_1}dA_2^R|_{M_2}, \quad dA_2^L|_{M_2} = \frac{r_2}{s_2} dA_1^R|_{M_2},
\end{equation}
which exactly correspond to first performing the general gauging \eqref{eq:genGauging} and then an S-transformation. By a similar calculation, the fusion rule involving $\cN_{2}$ in \eqref{eq:fusU1U1} can be reproduced using this explicit worldsheet action.  




\subsection{Product Theories}

Whenever a theory $Q$ has a $\mathbb{Z}_N^{(0)}$ symmetry without 't Hooft anomalies we can gauge it an consider the theory $\mathcal{Q}/\mathbb{Z}_N^{(0)}$ with a $\widehat{\mathbb{Z}}_N^{(1)}$ quantum symmetry\footnote{A similar setup has appeared in the appendix of \cite{Choi:2024rjm}.}. Although the two theories cannot be equivalent, due to the different symmetry content, we can consider the product\footnote{All the arguments in this subsection apply also to a theory $\mathcal{Q}$ with a $\mathbb{Z}_N^{(1)}$ symmetry. For example we could take, as starting point, $\mathcal{Q}'=\mathcal{Q}/\mathbb{Z}_N^{(0)}$ and consider $\mathcal{T}'=\mathcal{Q}'\times (\mathcal{Q}'/\widehat{\mathbb{Z}}_N^{(1)})$.}
\begin{equation}\label{eq.mathcalT}
    \mathcal{T}\equiv\mathcal{Q}\times(\mathcal{Q}/\mathbb{Z}_N^{(0)}).
\end{equation}
The resulting theory will have symmetries $\mathbb{Z}_N^{(0)}\times\widehat{\mathbb{Z}}_N^{(1)}$ free of mixed 't Hooft anomalies which can then be gauged simultaneously. Given that gauging the quantum symmetry gives the original theory
\begin{equation}
    (\mathcal{Q}/\mathbb{Z}_N^{(0)})/\widehat{\mathbb{Z}}_N^{(1)}\simeq\mathcal{Q}
\end{equation}
it is immediate to check that 
\begin{equation}\label{eq.mathcalT/()}
    \mathcal{T}/(\mathbb{Z}_N^{(0)}\times\widehat{\mathbb{Z}}_N^{(1)})\simeq\mathcal{T}.
\end{equation}
This relation can be employed to find theories with non-invertible symmetry defects as those described in \autoref{sec.2}.

\paragraph{Chern-Simons Matter Theories.}
As an example of the previous construction we can consider 
\begin{equation}\label{eq.CSSO(N)}
    \mathcal{Q}=SO(N)_K\text{ with }N_f\text{ adjoint scalars}.
\end{equation}
For $N,K,N_f=0\,\mmod\, 4$ this theory has a $\mathbb{Z}^{(1)}_2$ symmetry associated with the center of $SO(N)$, charge conjugation symmetry $\mathbb{Z}^C_2$ and magnetic symmetry $\mathbb{Z}^M_2$. As mentioned above, these three symmetries have a mixed 't Hooft anomaly which has been employed in \cite{Kaidi:2021xfk} to construct non-invertible symmetry defects gauging $\mathbb{Z}_2^M\times\mathbb{Z}_2^{(1)}$. This non-invertible symmetry defects satisfies the fusion rules of \autoref{sec.2.4}.

Instead, we consider generic values of $N,K,N_f$. The symmetry content of these theories may differ, however we are only interested in the magnetic $\mathbb{Z}_2^M$ symmetry which is present nevertheless. Hence, we consider a theory \eqref{eq.mathcalT}
\begin{equation}
    \mathcal{T}=(SO(N)_K\text{ with }N_f\text{ adjoint scalars})\times(Spin(N)_K\text{ with }N_f\text{ adjoint scalars}),
\end{equation}
where the second factor is obtained gauging $\mathbb{Z}_2^M$ in \eqref{eq.CSSO(N)} \cite{Cordova:2017vab} and its symmetry content contains $\widehat{\mathbb{Z}}_N^{(1)}$.

We can then gauge $\mathbb{Z}_2^M\times\widehat{\mathbb{Z}}_N^{(1)}$ in $\mathcal{T}$ in half of the spacetime. In this half, we find a theory isomorphic to the original one \eqref{eq.mathcalT/()}. Given that this operation involves half-space gauging it necessarily produces a non-invertible symmetry in the theory $\mathcal{T}$. The partition function of $\mathcal{T}/(\mathbb{Z}_2^M\times\widehat{\mathbb{Z}}_N^{(1)})$ can be obtained from that of $\mathcal{T}$ as in \eqref{Eq:sigmaG} and it is immediate to verify
\begin{equation}
    Z_{\mathcal{T}}=Z_{\mathcal{T}/(\mathbb{Z}_2^M\times\widehat{\mathbb{Z}}_N^{(1)})}.
\end{equation}
Finally, the theory possesses a non-invertible symmetry and the symmetry defect $\mathcal{N}_2$ implementing it satisfies the fusion rules studied in \autoref{sec.2.4}.

Note that we could have started with a theory 
\begin{equation}
    \mathcal{T}'=(SO(N)^g_K\text{ with }N_f\text{ adjoint scalars})\times(Spin(N)^{g'}_{K'}\text{ with }N'_f\text{ adjoint scalars}),
\end{equation}
where $g,g'$ are the values of the coupling constants in the two factors. The theory $\mathcal{T}'$ has the same $\mathbb{Z}_2^M\times\widehat{\mathbb{Z}}_N^{(1)}$ symmetries of $\mathcal{T}$ however it is immediate to check that it admits a non-invertible symmetry defect of the sort we just described only for $g=g'$, $K=K'$ and $N_f=N'_f$.

\subsection{Constraints on Trivially Gapped Phase}

The existence of duality or triality defects in a theory $\cQ_{\text{UV}}$ sometimes is not compatible with a trivially gapped phase. The idea is that if $\cQ_{\text{UV}}$ admits a duality defect, then it is self-dual under gauging $\bZ^{(0)}_N \times \bZ^{(1)}_N$, i.e. $\cQ_{\text{UV}} = \sigma \cQ_{\text{UV}}$. For any duality-preserving renormalization flows, one should be able to construct the same duality defect in infrared. Suppose $\cQ_{\text{UV}}$ flows to trivially gapped phase described by a SPT phase, then it should be invariant under gauging $\sigma$. This type of constraints on IR phases has been studied for duality defects in (1+1)d \cite{Chang:2018iay, Thorngren:2019iar, Thorngren:2021yso} and (3+1)d \cite{Choi:2021kmx,Choi:2022dju, Apte:2022xtu} by half spacetime gauging $\bZ^{(0)}_N $ and $\bZ^{(1)}_N$. We will perform this analysis for duality defects in (2+1)d with $N=p$ prime.  

The most general SPT phase with $\bZ^{(0)}_p \times \bZ^{(1)}_p$ symmetry is given by \cite{Wan:2018bns}
\begin{equation} \label{eq:sptsec4}
    Z_{\cP_{k_1,k_2}}[A,B] = \exp{\left(\frac{2\pi i }{p}\int_{M_3} k_1 AB + k_2 A dA \right)}
\end{equation}
where $A$, $B$ are background gauge fields for $\bZ^{(0)}_p$ and $\bZ^{(1)}_p$ and $k_1,k_2 \in \bZ_p$. Suppose $\cQ_{\text{UV}}$ has a non-invertible topological defect defined by first stacking an SPT phase $\cP_{k_1,k_2}$ and then gauging $\bZ^{(0)}_p \times \bZ^{(1)}_p$ in half of spacetime. Let us denote this combined operation as $\sigma \tau^{k'_1,k'_2}$. If $\cQ_{\text{UV}}$ is trivially gapped under a duality-preserving RG flow, we expect to find an SPT phase invariant under the same topological operation, i.e. $\cP_{k_1,k_2} = \sigma \tau^{k'_1,k'_2}\cP_{k_1,k_2} $. 

If we perform the operation $\sigma \tau^{k'_1,k'_2}$ to the most general SPT phase in \eqref{eq:sptsec4}, the partition function $ Z_{\cP_{k_1,k_2}}[A,B]$ becomes 
\begin{align}
     \frac{1}{|H^1(M_3;\mathbb{Z}_p)|}
    \sum_{\substack{a \in H^1(M_3;\mathbb{Z}_p),\\
    b \in H^2(M_3;\mathbb{Z}_p)}} \exp{\left(\frac{2\pi i}{p} \int_{M_3} a  B + b  A+ (k_1+k'_1)a  b + (k_2+k'_2)ada \right)}
\end{align}
where $a$,$b$ are dynamical gauge fields on $M_3$. Integrating out $b$, one obtains 
\begin{equation}
   Z_{\sigma \tau^{k'_1,k'_2}\cP_{k_1 k_2}}[A,B] = \exp{\left(\frac{2\pi i }{p}\int_{M_3} -\frac{AB}{k_1+k'_1}-\frac{k_2+k'_2}{(k_1+k'_1)^2}AdA\right)}.
\end{equation}
Comparing with\eqref{eq:sptsec4}, the invariance of $\cP_{k_1,k_2}$ requires that 
\begin{equation}
    k_1(k_1+k'_1)=-1\;\; \text{mod}\; p,\qquad k_2(k_1+k'_1)^2 = -(k_2+k'_2) \;\; \text{mod}\; p.
\end{equation}
Thus, if a (2+1)d theory $\cQ_{\text{UV}}$ has a topological defect defined by the operation $\sigma \tau^{k'_1,k'_2}$, it can be trivially gapped along a symmetry-preserving RG flow only if the conditions above are satisfied. 

For a duality defect with $k'_1=k'_2=0$, the invariant SPT phase exists if  
\begin{equation} \label{eq:cond1RG}
     k_1^2=-1,\quad \text{mod}\; p, 
\end{equation}
and $k_2$ can take any value in $\bZ_p$. A triality defect corresponds to $k'_1=1$ and $ k'_2=0$ as the SPT phase invariant under such operation needs to satisfy  
\begin{equation} \label{eq:cond2RG}
    k_1(k_1+1)=-1,\quad \text{mod}\; p
    \qquad k_2(k_1+1)= 0 \quad \text{mod}\; p.
\end{equation}
In this way, duality and triality defects puts constraints on the possible symmetry-preserving RG flows to trivially gapped phases.

\section{Examples from 6d $A_{N-1}$ Theory on Three Manifolds}\label{sec.5}

In this section we take as starting point a 6d $\mathcal{N}=(2,0)$ SCFTs of type $A_{N-1}$ on a manifold $M_6=M_3\times X_3$ and study its compactification on three manifolds $X_3$ \cite{Dimofte:2011ju,Terashima:2011qi}. These SCFTs are relative field theories \cite{Witten:1998wy,Witten:2009at,Freed:2012bs,Tachikawa:2013hya} which are specified by a partition vector, rather than by a partition function. Each element in this vector space can be understood as a state in the Hilbert space of a seven-dimensional TQFT on $M_7=M_4\times X_3$ with $\partial M_7=M_6$. The dimensional reduction produces a 4d bulk TQFT on $M_4$ coupled to a relative 3d theory on the boundary $M_3$. Similarly to compactifications on 2-manifolds \cite{Tachikawa:2013hya,Bashmakov:2022jtl,Bashmakov:2022uek} and on 4-manifolds \cite{Chen:2022vvd,Bashmakov:2023kwo}, to make the 3d theory absolute, we need to pick a maximal isotropic sublattice \cite{DelZotto:2015isa} in $X_3$. Such choice is associated with the existence of gapped topological boundary conditions for the 3d theory. Hence, this enables us to define the bulk theory on a slab and, for certain choices of $X_3$, to reproduce the setup of \autoref{sec.3}. 

The homology groups of the compactification manifold $X_3$ determine the SymTFT action in 4d. We will obtain both examples where there are no non-invertible symmetries\footnote{In these cases the SymTFT is a Dijkgraff-Witten theory \cite{Kaidi:2022cpf}.} and examples which allow us to reproduce the action for the SymTFT \eqref{eq.SymTFTaction} corresponding to a 3d theory which in principle can have intrinsically non-invertible symmetries. In the latter case we will be able to reproduce, via dimensional reduction, the duality defects studied in the previous sections. Moreover, we will show how different absolute theories are related to each other by gauging and dualities. We will focus on manifolds with free homology groups and leave the study of the torsion case for a future work.

\subsection{SymTFT from Dimensional Reduction}
We will be brief in presenting the 7d TQFT associated to the relative 6d $\mathcal{N}=(2,0)$ SCFTs of type $A_{N-1}$ and its compactification, more details can be found in \cite{Bashmakov:2022uek,Chen:2023qnv,Bashmakov:2023kwo}. 

\paragraph{6d SCFTs as Relative Theories.}
The action of the relevant 7d TQFT has been discussed in \cite{Witten:1998wy} and contains the following term
\begin{equation}\label{eq.7dTFT}
    S_{7d}=\frac{N}{4\pi}\int_{M_7}c\wedge\dd c,
\end{equation}
where $c\in H^3(M_7,U(1))$ is a 3-form field and $\partial M_7=M_6$. Associated to this field there are Wilson three-surfaces
\begin{equation}
    \Phi(Y_3)\equiv \exp\left(\ii\oint_{Y_3} c\right),\quad Y_3\in H_3(M_7,\mathbb{Z}_N),
\end{equation}
satisfying the equal-time commutation relation
\begin{equation}\label{eq.heisenberg}
    \Phi(Y_3)\Phi(Y_3')=e^{\langle Y_3,Y_3'\rangle}\phi(Y_3')\Phi(Y_3),
\end{equation}
where $\langle\cdot,\cdot\rangle$ is the intersection pairing.

The 6d SCFT, as a relative field theory, can be understood as the Hilbert space of the 7d TQFT. However, due to the Heisenberg algebra \eqref{eq.heisenberg}, a state in a representation is invariant at most under a maximal isotropic sublattice $\mathcal{L}\subset H_3(M_6,\mathbb{Z}_N)$ taking values in the homology groups of the manifold
\begin{equation}\label{eq.sublattice6d}
    \langle Y_3,Y_3'\rangle=0,\quad\forall\; Y_3,Y_3'\in\mathcal{L}.
\end{equation}
Hence, we define a state
\begin{equation}
    |\mathcal{L},0\rangle\quad\mbox{such that}\quad\Phi(Y_3)|\mathcal{L},0\rangle=|\mathcal{L},0\rangle,\quad\forall \;Y_3\in\mathcal{L}.
\end{equation}
Other states in the Hilbert space are obtained acting on $|\mathcal{L},0\rangle$ with elements of $\mathcal{L}^{\perp}\equiv H_3(M_6,\mathbb{Z}_N)/\mathcal{L}$.

It can be shown that for $N$ a perfect square, a choice of $\mathcal{L}$ determines an absolute theory in 6d via a choice of gapped boundary conditions \cite{Tachikawa:2013hya,Gukov:2020btk}. Instead, for $N$ not a perfect square there is no choice of topological gapped boundary conditions invariant under large diffeomorphisms of $M_6$. 

As we are interested in compactification on a three manifold $X_3$, we consider the product space $M_7=M_4\times X_3$ with $\partial M_4=M_3$. The theory obtained via dimensional reduction is a 3d relative theory, which we label $\mathcal{T}_N[X_3]$ living on the boundary of a 4d TQFT. A 3d absolute theory is specified by a choice of maximal isotropic sublattice in the internal geometry. Via the K\"unneth formula we can rewrite $\mathcal{L}$ \eqref{eq.sublattice6d} in terms of elements in the homology groups of $M_3$ and $X_3$ \cite{Eckhard:2019jgg}
\begin{equation}
    H_3(M_6,\mathbb{Z}_N)\supset H_1(M_3,\mathbb{Z}_N)\otimes H_2(X_3,\mathbb{Z}_N)\oplus H_2(M_3,\mathbb{Z}_N)\otimes H_1(X_3,\mathbb{Z}_N).
\end{equation}
Here, we are assuming that the torsion part of the homology groups is trivial and we neglect the contributions from the top homology groups. Moreover, given that the six dimensional manifold is a product space, every three-cycle $Y_3\in H_3(M_6,\mathbb{Z}_N)$ can be decomposed in one-cycles and two-cycles as follows
\begin{equation}
    Y_3=\Sigma\times\gamma'+\gamma\times\Sigma',
\end{equation}
where $\Sigma\in H_2(X_3,\mathbb{Z}_N)$, $\gamma\in H_1(X_3,\mathbb{Z}_N)$, $\Sigma'\in H_2(M_3,\mathbb{Z}_N)$ and $\gamma'\in H_1(M_3,\mathbb{Z}_N)$. The intersection form between three-cycles can be rewritten as
\begin{equation}\label{eq.decomposition.intersection}
    \langle Y_3,\widetilde{Y_3}\rangle=\langle\Sigma,\widetilde{\gamma}\rangle\times\langle\gamma',\widetilde{\Sigma}'\rangle+\langle\gamma,\widetilde{\Sigma}\rangle\times\langle\Sigma',\widetilde{\gamma}'\rangle.
\end{equation}
Thus, the choice of maximal isotropic lattice reduces to a choice of maximal isotropic sublattice $\Lambda\equiv\Lambda_1\oplus\Lambda_2\subset H_1(X_3,\mathbb{Z}_N)\oplus H_2(X_3,\mathbb{Z}_N)
$ taking values in the homology groups of $X_3$
\begin{equation}\label{eq.sublattice3d}
    \Lambda_1:\;\langle\Sigma,\widetilde{\gamma}\rangle=0,\quad\Lambda_2:\;\langle\gamma,\widetilde{\Sigma}\rangle=0,\quad\forall\,(\gamma,\Sigma),(\widetilde{\gamma},\widetilde{\Sigma})\in\Lambda.
\end{equation}
According to the decomposition of three-cycles on the product space \eqref{eq.decomposition.intersection}, $\Lambda$ immediately determines a maximal isotropic sublattice in $M_6$
\begin{equation}
    \mathcal{L}=(\Lambda_2\otimes H_1(M_3,\mathbb{Z}_N))\oplus(\Lambda_1\otimes H_2(M_3,\mathbb{Z}_N)),
\end{equation}
and thus the choice of absolute theory, which we denote by $\mathcal{T}_{N,\Lambda}[X_3]$. We will refer to this as a choice of \emph{polarization} \cite{Gukov:2020btk}. To specify the global variants we need to specify a representative in
\begin{equation}
     \Lambda^\perp\equiv\Lambda_1^\perp\oplus\Lambda_2^\perp= (H_1(X_3,\mathbb{Z}_N)/\Lambda_1)\oplus (H_2(X_3,\mathbb{Z}_N)/\Lambda_2).
\end{equation}
which in turn determines the complement of $\mathcal{L}$
\begin{equation}
    \mathcal{L}^\perp\equiv (\Lambda_2^\perp\otimes(H_1(M_3,\mathbb{Z}_N))\oplus (\Lambda_1^\perp\otimes H_2(M_3,\mathbb{Z}_N)),
\end{equation}
We denote different global variants by $\mathcal{T}_{N,\Lambda}[X_3,B]$, where $B\in \Lambda^\perp$. Note that a choice in $\Lambda^\perp$ denotes possible stacking on the 3d theory with SPT phases while elements in $H_1(M_3,\mathbb{Z}_N)\oplus H_2(M_3,\mathbb{Z}_N)$ determine the value of the background fields for zero and one form symmetries.

\paragraph{SymTFT in 4d.}
We now turn our attention the dimensional reduction of the action of the 7d TFT \eqref{eq.7dTFT}. Let $\{\zeta_i\}_{i=1}^{r_2}$ be a basis of $H_2(X_3,\mathbb{Z})$ and $\{\eta_j\}_{j=1}^{r_1}$ a basis of\footnote{As we are considering three-manifolds $X_3$ without torsion cycles, using Poincar\'e duality we assume that $r_1=r_2\equiv r$.} $H_1(X_3,\mathbb{Z})$. Then, the dimensional reduction gives
\begin{equation}\begin{split}\label{eq.4dTFT}
    S_{4d}=&\frac{N}{4\pi}\sum_{ij}Q^{ij}\int_{M_4}a_i\wedge\dd b_j\\
    =&\frac{\pi}{N}\sum_{ij}Q^{ij}\int_{M_4}a_i\cup\delta b_j, 
\end{split}\end{equation}
where $Q$, an $r\times r$ matrix, determines the intersection between one and two-cycles in $X_3$. 
Moreover, the three-form $c$ splits into 
\begin{equation}
    a_i=\int_{\zeta_i}c,\quad b_i=\int_{\eta_i}c
\end{equation}
that are, respectively, a one-form and a two-form gauge field. As in \autoref{sec.3} we used $\mathbb{Z}_N$-valued cochains to write the BF action, in the second line of \eqref{eq.4dTFT} we expressed one and two forms in terms of $\mathbb{Z}_N$-valued cochains appearing in \eqref{eq.SymTFTaction}, $a\rightarrow\frac{2\pi\ii}{N} a$ and same for $b$. From now one we will write all expressions in terms of cochains.

We can obtain line and surface operators from the 7d theory
\begin{equation}\label{eq.lines7dSymTFT}
    L_{\Vec{l}}(\gamma')=\exp{\left( \frac{2\pi i}{N} \oint_{\Sigma\times\gamma'} c\right)}=\exp{\left( \frac{2\pi i}{N}\oint_{\gamma'}\Vec{l}\cdot\Vec{a}\right)}, \quad \Vec{l}\in (\bZ_N)^{r},
\end{equation}
\begin{equation}\label{eq.surfaces7dSymTFT}
    S_{\Vec{s}}(\Sigma')=\exp{\left( \frac{2\pi i}{N} \oint_{\gamma\times\Sigma'} c\right)}=\exp{\left( \frac{2\pi i}{N}\oint_{\Sigma'}\Vec{s}\cdot\Vec{b}\right)}, \quad \Vec{s}\in (\bZ_N)^{r}.
\end{equation}
For specific choices of $Q$, these operators will reproduce line and surface operators \eqref{eq.SymTFT.line}-\eqref{eq.SymTFT.surface} for the 4d SymTFT in \autoref{sec.3}.

Hence, the dimensional reduction of the 7d TFT gives us a 4d TFT with a 3d relative theory on the boundary. Ultimately, we are interested in defining a SymTFT in 4d which requires the existence also of a topological boundary. As shown in \cite{Gukov:2020btk,Bashmakov:2022uek,Chen:2023qnv,Bashmakov:2023kwo}, a choice a maximal isotropic lattice in $\mathcal{L}$ allows to define gapped boundary boundary conditions and to consider the theory on a slab $M_4=M_3\times I_{[0,\epsilon]}$. This determines the SymTFT of $\mathcal{T}_{N,\Lambda}[X_3,B]$.

\paragraph{Topological Manipulations and Mapping Class Group.}
There are three different topological manipulations that transform among different global variants of $\mathcal{T}_N[X_3]$. First, when the 3d theories we will obtain via dimensional reductions have a non-anomalous discrete symmetry $G$, this can be gauged. We denote this operation as follows
\begin{equation}
    \sigma(G)\mathcal{T}_{N,\Lambda}[X_3]\equiv\mathcal{T}_{N,\Lambda}[X_3]/G.
\end{equation}
The action on the partition function is that we wrote in \eqref{Eq:sigmaG} for $G=\mathbb{Z}_N^{(0)}\times\mathbb{Z}_N^{(1)}$. All absolute theories $\mathcal{T}_{N,\Lambda}[X_3]$ can be obtained in this way. Moreover, to obtain all the global variants, we should also consider stacking with SPT phases. Due to the rich landscape of SPT phases in 3d we will leave a study of these for future work. Combining these two operation together with all the global theories, one obtains the \emph{orbifold groupoid} \cite{Gaiotto:2020iye}. There are also permutations of symmetry lines and surfaces which change how $\mathcal{T}_{N,\Lambda}[X_3,B]$ couples to background fields for the zero and one form symmetries.

Finally, the mapping class group of $X_3$ is defined as the group of diffeomorphisms of $X_3$ connected to the identity. Thus, the group is determined by the matrices that preserve the intersection paring matrix $Q$
\begin{equation}
    P^\intercal QP=Q,\quad P\in GL(r,\mathbb{Z}_N)
\end{equation}
Recall that $r$ is the rank of $Q$. This will give rise to Montonen-Olive-like dualities in the 3d theories. Crucially, an element of the mapping class group gives rise to an operator in the SymTFT which, upon shrinking the slab, creates an interface
between two dual theories.

\subsection{$S^2\times S^1$}

Let us start from $X_3=S^2\times S^1$ which is the simplest 3-manifold with non-trivial $H_1(X_3,\bZ_N)$. Compactifying a 6d $A_{N-1}$ theory on it leads to a relative theory $T_N[S^2\times S^1]$ \footnote{The infrared phase of $T_N[S^2\times S^1]$ is determined by a vanishing 3d index and spontaneously broken supersymmetry \cite{Choi:2022dju}.}. We will setup the notation and illustrate the ideas presented in the previous subsection in this example. The homology groups of $S^2 \times S^1$ are 
\begin{equation}
        H_*(S^2 \times S^1,\bZ) = \{\bZ, \bZ,\bZ , \bZ\}.
\end{equation}
Let $l$ and $s$ be the generators of $H_1(S^2 \times S^1,\bZ)$ and $H_2(S^2\times S^1,\bZ)$. The intersection pairing is a map
\begin{equation}
    Q: H_2(S^2\times S^1,\mathbb{Z}_N)\times H_1(S^2\times S^1,\mathbb{Z}_N)\rightarrow \mathbb{Z}_N,
\end{equation}
given by $Q(l,s)=1$ and, more in general, by:
\begin{equation}
    Q(ml,ns)=mn,\;\;\; \text{mod} \; N.
\end{equation}
Due to this nontrivial intersection, the 3d effective theory is relative.

The defect group \cite{DelZotto:2015isa} is 
\begin{equation}
     D = H^1(S^2\times S^1,\bZ_N)\oplus H^2(S^2 \times S^1,\bZ_N)=\bZ_N\oplus\bZ_N.
\end{equation}
To obtain an absolute theory, one needs to choose a  maximal isotropic sublattice or polarization, i.e. a subset $\Lambda \subset D$ such that 
\begin{equation}
    Q(a,b) = 0, \quad a,b \in \Lambda.
\end{equation}
Moreover, the intersection matrix determines the action \eqref{eq.4dTFT} for the 4d SymTFT
\begin{equation}\label{eq.4dTFT.S2S1}
     S_{4d}=\frac{4\pi}{N}\int_{M_4}a\cup\delta b
\end{equation}
Thus, we obtain a different BF theory than the one we described in \autoref{sec.3}. Lines and surfaces operators are given by:
\begin{equation}
     L_{l}(\gamma')=\exp{\left(\frac{2\pi i}{N}\oint_{\gamma'}la\right)}, \quad S_{s}(\Sigma')=\exp{\left(\frac{2\pi i}{N}\oint_{\Sigma'}sb\right)}.
\end{equation}

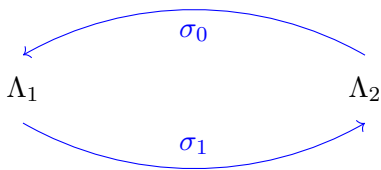
\begin{figure}
\centering
\begin{tikzpicture}[scale=1.5]
\draw node at (0,0) {$\Lambda_1$};
\draw node at (3,0) {$\Lambda_2$};
\draw [<-,blue] (0,.3) arc (120:60:3);
\draw [->,blue] (0,-.3) arc (-120:-60:3);
 \draw node at (1.5,-0.5) {${\color{blue} \sigma_1}$};
 \draw node at (1.5,.5) {${\color{blue} \sigma_0}$};
\end{tikzpicture}
\caption{The global forms of $T_p[S^2\times S^1]$ with different choices of polarization $\Lambda_1$ and $\Lambda_2$. Gauging $\mathbb{Z}_p^{(1)}$ and $\mathbb{Z}_p^{(0)}$ is labeled, respectively, by $\sigma_1$ and $\sigma_0$.}
\label{fig.S2xS1prime}
\end{figure}

Assuming $N=p$ prime, we introduce the following notation
\begin{equation}
    \langle l \rangle =\{0_l,l,2l,\ldots, (p-1)l\}, \quad \langle s \rangle =\{0_s,s, 2s,\ldots, (p-1)s\}
\end{equation}
to represent the sets of lines and surfaces generated by $l$ and $s$. There are two choices of polarization
\begin{equation} 
\begin{array}{cccc}
\Lambda_1= \langle l \rangle &\to \; &\bZ^{(1)}_p\\
\Lambda_2= \langle s \rangle &\to \; &\bZ^{(0)}_p
\end{array}
\end{equation}
which give rise to theories with either a 0-form or a 1-form symmetry. Note that the symmetry is given by the complement of the polarization, i.e. $\Lambda^\perp$. For example, for $\Lambda_1^\perp = \langle s \rangle$, the Wilson three-surfaces wrapping two-cycles in $S^2\times S^1$ lead to line operators generating $\mathbb{Z}_p^{(1)}$ symmetries in 3d. The two different absolute theories are related to each other by gauging the $\bZ_p$ symmetry, as shown in \autoref{fig.S2xS1prime}. 

\subsection{$(S^2\times S^1)\#(S^2\times S^1)$}

The second example we consider is the connected sum of two copies of $S^2\times S^1$, i.e. $X_3 = (S^2\times S^1)\#(S^2\times S^1)$. The 3d effective theory after twisted compactification is expected to flow to a 3d $\cN=2$ SCFT \footnote{Connected sum of 3-manifolds belong to the ``infinite class" and the computation of 3d index suggests that this type of theories will flow to 3d $\cN=2$ SCFTs with possible accidental flavor symmetry  \cite{Choi:2022dju}.}. Due to non-trivial $H^2(X_3, \bZ_N)$, $T_N[X_3]$ is relative. We will enumerate its global forms by classifying the polarizations for $N=p$ prime and show that two of them admit duality defects. 

For generic $N$ the homology groups of $X_3$ are
\begin{equation}
        H_*(X_3,\bZ) = \{\bZ, \bZ^2,\bZ^2 , \bZ\}.
\end{equation}
Let $l_1,l_2$ be the generators of 1-cycles and $s_1,s_2$ be the generators of 2-cycles. The nontrivial intersections are 
\begin{equation}
    l_i \cdot s_j = s_i \cdot l_j = \delta_{ij}, \qquad i,j=1,2. 
\end{equation}
The action for the SymTFT \eqref{eq.4dTFT} is
\begin{equation}\label{eq.4dTFT.S2S1sum}
    S_{4d}=\frac{4\pi}{N}\int_{M_4}\left(a_1\cup\delta b_1+a_2\cup\delta b_2\right)
\end{equation}
This reproduces the BF action \eqref{eq.SymTFTaction} we discussed earlier. We will now show that indeed the 3d theory admits non-invertible symmetry defects as suggested by the action \eqref{eq.4dTFT.S2S1sum}.

\begin{figure}
\centering\[
\begin{tikzpicture}[scale=1.5]
\draw node at (0,0) {$\Lambda_3$};
\draw node at (3,0) {$\Lambda_4$};
\draw [<->,blue] (0,.3) arc (120:60:3);
\draw [<->,orange] (0,-.3) arc (-120:-60:3);
 \draw node at (1.5,-0.5) {${\color{orange} S}$};
 \draw node at (1.5,.5) {${\color{blue} \sigma}$};
\end{tikzpicture}
\]
\caption{Global forms of $T_p[(S^2\times S^1) \# (S^2\times S^1)]$ for the polarizations $\Lambda_3$ and $\Lambda_4$. Here, blue lines represent gauging $\bZ_p^{(1)} \times \bZ_p^{(0)}$ and orange lines represent the duality $s$ defined in \eqref{eq:6dex2S}.}
\label{Fig:GFsigmaS2S2}
\end{figure}
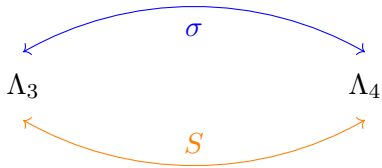

When the number of M5 branes is prime $N=p$, the defect group is 
\begin{equation}
    D = \bZ_p^{(1)} \oplus \bZ_p^{(1)} \oplus \bZ_p^{(2)} \oplus \bZ_p^{(2)}
\end{equation}
Consider the following four obvious polarizations\footnote{Besides these ones, there are more polarizations which are related with each other by gauging and stacking SPT phases. We will study them in future work.}
\begin{equation} 
\begin{array}{cccc}
\Lambda_1 = \langle l_1,l_2 \rangle &\;\to \;\;&\bZ^{(1)}_p \times \bZ^{(1)}_p \\
\Lambda_2 = \langle s_1,s_2 \rangle &\;\to \;\; &\bZ^{(0)}_p \times \bZ^{(0)}_p \\
\Lambda_3 = \langle l_1,s_2 \rangle &\; \to \;\; &\bZ^{(1)}_p \times \bZ^{(0)}_p \\
\Lambda_4 = \langle l_2,s_1 \rangle &\; \to \;\; &\bZ^{(0)}_p \times \bZ^{(1)}_p 
\end{array}
\end{equation}
The first two polarizations give rise to 3d absolute theories with, respectively, only 1-form symmetries $\bZ^{(1)}_p \times \bZ^{(1)}_p$ and 0-form symmetries $\bZ^{(0)}_p \times \bZ^{(0)}_p$ symmetries. More interesting are the third and fourth choices of polarization, which lead to two distinct 3d absolute theories with both 0-form and 1-form symmetries $\bZ^{(0)}_p \times \bZ^{(1)}_p $. The absolute theories are related by discrete gauging as shown in \autoref{Fig:GFsigmaS2S2}.

It is known that $(S^2\times S^1)\#(S^2\times S^1)$ has nontrivial mapping class group 
\begin{equation}
    (\bZ_2^2 \oplus  \bZ_2^2) \times \bZ_2
\end{equation}
where $\MCG( S^2\times S^1) = \bZ_2^2$ and the overall $\bZ_2$ factor switches the two $S^2\times S^1$ components 
\begin{equation} \label{eq:6dex2S}
    S:  (S^2 \times S^1)_1 \leftrightarrow (S^2 \times S^1)_2.
\end{equation}
The absolute theories specified by $\Lambda_1$ and $\Lambda_2$ are invariant under $S$ while the other two labelled by $\Lambda_3$ and $\Lambda_4$ transform into each other. 

\begin{figure}[!tbp]
\centering
\begin{tikzpicture}[baseline=19,scale=1]
\draw[thick,blue] (0,-0.2)--(0,3.2);
\draw[thick,orange] (3,-0.2)--(3,3.2);
\draw[thick,  ->] (6,1.5) -- (7,1.5);
\draw[brown, thick] (10,-0.2)--(10,3.2);
\draw[ dashed] (-2.5,0) -- (5.5,0);
\draw[ dashed] (7.5,0) -- (12.5,0);


\node[left] at (-0.7,1.5) {$\Lambda_3[\tau]$};
    \node at (1.5,1.5) {$\Lambda_4[\tau]$};
    \node[right] at (3.7,1.5) {$\Lambda_3[\frac{1}{\tau}]$};
      \node[left,blue] at (0,3) {$\sigma$};
          \node[right,orange] at (3,3) {$S$}; 
  \node[left] at (9.3,1.5) {$\Lambda_3[\tau]$};
    \node[right] at (10.7,1.5) {$\Lambda_3[\frac{1}{\tau}]$};
      \node[left,brown] at (10,3) {$\cN_2$};
\end{tikzpicture}
\caption{At $\tau=1$, the $T_{p,\Lambda_3}[(S^2\times S^1)\#(S^2\times S^1)]$ absolute theory has a duality defect $\mathcal{N}_2=\sigma S$. To simplify the notation, only polarization and coupling are included explicitly.}
\label{Fig:NFG}
\end{figure}
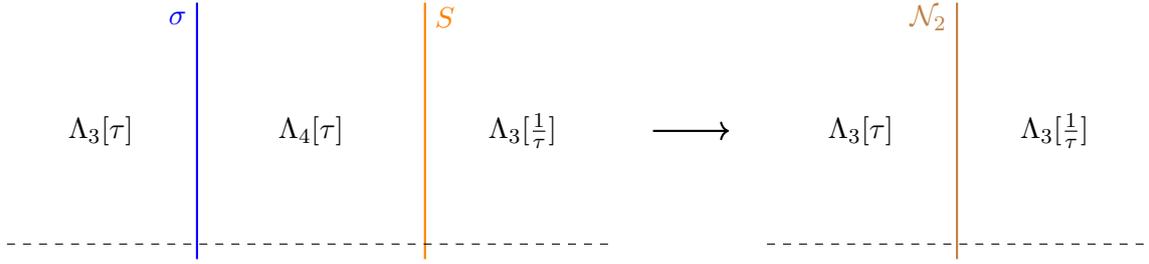

Despite the combined action of gauging $\sigma$ and $S$ on, say, $\Lambda_3$ gives back the same absolute theory, it might affect the gauge coupling thus behaving as an interface between to inequivalent theories. The reason for this is that the coupling constant of the absolute theory can be defined by the ratio of the volume of two components 
\begin{equation}
    \tau = \frac{ \text{vol}(S^2 \times S^1)_1}{\text{vol}(S^2 \times S^1)_2}.
\end{equation}
Thus, one can construct the duality defect only at the self-duality parameter of the theory, depending on the volume of the two $S^2\times S^1$. If we take the volume of both $S^2\times S^1$ to be the same, then the combined action of $\sigma$ and $s$ defines the duality defect $\cN_2=\sigma s$, see \autoref{Fig:NFG}. The fusion rules are those studied in \autoref{sec.2}.

\subsection{$T^3$}

Finally, let us consider the toroidal compactification of 6d $A_{N-1}$ theories on $T^3$. The theory after reduction $\cT_N[T^3]$ is a 3d $\cN=8$ $SU(N)$ super Yang-Mills theory, which is expected to flow to a 3d SCFT.
We will study the absolute theories and the symmetries of $\cT_N[T^3]$ using the SymTFT. Along the way, we find that for special couplings, $\cT_N[T^3]$ admits duality defect. 

Let us denote each one-cycle in $T^3$ as $\{l_1,l_2,l_3\}$ and product of pairs of one-cycles give two-cycles $s_{ij}=l_i \times l_j$ for $i\neq j$. For a fixed orientation, let us denote them by $(s_{23},s_{31},s_{12})$. Thus, the complete homology group of 3-torus is 
\begin{equation}
        H_*(T^3,\bZ) = \{\bZ, \bZ^3,\bZ^3 , \bZ\}.
\end{equation}
The non-trivial intersections between one/two-cycles are 
\begin{equation}
    l_i \cdot s_{jk} =  s_{ij} \cdot l_i = \epsilon_{ijk}, \qquad i,j,k=1,2,3.
\end{equation}
This can be confirmed from the cohomology ring by Poincar\'e duality. Thus, the action for the SymTFT \eqref{eq.4dTFT} is
\begin{equation}
    S_{4d}=\frac{4\pi}{N}\int_{M_4}\left(a_1\cup\delta b_1+a_2\cup\delta b_2+a_3\cup\delta b_3\right)
\end{equation}
which does not reproduce the BF action \eqref{eq.SymTFTaction} of the SymTFT in \autoref{sec.3}.

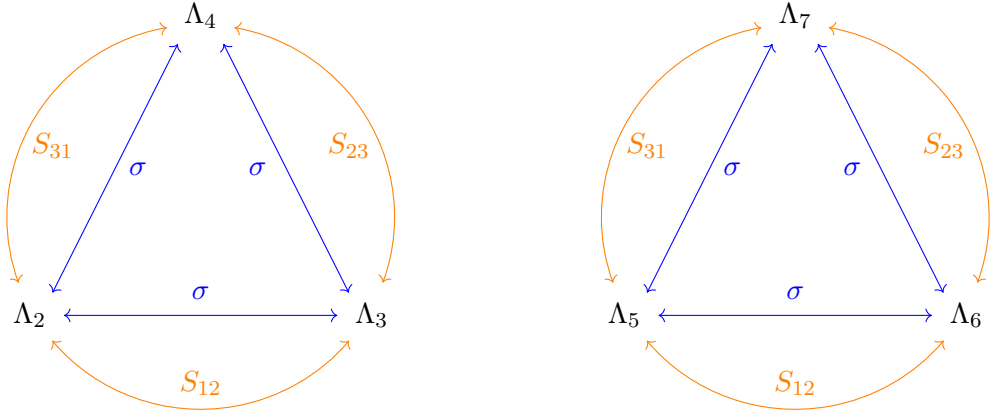
\begin{figure}
\centering\[
\begin{tikzpicture}[scale=1.5]
\draw node at (0,0) {$\Lambda_2$};
\draw node at (3,0) {$\Lambda_3$};
\draw node at (1.5,2.65) {$\Lambda_4$};
\draw [<->,blue] (.3,0) -- (2.7,0);
\draw [<->,blue] (.2,0.2) -- (1.3,2.4);
\draw [<->,blue] (2.8,0.2) -- (1.7,2.4);
\centerarc[orange,<->](1.5,0.87)(-20:80:1.7)
\centerarc[orange,<->](1.5,0.87)(100:200:1.7)
\centerarc[orange,<->](1.5,0.87)(220:320:1.7)
\draw node at (1.5,.2) {${\color{blue} \sigma}$};
\draw node at (0.95,1.3) {${\color{blue} \sigma}$};
\draw node at (2.,1.3) {${\color{blue} \sigma}$};
\draw node at (1.5,-0.6) {${\color{orange} S_{12}}$};
\draw node at (0.2,1.5) {${\color{orange} S_{31}}$};
\draw node at (2.8,1.5) {${\color{orange} S_{23}}$};
\end{tikzpicture}
\hspace{2.5cm} 
\begin{tikzpicture}[scale=1.5]
\draw node at (0,0) {$\Lambda_5$};
\draw node at (3,0) {$\Lambda_6$};
\draw node at (1.5,2.65) {$\Lambda_7$};
\draw [<->,blue] (.3,0) -- (2.7,0);
\draw [<->,blue] (.2,0.2) -- (1.3,2.4);
\draw [<->,blue] (2.8,0.2) -- (1.7,2.4);
\centerarc[orange,<->](1.5,0.87)(-20:80:1.7)
\centerarc[orange,<->](1.5,0.87)(100:200:1.7)
\centerarc[orange,<->](1.5,0.87)(220:320:1.7)
\draw node at (1.5,.2) {${\color{blue} \sigma}$};
\draw node at (0.95,1.3) {${\color{blue} \sigma}$};
\draw node at (2.,1.3) {${\color{blue} \sigma}$};
\draw node at (1.5,-0.6) {${\color{orange} S_{12}}$};
\draw node at (0.2,1.5) {${\color{orange} S_{31}}$};
\draw node at (2.8,1.5) {${\color{orange} S_{23}}$};
\end{tikzpicture}
\]
\caption{Global forms of $T_p[T^3, \Lambda_j]$ for $j=2,3,\ldots, 7$. Blue lines represents gauging $\bZ_p^{(1)} \times \bZ_p^{(0)}$ while orange lines are the duality defined in \eqref{eq:6dS3}. }
\label{Fig:T3}
\end{figure}

The defect group of $T_N[T^3]$ for prime $N=p$ is given by
\begin{equation}
    D = H_1(T^3,\bZ_p) \oplus H_2(T^3,\bZ_p) = \bZ_p^3 \oplus \bZ_p^3.
\end{equation}
Hence, to find absolute theories we need to choose a polarization. When $N=p$ is prime, there are the following eight different polarizations
\begin{equation} 
\begin{array}{cccc}
\Lambda_1 = \langle l_1,l_2, l_3 \rangle &\to \; \; &\bZ^{(1)}_p \times \bZ^{(1)}_p \times \bZ^{(1)}_p \\
\Lambda_2 = \langle s_{23},l_2, l_3 \rangle &\to \; \; &\bZ^{(0)}_p \times \bZ^{(1)}_p \times \bZ^{(1)}_p \\
\Lambda_3 = \langle l_1,s_{31}, l_3 \rangle &\to \; \; &\bZ^{(1)}_p \times \bZ^{(0)}_p \times \bZ^{(1)}_p \\
\Lambda_4 = \langle l_1,l_2, s_{12} \rangle &\to \; \; &\bZ^{(1)}_p \times \bZ^{(1)}_p \times \bZ^{(0)}_p \\
\Lambda_5 = \langle l_1,s_{31}, s_{12} \rangle &\to \; \; &\bZ^{(1)}_p \times \bZ^{(0)}_p \times \bZ^{(0)}_p \\
\Lambda_6 = \langle s_{23},l_2, s_{12} \rangle &\to \; \; &\bZ^{(0)}_p \times \bZ^{(1)}_p \times \bZ^{(0)}_p \\
\Lambda_7 = \langle s_{23},s_{31}, l_3 \rangle &\to \; \; &\bZ^{(0)}_p \times \bZ^{(0)}_p \times \bZ^{(1)}_p \\
\Lambda_8 = \langle s_{23},s_{31}, s_{12} \rangle &\to \; \; &\bZ^{(0)}_p \times \bZ^{(0)}_p \times \bZ^{(0)}_p \\
\end{array}
\end{equation}
Each of them defines an absolute theory $T_{p,\Lambda_j}[T^3]$ for $j=1,2,\ldots, 8$. The global symmetries are listed on the right. Different absolute theories are related among each other by gauging either $\bZ^{(1)}_p$ or $\bZ^{(0)}_p$. We will focus on the operation of gauging $\bZ^{(1)}_p \times \bZ^{(0)}_p$ denoted by $\sigma$. It transforms between different global forms as plotted in \autoref{Fig:T3}.

The mapping class group of $T^3$ is known to be $SL(3,\bZ)$ with two generators given by \cite{Wang:2014oya,Wang:2016qhf,Wang:2019diz} 
\begin{equation}
S_{123}=\left(
\begin{array}{ccc}
 0 & 0 & 1 \\
 1 & 0 & 0 \\
 0 & 1 & 0 
\end{array}
\right), 
\qquad 
T_{12}=\left(
\begin{array}{ccc}
 1 & 1 & 0 \\
 0 & 1 & 0 \\
 0 & 0 & 1
\end{array}
\right).
\end{equation}
The modular group of the 2-torus is the $SL(2,\bZ)$ subgroup. In particular, the S-transformations of each 2-torus $s_{23}$, $ s_{31}$ and $s_{12}$ inside $T^3$ are given by 
\begin{equation} \label{eq:6dS3}
S_{23}=\left(
\begin{array}{ccc}
 1 & 0 & 0 \\
 0 & 0 & -1 \\
 0 & 1 & 0
\end{array}
\right), 
\qquad 
S_{31}=\left(
\begin{array}{ccc}
 0 & 0 & 1 \\
 0 & 1 & 0 \\
 -1 & 0 & 0
\end{array}
\right),
\qquad 
S_{12}=\left(
\begin{array}{ccc}
 0 & -1 & 0 \\
 1 & 0 & 0 \\
 0 & 0 & 1
\end{array}
\right).
\end{equation}
Moreover, for each $s_{ij}$, the S-matrix $S_{ij}$ transforms the corresponding one-cycles as follows
\begin{equation}
    l_i \to -l_j,\qquad l_j \to l_i, 
\end{equation}
with $i,j=1,2,3$. They generate non-trivial dualities that transform among different absolute theories\footnote{For each 2-torus, there are also T-transformations. However, they do not transform among different absolute theories and we will not discuss them here.}. For example, the action of $S_{12}$ on the absolute theory $\Lambda_2$ is 
\begin{equation}
   S_{12}:\;\;\; \langle s_{23},l_2, l_3 \rangle \to \langle s_{13},l_1, l_3 \rangle, 
\end{equation}
which is the absolute theory $\Lambda_3$ up to a flip of orientation of $s_{13}$. Note that polarization is not sensitive to the choice of orientation because they are valued on a finite integer lattice. Hence, $s_{13} =-s_{31}$ together with $l_1$ and $l_3$ generate the polarization $\Lambda_3$. Similarly, we study how the global forms transform under other S-transformations and plot the result in \autoref{Fig:T3}. 

As in the previous example, to find duality defects we need to take into account the value of the coupling constants of the absolute theories. Let $r_i$ be the radius of $l_i = S^1$ with $i=1,2,3$. There are two couplings in the absolute theories $T_{p,\Lambda_j}[T^3]$ given by 
\begin{equation}
    \tau_1 = \frac{r_1}{r_3}, \qquad \tau_2 = \frac{r_2}{r_3}
\end{equation}
By construction, they transform under the dualities as 
\begin{equation}
\begin{split}
    & S_{12}:\qquad  (\tau_1,\tau_2) \;\; \to \;\;(\tau_2,\tau_1) \\
    & S_{23}:\qquad (\tau_1,\tau_2) \;\;\to \;\;(\tau_1/\tau_2,\tau_2^{-1}) \\
    & S_{31}:\qquad (\tau_1,\tau_2) \;\;\to \;\;(\tau_1^{-1},\tau_2/\tau_1) 
\end{split}
\end{equation}
Note that these dualities act on the three-dimensional lattice in $T^3$ by switching between different one-cycles. Although the direction of these 1-cycles may change, their radius remains the same. 

We will show that many of these absolute theories admit a duality defect. For example, the absolute theory specified by $\Lambda_2$ is invariant under the composition of gauging $\bZ_p^{(1)} \times \bZ_p^{(0)}$ and duality $S_{12}$. However, couplings are changed from $(\tau_1,\tau_2) $ to $(\tau_2,\tau_1)$. The combined operation $\sigma S_{12}$ becomes duality defect if these couplings are set to be $\tau_1 = \tau_2$. Equivalently, this requires $r_1=r_2$, i.e. $l_1$ and $l_2$ cycles have the same length. This condition also appears in the geometric engineer of duality defects in 4d $\cN=4$ SYM in \cite{Bashmakov:2022uek}. 

From \autoref{Fig:T3}, one can find other operations containing both $\sigma$ and $S_{ij}$ which keep a polarization invariant. Hence, it is possible to define duality defects in these absolute theories. The result is summarised below 
\begin{itemize}
\item 
Duality defect $\cN_2=\sigma S_{12}$ exists in absolute theories
\begin{equation}
 \{\Lambda_2,\Lambda_3,\Lambda_5,\Lambda_6 \},  \qquad \qquad  \text{at} \; (\tau, \tau), \quad  \tau \in \bR
\end{equation}
\item 
Duality defect $\cN_2=\sigma S_{31}$ exists in absolute theories
\begin{equation}
 \{\Lambda_2,\Lambda_4,\Lambda_5,\Lambda_7 \},  \qquad \qquad  \text{at} \; (\tau, 1), \quad  \tau \in \bR
\end{equation}
\item 
Duality defect $\cN_2=\sigma S_{23}$ exists in absolute theories
\begin{equation}
 \{\Lambda_3,\Lambda_4,\Lambda_6,\Lambda_7 \},  \qquad \qquad  \text{at} \; (1, \tau), \quad  \tau \in \bR
\end{equation}
\end{itemize}

In conclusion, $T_p[T^3]$ has eight absolute theories. Except for those labelled by $\Lambda_1$ and $\Lambda_8$, we find duality defect at special couplings in all other six different absolute theories. It is straightforward to generalize the analysis to $X_3=C_g \times S^1$ where $C_g$ is a Riemann surface with $g>1$. We expect the existence of duality defects also for $T_p[C_g \times S^1]$.  


\section{Conclusion and Outlook}\label{sec.6}

In this paper we have studied non-invertible topological surface defects of (2+1)d QFTs by gauging $\bZ_N^{(0)} \times \bZ_N^{(1)}$ symmetry in half of the spacetime. This includes duality defect, triality defect and, possibly, N-ality defect. The fusion rule involving duality defects and triality defects are derived and they are shown to form a fusion 2-category. We have further found a (3+1)d BF theory on a slab that is able to describe the duality interfaces on its boundary upon shrinking the slab. The SymTFT describing the duality defects is obtained by gauging the $\bZ_4^\text{EM}$ electric-magnetic symmetry in the (3+1)d BF theory.

We have also provided concrete examples of theories with such duality defects. In particular we discussed how duality defects are realized in a $U(1) \times U(1)$ gauge theory. We found that duality defects are common in product theories and provided a general way to construct them. Finally, we studied the absolute theories and duality defects in non-Lagrangian theories from twisted compactification of 6d SCFTs on 3-manifolds including $T^3$ and $ (S^2\times S^1) \# (S^2\times S^1)$. 

There are a certain interesting direction for future research. First, it would be interesting to find more examples in 3d QFTs beyond those explored in \autoref{sec.4}, possibly in theories which are not obtained taking the product of two more elementary theories. Second, determining the F-symbol of the fusion 2-category would give us a deeper understanding of $\text{TY}(\bZ^{(0)}_N\times \bZ^{(1)}_N )$. Third, for a theory 2+1d $\mathcal{Q}$ defined on a manifold with a 2d boundary, it would be interesting to study how the non-invertible topological defect $\mathcal{N}_2$ affects the boundary conditions. Finally, it would also be interesting to find criteria which allow us to determine when a give duality defect is intrinsically non-invertible by studying the quantum dimension of non-invertible symmetry operators as in \cite{Kaidi:2022cpf, Sun:2023xxv}.

\paragraph*{Acknowledgments}
We would like to thank Sergio Cecotti, Yi Huang, Tomoki Nosaka, Zheyan Wan for valuable discussions. 
BH is supported by the NSFC grant 12250610187. The work of WC is supported by the fellowship
of China Postdoctoral Science Foundation NO.2022M720507 and in part by the Beijing Postdoctoral Research Foundation. LR acknowledges support from the Shuimu Tsinghua Scholar Program.

\appendix

\section{Details on Fusion Rules of Duality Defects}\label{app.A}

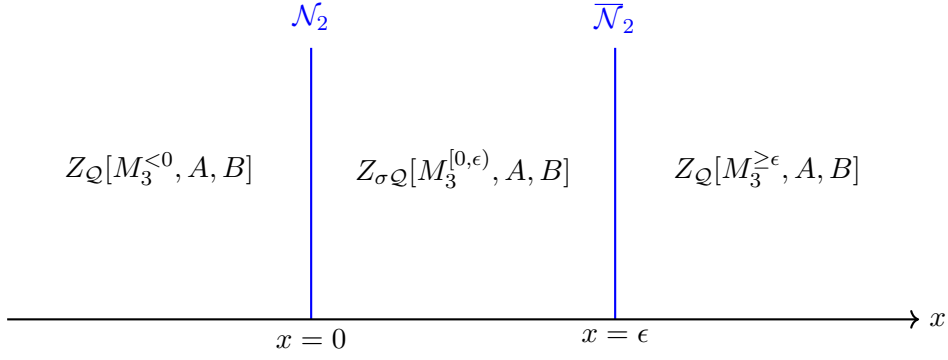
\begin{figure}[!tbp]
\centering
\begin{tikzpicture}[baseline=19,scale=1]
\draw[blue, thick] (2,-0)--(2,3.6);
\draw[blue, thick] (6,-0)--(6,3.6);
\node[below] at (2,0) {$x=0$};
\node[below] at (6,0) {$x=\epsilon$};
\draw[ thick,->] (-2,0) -- (10,0);
\node[right] at (10,0) {$x$};
\node[blue] at (2,4) {$\cN_2$};
\node[blue] at (6,4) {$\overline{\cN}_2$};
\node at (0,2) {$Z_{\cQ}[M_3^{<0},A,B]$};
\node at (4,2) {$Z_{\sigma \cQ}[M_3^{[0,\epsilon)},A,B]$};
\node at (8,2) {$Z_{\cQ}[M_3^{\geq \epsilon},A,B]$};
\end{tikzpicture}
\caption{Fusion of two (2+1)d duality defects. }
\label{Fig:fusPartition}
\end{figure}

We will re-derive the fusion rules between two duality defects presented in \eqref{eq.fusionD1} following the approach in \cite{Kaidi:2022cpf}. Suppose the theory $\cQ$ has a duality defect $\cN_2$ defined by gauging $\bZ_N^{(0)} \times \bZ_N^{(1)}$. Consider the configuration in \autoref{Fig:fusPartition} where $\cN_2$ is inserted at $x=0$ and $\overline{\cN}_2$ is inserted at $x=\epsilon$. Let $Z_{\cQ}[A,B]$ be the partition function of $\cQ$ with background gauge field $A$ and $B$ for $\bZ_N^{(0)}$ and $ \bZ_N^{(1)}$. The partition function in the regions $x\in [0,\epsilon)$ and $x>\epsilon$ is as follows
\begin{equation}\label{eq:NN4d}
\begin{split}
    &  \frac{1}{|H^1(M_3^{\geq 0}, M_2|_0, \bZ_N)|}\frac{1}{|H^1(M_3^{\geq \epsilon}, M_2|_\epsilon, \bZ_N)|} \times 
    \\& \sum_{\substack{a \in H^1(M_3^{\geq 0}, M_2|_0, \bZ_N)\\ b \in H^2(M_3^{\geq 0}, M_2|_0, \bZ_N) \\  \widetilde a \in H^1(M_3^{\geq \epsilon}, M_2|_\epsilon, \bZ_N)\\\widetilde b \in H^2(M_3^{\geq \epsilon}, M_2|_\epsilon, \bZ_N)}} 
    Z_{\cQ}[M_3^{\geq 0}, a,b]\, \exp{\left({2 \pi i \over N}\int_{M_3^{[0,\epsilon)}} (b A + a B) + {2 \pi i \over N}\int_{M_3^{\geq \epsilon}} (b-B) \widetilde a+ (a-A) \widetilde b\right)}~.
\end{split}
\end{equation}
Notice that $a$,$b$ and $\widetilde a$,$\widetilde b$ are valued in relative cohomologies. 

One can convert these relative cohomologies to co-chains by introducing the extra dynamical fields $\alpha \in C^1(M_3^{\geq 0},\bZ_N)$, $\widetilde \alpha \in C^1(M_3^{[0,\epsilon)},\bZ_N)$, $\phi \in C^0(M_3^{\geq 0},\bZ_N)$ and $\widetilde \phi \in C^0(M_3^{[0,\epsilon)},\bZ_N)$. Then we also introduce the couplings $\alpha a$, $\phi b$ on $M_3|_0$ and $\widehat \alpha \widehat a$, $\widehat \phi \widehat b$ on $M_3|_{\epsilon}$. After integrating out all these auxiliary fields and $\widetilde a$, $\widetilde b$ the equation above becomes 
\begin{equation}\label{eq:NN4d1}
\begin{split}
& \frac{|H^1(M_3^{\geq \epsilon}, \bZ_N)|}{|H^1(M_3^{\geq 0}, M_2|_0, \bZ_N)|}\frac{|H^2(M_3^{\geq \epsilon}, \bZ_N)|}{|H^1(M_3^{\geq \epsilon}, M_2|_\epsilon, \bZ_N)|} \times 
\\ & \sum_{\substack{a \in H^1(M_3^{[0,\epsilon)}, M_2|_0 \cup M_2|_{\epsilon}, \bZ_N)\\ b \in H^2(M_3^{[0,\epsilon)}, M_2|_0 \cup M_2|_{\epsilon}, \bZ_N) }} 
    Z_{\cQ}[M_3^{\geq 0}, a+A|_{M_3^{\geq \epsilon}},b+B|_{M_3^{\geq \epsilon}}]\, \exp{\left({2 \pi i \over N}\int_{M_3^{[0,\epsilon)}} (b A + a B) \right)}.
\end{split}
\end{equation}
Here, we have the following identities 
\begin{equation}
    |C^n(M_3^{\geq 0}, M_2|_0, \bZ_N)| = |C^n(M_3^{[0,\epsilon)}, M_2|_0 \cup M_2|_{\epsilon}, \bZ_N)||C^n(M_3^{\geq \epsilon}, \bZ_N)|
\end{equation}
and $|C^n(M_3,M_2,\bZ_N)|=|C^{3-n}(M_3,\bZ_N)|$.

Now we shrink the slab and take $\epsilon \to 0$. The region becomes $M_3^{[0,\epsilon)} \to M_2$ at $x=0$ and the relative cohomology becomes absolute. Thus, the partition function is 
\begin{equation}\label{eq:NN4d2}
\begin{split}
    & \frac{1}{|H^0(M_2, \bZ_N)|} \sum_{\substack{a \in H^1(M_2, \bZ_N)\\ b \in H^2(M_2, \bZ_N) }} 
    Z_{\cQ}[M_3^{\geq 0}, A,B]\, \exp{\left({2 \pi i \over N}\int_{M_2} (b A + a B) \right)}.
\end{split}
\end{equation}
From this partition function one can read off the fusion of the duality defect and its orientation-reversal on $M_2$  
\begin{align}
    \overline{\cN}_2 \times \cN_2 =  \cN_2 \times \overline{\cN}_2 = C \sum_{i=0}^{N-1} \left(\eta_0\right)^{i}
\end{align}
which is consistent with the result in \eqref{eq.fusionD1}.

\section{Boundary of the Topological Action}\label{app.ta}

The topological action in \autoref{sec.2.4}  
\begin{equation}
    \frac{2\pi p}{N} \int_{M_3} a \cup b, \qquad p \in \bZ_N
\end{equation}
has an equivalent description \cite{Kapustin:2014gua, Gaiotto:2014kfa}
\begin{equation} \label{eq.appB1}
    S= \frac{N}{2\pi}\int_{M_3} p a b +  a d \Tilde{a} +  b d\phi
\end{equation}
where $a,\Tilde{a} \in H^1(M_3,U(1))$, $b \in H^2(M_3,U(1))$ and $\phi$ is a compact scalar with $\phi = \phi + 2\pi$. Here $\tla$ and $\phi$ act as Lagrangian multiplier constraining $a$ and $b$ to be valued in $\bZ_N$.

The topological theory \eqref{eq.appB1} on a closed manifold $M_3$ is invariant under the following 0-form and 1-form gauge transformations 
\begin{equation}
\begin{split}
   &a\to a-d\lambda_0,\quad \phi \to \phi +p \lambda_0, \\
   &b \to b-d\lambda_1,\quad \tla \to \tla +p \lambda_1.
\end{split}
\end{equation}
The gauge invariant operators include the following surface and line operators   
\begin{equation}
    W(\sigma) = \exp{\left({i\oint_{\sigma}b}\right)} ,\quad U(\gamma) = \exp{\left({i\oint_{\gamma}a}\right)}.
\end{equation}

Besides these, there are also gauge invariant line and local operators defined by ending them on an open surface or line operator 
\begin{equation}
    \hat{U}(\gamma) = \exp{\left( i\oint_{\gamma}\Tilde{a}+ ip\int_{\sigma'} b \right)} ,\quad \hat{\Phi}(q) = \exp{\left({i\phi(q)} + i p\int_{\gamma'} a \right)}
\end{equation}
where $\gamma=\partial \sigma'$ is a curve and $ q=\partial \gamma'$ is a point in $M_3$. To make them bulk independent, we will consider the operators $\Tilde{U}(\gamma)= \hat{U}^L(\gamma)$ and $  \Phi(q)=\hat{\Phi}^L(q) $ with $L=\text{gcd}(N,p)$. By a similar argument as in \cite{Kapustin:2014gua, Gaiotto:2014kfa}, one can show that
\begin{equation}
   \Tilde{U}^{N/L}(\gamma)= \Phi^{N/L}(q) = 1,\quad  W^{N/L}(\sigma) = U^{N/L}(\gamma) =1,
\end{equation}
which generate the $\bZ^{(0)}_{N/L}$, $\bZ^{(1)}_{N/L} \times \bZ^{(1)}_{N/L}$ and $\bZ^{(2)}_{N/L}$ symmetries in the bulk theory. The charged objects are $\Phi(q)$, $\Tilde{U}(\gamma)U(\gamma)$ and $ W(\sigma)$ respectively.

When the spacetime has a boundary, the gauge transformations give a non-trivial boundary term 
\begin{equation}
    - \frac{N}{2 \pi} \int_{\partial M_3} \left( \phi d\lambda_1 + \tla d\lambda_0 \right).
\end{equation}
To make the theory gauge invariant, one can add the following terms on the boundary 
\begin{equation}
    S_{\text{bdry}} = \frac{N}{2 \pi} \int_{\partial M_3} ( ydx + b y + a x - b\phi  -a \tla  )
\end{equation}
where $x \in H^1(M_2,U(1))$ and $y\in H^0(M_2,U(1))$ are dynamical gauge fields on the boundary. They transform as 
\begin{equation}
    x\to x + \lambda_1,\quad y \to y+ \lambda_0.
\end{equation}
One can check that now the action $S+S_{\text{bdry}}$ is gauge invariant on $M_3$. 

For the case of $p=1$, one can see that the bulk is trivial and the boundary theory with Dirichlet boundary conditions $a|_{\partial M_3}=b|_{\partial M_3}=0$ becomes 
\begin{equation}
    S_{\text{bdry}} =   \frac{N}{2 \pi} \int_{\partial M_3}  ydx.
\end{equation}
This gives the 2d $\bZ_N$ gauge theory on the boundary.

\section{Correlation Functions of Line and Surface Operators in the SymTFT}\label{app.B}

In this appendix, we calculate the correlation functions involving line and surface operators in the SymTFT. 
Consider the insertion of a line operator $L_{(l_1,l_2)}(\gamma)$ along a curve $\gamma \subset M_4$ and a surface operator $S_{(s_1,s_2)}(\sigma)$ along a surface $\sigma \subset M_4$. 
The action becomes 
\begin{equation}
\begin{split}
    S  &= \frac{2\pi}{N} \int_{M_4} b_1\delta a_1 + b_2\delta a_2 + \frac{2\pi}{N} \int_{\gamma}  l_1 a_1 + l_2 a_2 + \frac{2\pi}{N} \int_{\sigma} s_1 b_1 + s_2 b_2 \\
    &= 
    \frac{2\pi}{N} \int_{M_4}
     (l_1 a_1 + l_2 a_2)\text{PD}(\gamma) 
     + 
     \frac{2\pi}{N} \int_{M_4} b_1(\delta a_1 + s_1 \text{PD}(\sigma)) + b_2(\delta a_2+  s_2 \text{PD}(\sigma)) 
\end{split}
\end{equation}
where $\text{PD}$ represents the Poincar\'e dual.

Integrating out $b_1$, $b_2$ gives the constraints 
\begin{equation}
    \delta a_i + s_i \text{PD}(\sigma) =0 , \qquad i=1,2.
\end{equation}
Equivalently, one can write $a_i = -s_i\text{PD}(V)$ with $\partial V = \sigma$. Plugging this into the action above, we have 
\begin{equation}
 -\frac{2\pi}{N} 
     (l_1s_1 + l_2s_2)\text{link}(\sigma , \gamma)
\end{equation}
where $\text{link}(\sigma, \gamma)$ is defined by the intersection between $V$ and $\gamma$. 

Thus, the correlation function is 
\begin{equation}
    \langle L_{(l_1,l_2)}(\gamma) S_{(s_1,s_2)}(\sigma) \cdots \rangle = \exp{\left(-\frac{2\pi i}{N}(l_1s_1+l_2s_2) \text{link}(\gamma, \sigma)\right)}\langle\cdots\rangle
\end{equation}
This reproduces \eqref{eq.correlation.sigmagamma}. Similarly, we can compute the correlation functions involving two line operators or two surface operators. However, since the linking number is trivial, it does not contribute to the correlation function.

\section{Fusion Rules Involving Condensation Defects}\label{app.fusionCon}

In this appendix, we will derive the fusion rule of condensation defects presented in \autoref{sec.3}. The definition of condensation defects depends on the value of $N$. As discussed in \eqref{eq.condesN}, higher form symmetries condensed along $M_3$ are different when $N$ is odd, or even and $N=2$. 
We will first study the fusion rule involving condensation defect for $N=2$ and then consider the generic case.

\subsection{$N=2$}

The condensation defect for $N=2$ is defined as \eqref{eq.condefect.2} 
\begin{equation}
    D(M_3) = \frac{1}{|H^1(M_3,\bZ_2)|}\sum_{\substack{\gamma \in H_1(M_3,\mathbb{Z}_2),\\
    \sigma \in H_2(M_3,\mathbb{Z}_2)}} e^{-\pi i \langle \sigma,\gamma \rangle } \; S_{(1,-1)}(\sigma) L_{(1,-1)}(\gamma).
\end{equation}
Below, we will calculate the fusion of $D(M_3)$ with the line operator $L_{(l_1,l_2)}$, the surface operator $S_{(s_1,s_2)}$ and $D(M_3)$ itself. 

\paragraph{Fusion with $L_{(l_1,l_2)}$.}

The fusion between the line operator $L_{(l_1,l_2)}$ and $D(M_3)$ is   
\begin{align*}
\begin{split}
     & \;\;\;\;\; L_{(l_1,l_2)}(M_1) \times D(M_3) \\
     & = \frac{1}{|H^1(M_3,\bZ_2)|}\sum_{\substack{\gamma \in H_1(M_3,\mathbb{Z}_2),\\
    \sigma \in H_2(M_3,\mathbb{Z}_2)}}  L_{(l_1,l_2)}(M_1) S_{(1,-1)}(\sigma) L_{(1,-1)}(\gamma) e^{-\pi i \langle \sigma,\gamma \rangle} \\
    & = \frac{1}{|H^1(M_3,\bZ_2)|}\sum_{\substack{\gamma \in H_1(M_3,\mathbb{Z}_2),\\
    \sigma \in H_2(M_3,\mathbb{Z}_2)}}   S_{(1,-1)}(\sigma) L_{(1,-1)}(\gamma) e^{-\pi i (\langle \sigma,\gamma \rangle+(l_1-l_2)\langle \sigma, M_1 \rangle)} L_{(l_1,l_2)}(M_1) \\ 
    & = \frac{1}{|H^1(M_3,\bZ_2)|}\sum_{\substack{\gamma \in H_1(M_3,\mathbb{Z}_2),\\
    \sigma \in H_2(M_3,\mathbb{Z}_2)}}   S_{(1,-1)}(\sigma) L_{(1,-1)}(\gamma+(l_1-l_2)M_1) e^{-\pi i (\langle \sigma,\gamma+(l_1-l_2)M_1 \rangle)} L_{(l_2,l_1)}(M_1)  \\
    & =  D(M_3) \times L_{(l_2,l_1)}(M_1).
\end{split}
\end{align*}
The equal-time commutation relation \eqref{eq.comuteLS} has been used in the second line, and both the quantum torus algebra \eqref{eq.qtaLS} and 
\begin{equation} \label{eq.LLL}
     L_{(l_1,l_2)}(M_1) = L_{(l_2,l_1)}(M_1) \times L^{(l_1-l_2)}_{(1,-1)}(M_1)
\end{equation}
have been used in the third line. Because $l_2$ is a $\bZ_2$ valued cycle, $l_2=-l_2$. Therefore, the fusion rule reproduces \eqref{eq.FusionD1}. 

\paragraph{Fusion with $S_{(s_1,s_2)}$.}

The fusion between the line operator $S_{(s_1,s_2)}$ and $D(M_3)$ is 
\begin{align*}
\begin{split}
     & \;\;\;\;\; S_{(s_1,s_2)}(M_2) \times D(M_3) \\
     & = \frac{1}{|H^1(M_3,\bZ_2)|}\sum_{\substack{\gamma \in H_1(M_3,\mathbb{Z}_2),\\
    \sigma \in H_2(M_3,\mathbb{Z}_2)}}  S_{(s_1,s_2)}(M_2) S_{(1,-1)}(\sigma) L_{(1,-1)}(\gamma) e^{-\pi i \langle \sigma,\gamma \rangle} \\
    & = \frac{1}{|H^1(M_3,\bZ_2)|}\sum_{\substack{\gamma \in H_1(M_3,\mathbb{Z}_2),\\
    \sigma \in H_2(M_3,\mathbb{Z}_2)}}   S_{(1,-1)}(\sigma+(s_1-s_2)M_2) S_{(s_2,s_1)}(M_2)L_{(1,-1)}(\gamma) e^{-\pi i \langle \sigma,\gamma \rangle}  \\ 
    & = \frac{1}{|H^1(M_3,\bZ_2)|}\sum_{\substack{\gamma \in H_1(M_3,\mathbb{Z}_2),\\
    \sigma \in H_2(M_3,\mathbb{Z}_2)}}   S_{(1,-1)}(\sigma+(s_1-s_2)M_2) L_{(1,-1)}(\gamma) e^{-\pi i (\langle \sigma+(s_1-s_2)M_2,\gamma \rangle)} S_{(s_2,s_1)}(M_2)  \\
    & =  D(M_3) \times S_{(s_2,s_1)}(M_2)
\end{split}
\end{align*}
The quantum torus algebra \eqref{eq.qtaLS} and a similar relation to \eqref{eq.LLL} for $S_{(s_1,s_2)}(M_2)$ have been applied in the second line. Moreover, the commutation relation \eqref{eq.comuteLS} has been used in the third line. 

\paragraph{Fusion with $D(M_3)$.}

The fusion between two condensation defects is given by:
\begin{align*}
\begin{split}
      & \;\; \;\; \; D(M_3) \times D(M_3) \\  &= \frac{1}{|H^1(M_3,\bZ_2)|^2}\sum_{\substack{\gamma,\gamma' \in H_1(M_3,\mathbb{Z}_2),\\
    \sigma,\sigma' \in H_2(M_3,\mathbb{Z}_2)}} S_{(1,-1)}(\sigma) L_{(1,-1)}(\gamma)S_{(1,-1)}(\sigma') L_{(1,-1)}(\gamma') e^{-\pi i (\langle \sigma,\gamma \rangle+\langle \sigma',\gamma' \rangle)} \\
    & = \frac{1}{|H^1(M_3,\bZ_2)|^2}\sum_{\substack{\gamma,\gamma' \in H_1(M_3,\mathbb{Z}_2),\\
    \sigma,\sigma' \in H_2(M_3,\mathbb{Z}_2)}} S_{(1,-1)}(\sigma+\sigma') L_{(1,-1)}(\gamma+\gamma')  e^{-\pi i \left(\langle \sigma,\gamma \rangle+\langle \sigma',\gamma' \rangle + 2 \langle \sigma',\gamma \rangle \right)} \\
    & = \frac{1}{|H^1(M_3,\bZ_2)|^2}\sum_{\substack{\gamma,\gamma' \in H_1(M_3,\mathbb{Z}_2),\\
    \sigma,\sigma' \in H_2(M_3,\mathbb{Z}_2)}} S_{(1,-1)}(\sigma) L_{(1,-1)}(\gamma)  e^{-\pi i  \left(\langle \sigma,\gamma \rangle + \langle \sigma',\gamma \rangle - \langle \sigma,\gamma' \rangle \right)}.
    = 1,
\end{split}
\end{align*}
The quantum torus algebra \eqref{eq.qtaLS} and the commutation relation \eqref{eq.comuteLS} have been used in the second line. We also made a change of variables $\sigma \to \sigma -\sigma'$ and $\gamma \to \gamma - \gamma'$ and further integrated out both $\sigma'$ and $\gamma'$ in the third line. This shows the invertibility of the condensation defect in \eqref{eq.fusionD5}.

\subsection{$N>2$}

The condensation defect for $N>2$ is defined in \eqref{eq.condefect.N} and it is given by
\begin{equation}
\begin{split}
    D^i(M_3)=\frac{1}{|H^1(M_3,\Xi^i)|}\sum_{\substack{(\gamma,\gamma')\in H_1(M_3,\Xi^i),\\
    (\sigma,\sigma')\in H_2(M_3,\Xi^i)}}&\exp{\left(-\frac{2\pi i}{N} (\langle \sigma,\gamma-\gamma'\rangle+\langle \sigma',\gamma+\gamma'\rangle)\right)} 
    \\
    &\times S_{(1,-1)}(\sigma)S_{(1,1)}(\sigma')L_{(1,-1)}(\gamma)L_{(1,1)}(\gamma'), 
\end{split}
\end{equation}
where $\Xi^1=\mathbb{Z}_N\times\mathbb{Z}_N$ for odd $N$ and $\Xi^2=(\mathbb{Z}_N\times\mathbb{Z}_N)/\mathbb{Z}_2$ for even $N$. 
We will study the fusion of $D(M_3)$ with the line operator $L_{(l_1,l_2)}$, the surface operator $S_{(s_1,s_2)}$. We will also define the charge conjugation defect and confirm the invertibility of $D(M_3)$.

\paragraph{Fusion with $L_{(l_1,l_2)}$.}

The fusion between the line operator $L_{(l_1,l_2)}(M_1)$ and $D(M_3)$ is  
\begin{equation}
\begin{split}
     & \;\;\;\;\; L_{(l_1,l_2)}(M_1) \times D^i(M_3) \\
     & = \frac{1}{|H^1(M_3,\Xi^i)|}\sum_{\substack{(\gamma,\gamma')\in H_1(M_3,\Xi^i),\\
    (\sigma,\sigma')\in H_2(M_3,\Xi^i)}}  \exp{\left(-\frac{2\pi i}{N} (\langle \sigma,\gamma-\gamma' \rangle + \langle \sigma', \gamma+\gamma' \rangle) \right)} 
    L_{(l_1,l_2)}(M_1) S_{(1,-1)}(\sigma) \\ 
    & \hspace{5.2cm} \times S_{(1,1)}(\sigma') 
      L_{(1,-1)}(\gamma) L_{(1,1)}(\gamma') \\\\
    & = \frac{1}{|H^1(M_3,\Xi^i)|}\sum_{\substack{(\gamma,\gamma')\in H_1(M_3,\Xi^i),\\
    (\sigma,\sigma')\in H_2(M_3,\Xi^i)}}  \exp{\left(-\frac{2\pi i}{N} (\langle \sigma,\gamma-\gamma' + (l_1-l_2)M_1 \rangle + \langle \sigma', \gamma+\gamma' + (l_1+l_2)M_1\rangle ) \right)} 
     \\ 
    & \hspace{5.2cm} \times S_{(1,-1)}(\sigma) S_{(1,1)}(\sigma') 
      L_{(1,-1)}(\gamma) L_{(1,1)}(\gamma') L_{(l_1,l_2)}(M_1) \\\\
       & = \frac{1}{|H^1(M_3,\Xi^i)|}\sum_{\substack{(\gamma,\gamma')\in H_1(M_3,\Xi^i),\\
    (\sigma,\sigma')\in H_2(M_3,\Xi^i)}}  \exp{\left(-\frac{2\pi i}{N} (\langle \sigma,\gamma-\gamma' + (l_1-l_2)M_1 \rangle + \langle \sigma', \gamma+\gamma' + (l_1+l_2)M_1\rangle ) \right)} 
     \\ 
    & \hspace{5.2cm} \times S_{(1,-1)}(\sigma) S_{(1,1)}(\sigma') 
      L_{(1,-1)}(\gamma+l_1M_1) L_{(1,1)}(\gamma'+l_2M_1) L_{(-l_2,l_1)}(M_1) \\\\
      & = D^i(M_3) \times L_{(-l_2,l_1)}(M_1).
\end{split}
\end{equation}
The equal-time commutation relation \eqref{eq.comuteLS} has been used in the second line, and the quantum torus algebra \eqref{eq.qtaLS} and 
\begin{equation} \label{eq.LLL2}
     L_{(l_1,l_2)}(M_1) = L_{(-l_2,l_1)}(M_1) \times L_{(l_1+l_2,l_2-l_1)}(M_1) = L_{(-l_2,l_1)}(M_1) \times L^{l_1}_{(1,-1)}(M_1)L^{l_2}_{(1,1)}(M_1)
\end{equation}
have been used in the third line. Therefore, the fusion rule reproduces \eqref{eq.FusionD1}.

\paragraph{Fusion with $S_{s_1,s_2}$.}

The fusion between the surface operator $S_{(s_1,s_2)}(M_2)$ and $D^i(M_3)$ is  
\begin{equation}
\begin{split}
     & \;\;\;\;\; S_{(s_1,s_2)}(M_2) \times D^i(M_3) \\
     & = \frac{1}{|H^1(M_3,\Xi^i)|}\sum_{\substack{(\gamma,\gamma')\in H_1(M_3,\Xi^i),\\
    (\sigma,\sigma')\in H_2(M_3,\Xi^i)}}  \exp{\left(-\frac{2\pi i}{N} (\langle \sigma,\gamma-\gamma' \rangle + \langle \sigma', \gamma+\gamma' \rangle) \right)} 
    S_{(s_1,s_2)}(M_2) S_{(1,-1)}(\sigma) \\ 
    & \hspace{5.2cm} \times S_{(1,1)}(\sigma') 
      L_{(1,-1)}(\gamma) L_{(1,1)}(\gamma') \\\\
    & = \frac{1}{|H^1(M_3,\Xi^i)|}\sum_{\substack{(\gamma,\gamma')\in H_1(M_3,\Xi^i),\\
    (\sigma,\sigma')\in H_2(M_3,\Xi^i)}}  \exp{\left(-\frac{2\pi i}{N} (\langle \gamma,\sigma+\sigma' \rangle + \langle \gamma', \sigma'-\sigma \rangle) \right)} S_{(1,-1)}(\sigma+s_1 M_2) \\ 
    & \hspace{5.2cm} \times S_{(1,1)}(\sigma'+l_2 M_2) 
     S_{(-s_2,s_1)}(M_2) L_{(1,-1)}(\gamma) L_{(1,1)}(\gamma') \\\\     
     & = \frac{1}{|H^1(M_3,\Xi^i)|}\sum_{\substack{(\gamma,\gamma')\in H_1(M_3,\Xi^i),\\
    (\sigma,\sigma')\in H_2(M_3,\Xi^i)}}  \exp{\left(-\frac{2\pi i}{N} (\langle \gamma,\sigma+\sigma' + (s_1 + s_2) M_2 \rangle + \langle \gamma', \sigma'-\sigma + (s_2-s_1)M_2 \rangle ) \right)}  \\ 
    & \hspace{5.2cm} \times S_{(1,-1)}(\sigma+s_1 M_2) S_{(1,1)}(\sigma'+l_2 M_2) 
      L_{(1,-1)}(\gamma) L_{(1,1)}(\gamma') S_{(-s_2,s_1)}(M_2) \\\\ 
      & =D^i_3(M_3) \times S_{(-s_2,s_1)}(M_2).
\end{split}
\end{equation}
The quantum torus algebra \eqref{eq.qtaLS} and a similar relation to \eqref{eq.LLL} for $S_{(s_1,s_2)}(M_2)$ have been applied in the second line. The commutation relation \eqref{eq.comuteLS} has been used in the third line. These fusion rules give the expected action of the $\bZ_4^{\text{EM}}$ symmetry on line and surface defects \eqref{eq.emls}. In this way, we have obtain the fusion rule in \eqref{eq.fusionD2}. 

\paragraph{Charge Conjugation Defect.}

The fusion of the condensation defect $D^i(M_3)$ with itself defines the charge conjugation defect $C(M_3)$. This is given by 
\begin{equation}
\begin{split}
     & \;\;\;\;D^i(M_3) \times D^i(M_3) \\
     & = \frac{1}{|H^1(M_3,\Xi^i)|^2}\sum_{\substack{(\gamma_1,\gamma'_1),(\gamma_2,\gamma'_2) \in H_1(M_3,\Xi^i),\\
    (\sigma_1,\sigma_1'),(\sigma_2,\sigma_2') \in H_2(M_3,\Xi^i)}} 
    S_{(1,-1)}(\sigma_1) S_{(1,1)}(\sigma'_1)  L_{(1,-1)}(\gamma_1) L_{(1,1)}(\gamma'_1) 
    S_{(1,-1)}(\sigma_2) S_{(1,1)}(\sigma'_2) 
    \\ 
    &  \hspace{0.5cm} \times    L_{(1,-1)}(\gamma_2) L_{(1,1)}(\gamma'_2) 
    \exp{\left(-\frac{2\pi i}{N} (\langle \sigma_1,\gamma_1-\gamma'_1 \rangle + \langle \sigma'_1, \gamma_1+\gamma'_1 \rangle+ \langle \sigma_2,\gamma_2-\gamma'_2 \rangle + \langle \sigma'_2, \gamma_2+\gamma'_2 \rangle) \right)} \\\\
    & = \frac{1}{|H^1(M_3,\Xi^i)|}\sum_{\substack{(\gamma_1,\gamma'_1) \in H_1(M_3,\Xi^i),\\
    (\sigma_1,\sigma_1') \in H_2(M_3,\Xi^i)}} 
    \exp{\left(-\frac{2\pi i}{N} (\langle \sigma_1,\gamma_1 \rangle+\langle \sigma'_1,\gamma'_1 \rangle) \right)}
    S_{(1,-1)}(\sigma_1) S_{(1,1)}(\sigma'_1)  \\
    &\hspace{5.8cm} \times L_{(1,-1)}(\gamma_1) L_{(1,1)}(\gamma'_1) 
\end{split}
\end{equation}
The quantum torus algebra \eqref{eq.qtaLS} and the commutation relation \eqref{eq.comuteLS} have been used in the second line. By changing variables $\sigma_1 \to \sigma_1 -\sigma_2$, $\gamma_1 \to \gamma_1 - \gamma_2$, $\sigma'_1 \to \sigma'_1 -\sigma'_2$, $\gamma'_1 \to \gamma'_1 - \gamma'_2$, and integrating out $\sigma_2$, $\sigma'_2$,  $\gamma_2$ and $\gamma'_2$, we obtain the definition of the charge conjugation defect in \eqref{eq.symTFTC}. We have shown the fusion rule presented in \eqref{eq.fusionD3}. 

\paragraph{Invertibility of $D(M_3)$.}

We will study the fusion of four copies of the condensation defect $D(M_3)$. Equivalently, we consider the fusion between two charge conjugation defects 
\begin{equation}
\begin{split}
     & \;\;\;\;C(M_3) \times C(M_3) \\
    & = \frac{1}{|H^1(M_3,\Xi^i)|^2}\sum_{\substack{(\gamma_1,\gamma'_1),(\gamma_2,\gamma'_2) \in H_1(M_3,\Xi^i),\\
    (\sigma_1,\sigma_1'),(\sigma_2,\sigma_2') \in H_2(M_3,\Xi^i)}} 
    \exp{\left(-\frac{2\pi i}{N} (\langle \sigma_1,\gamma_1 \rangle+\langle \sigma'_1,\gamma'_1 \rangle+\langle \sigma_2,\gamma_2 \rangle+\langle \sigma'_2,\gamma'_2 \rangle) \right)} \\
    &\hspace{0.5cm} \times S_{(1,-1)}(\sigma_1) S_{(1,1)}(\sigma'_1)  L_{(1,-1)}(\gamma_1) L_{(1,1)}(\gamma'_1) S_{(1,-1)}(\sigma_2) S_{(1,1)}(\sigma'_2)  L_{(1,-1)}(\gamma_2) L_{(1,1)}(\gamma'_2)\\\\ 
     & = \frac{1}{|H^1(M_3,\Xi^i)|^2}\sum_{\substack{(\gamma_1,\gamma'_1),(\gamma_2,\gamma'_2) \in H_1(M_3,\Xi^i),\\
    (\sigma_1,\sigma_1'),(\sigma_2,\sigma_2') \in H_2(M_3,\Xi^i)}} S_{(1,-1)}(\sigma_1+\sigma_2) S_{(1,1)}(\sigma'_1+\sigma'_2)  L_{(1,-1)}(\gamma_1+\gamma_2)
    \\
    &\hspace{0.5cm} \times  L_{(1,1)}(\gamma'_1+\gamma'_2)  \exp{\left(-\frac{2\pi i}{N} (\langle \sigma_1,\gamma_1 \rangle+\langle \sigma'_1,\gamma'_1 \rangle+\langle \sigma_2,\gamma_2 \rangle+\langle \sigma'_2,\gamma'_2 \rangle+ 2\langle \sigma_2,\gamma_1 \rangle + 2\langle \sigma'_2,\gamma'_1 \rangle) \right)}\\\\
     & = \frac{1}{|H^1(M_3,\Xi^i)|^2}
     \sum_{\substack{(\gamma_1,\gamma'_1),(\gamma_2,\gamma'_2) \in H_1(M_3,\Xi^i),\\
    (\sigma_1,\sigma_1'),(\sigma_2,\sigma_2') \in H_2(M_3,\Xi^i)}}
    \exp{\left(-\frac{2\pi i}{N} (\langle \sigma_1, \gamma_1-\gamma_2 \rangle+ \langle \sigma'_1, \gamma'_1-\gamma'_2 \rangle + \langle \sigma_2,\gamma_1\rangle + \langle \sigma'_2,\gamma'_1 \rangle) \right)}
    \\
    &\hspace{0.5cm} \times   S_{(1,-1)}(\sigma_1) S_{(1,1)}(\sigma'_1)  L_{(1,-1)}(\gamma_1)L_{(1,1)}(\gamma'_1) 
    = 1.
\end{split}
\end{equation}
Again, the quantum torus algebra \eqref{eq.qtaLS} and the commutation relation \eqref{eq.comuteLS} have been used in the second line. After the following change of variables $\sigma_1 \to \sigma_1 -\sigma_2$, $\gamma_1 \to \gamma_1 - \gamma_2$, $\sigma'_1 \to \sigma'_1 -\sigma'_2$, $\gamma'_1 \to \gamma'_1 - \gamma'_2$, and integrating out $\sigma_2$, $\sigma'_2$,  $\gamma_2$ and $\gamma'_2$, we confirm that the condensation defect defined in \eqref{eq.condefect.N} satisfies
\begin{equation}
    D^i(M_3)\times D^i(M_3)\times D^i(M_3)\times D^i(M_3)=1
\end{equation}
and thus realizes the $\mathbb{Z}_4^\text{EM}$ symmetry of \eqref{eq.emsymmetry}.

\paragraph{Orientation Reversal of $D^i(M_3)$.}

The orientation reversal of the condensation defect is 
\begin{equation}
\begin{split}
    \overline{D}^i(M_3) 
    = \frac{1}{|H^1(M_3,\Xi^i)|}\sum_{\substack{(\gamma,\gamma')\in H_1(M_3,\Xi^i),\\
    (\sigma,\sigma')\in H_2(M_3,\Xi^i)}} & \exp{\left(-\frac{2\pi i}{N} (\langle \sigma,\gamma+\gamma' \rangle + \langle \sigma', \gamma'-\gamma \rangle ) \right)}S_{(1,-1)}(\sigma)   \\
    &  \times  S_{(1,1)}(\sigma') L_{(1,-1)}(\gamma) L_{(1,1)}(\gamma').
\end{split}
\end{equation}
With this definition of $\overline{D}^i(M_3)$, one can show that 
\begin{equation}
\begin{split}
     & \;\;\;\;\overline{D}^i(M_3) \times D^i(M_3) \\
    & = \frac{1}{|H^1(M_3,\Xi^i)|^2}\sum_{\substack{(\gamma_1,\gamma'_1),(\gamma_2,\gamma'_2) \in H_1(M_3,\Xi^i),\\
    (\sigma_1,\sigma_1'),(\sigma_2,\sigma_2') \in H_2(M_3,\Xi^i)}} S_{(1,-1)}(\sigma_1) S_{(1,1)}(\sigma'_1)  L_{(1,-1)}(\gamma_1) L_{(1,1)}(\gamma'_1)S_{(1,-1)}(\sigma_2) S_{(1,1)}(\sigma'_2)
     \\
    & \times    L_{(1,-1)}(\gamma_2) L_{(1,1)}(\gamma'_2)
    \exp{\left(-\frac{2\pi i}{N} (\langle \sigma_1,\gamma_1+\gamma'_1 \rangle + \langle \sigma'_1, \gamma'_1-\gamma_1 \rangle + \langle \sigma_2,\gamma_2-\gamma'_2 \rangle + \langle \sigma'_2, \gamma_2+\gamma'_2 \rangle) \right)}
    \\\\ 
    & = \frac{1}{|H^1(M_3,\Xi^i)|^2}\sum_{\substack{(\gamma_1,\gamma'_1),(\gamma_2,\gamma'_2) \in H_1(M_3,\Xi^i),\\
    (\sigma_1,\sigma_1'),(\sigma_2,\sigma_2') \in H_2(M_3,\Xi^i)}} S_{(1,-1)}(\sigma_1+\sigma_2) S_{(1,1)}(\sigma'_1+ \sigma'_2)  L_{(1,-1)}(\gamma_1+\gamma_2) L_{(1,1)}(\gamma'_1+\gamma'_2)
     \\
    & \times    
    \exp{\left(-\frac{2\pi i}{N} (\langle \sigma_1,\gamma_1+\gamma'_1 \rangle + \langle \sigma'_1, \gamma'_1-\gamma_1 \rangle + \langle \sigma_2,\gamma_2-\gamma'_2+ 2 \gamma_1 \rangle + \langle \sigma'_2, \gamma_2+\gamma'_2 + 2\gamma'_1 \rangle ) \right)}=1.
    \\\\ 
\end{split}
\end{equation}
In this way, we have derived the fusion rule in \eqref{eq.fusionD5}.

\bibliographystyle{JHEP}
\bibliography{main}

\end{document}